\documentclass{ws-procs975x65}

\renewcommand{\theequation}{\thesection.\arabic{equation}}
\begin{document}

\title{
\vspace{-3mm}
\rightline{\scriptsize HU-EP-04/35,
DFUPG-2004/98}
\vspace{8mm}
3D Georgi--Glashow model and confining strings\\[1mm] at zero and 
finite temperatures}

\author{Dmitri Antonov\footnote{\uppercase{P}ermanent address: 
        \uppercase{ITEP}, \uppercase{B}. \uppercase{C}heremushkinskaya 25, 
\uppercase{RU}-117 218 \uppercase{M}oscow, \uppercase{R}ussia.}
}

\address{
        Institute of Physics, Humboldt University of Berlin, \\
        Newtonstr. 15, 12489 Berlin, Germany}

\author{Maria Cristina Diamantini}

\address{Dipartimento di Fisica dell'Universit\`a 
        di Perugia and \\ 
        I.N.F.N. sezione di Perugia, \\
        Via A. Pascoli, 06123 Perugia, Italy} 

\maketitle

\abstracts{
In this review, we  discuss the confining and finite-temperature properties 
of the 3D SU($N$) Georgi--Glashow model, and of 4D compact QED. 
At zero temperature, we derive string representations of both theories, thus constructing 
the SU($N$)-version of  Polyakov's theory of confining strings. We discuss the 
geometric properties of confining strings, as well as the 
appearance of the string $\theta$-term from the field-theoretical one in 4D, and $k$-string 
tensions at $N>2$.
In particular, we point out the relevance of negative stiffness for stabilizing
confining strings, an effect recently re-discovered in material science.
At finite temperature, we present a derivation of the confining-string 
free energy and show that, at the one-loop level and 
for a certain class of string models in the large-D limit, it matches that of QCD 
at large $N$. 
This crucial matching is again a consequence of the negative stiffness.
In the discussion of the finite-temperature properties of the 3D Georgi--Glashow model, 
in order to be closer to QCD,
we mostly concentrate at the effects produced by some extensions
of the model by external matter fields, such as dynamical fundamental quarks or photinos,
in the supersymmetric generalization of the model. 
}

\newpage

\tableofcontents

\newpage

{\sl %
Many papers related to this review would have never been written without Ian Kogan's seminal
ideas on the finite-temperature phase transitions in string theory and in the 3D Georgi--Glashow model. 
The sudden death of Ian Kogan was a loss for physics and a 
personal tragedy for everybody who had known him, in particular for the 
authors of this manuscript. In the personal sense, Ian was a very bright man with a warm heart, and he will be 
deeply missed as a good friend. We nevertheless strongly 
believe that Ian's ideas will survive over decades, inspiring many new
generations of physicists.
}%

\vspace{-3mm}
\setcounter{equation}{0}
\section{Introduction}

During last 30 years, 
the problem of quark confinement remains as a great challenge not only to 
QCD theorists, but to the whole theoretical high-energy physics community. 
In general, confinement can be defined as 
the absence in the spectrum   
of a certain field theory of physical $\left|{\rm in}\right>$ and  
$\left|{\rm out}\right>$ states of  
some particles, whose fields are 
nevertheless present in the Lagrangian of that theory (see e.g.~\cite{revs, digiacomo, polbook}
for reviews and books on confinement). With regard to QCD, this definition means the 
absence of asymptotic quark and gluon states, i.e. states which carry color.\footnote{\,An attempt to construct a momentum-space interpretation of 
this phenomenon has been done 
(see e.g. Ref.~\cite{nishi}). In this review, we will, however, prefer to deal with 
the conventional space-time picture, which enables one to directly operate with such notions as
the vacuum correlation length or potential between color particles.} 
This fact is reflected in the linear growth with the distance of the potential 
between two color particles, as well as in the logarithmic growth of the strong-coupling constant.
The latter makes the standard perturbative diagrammatic techniques inapplicable 
at distances larger than ${\mathcal O}\left(\Lambda_{\rm QCD}^{-1}\right)$. However, such distances 
are of primary interest, since only there physically observable colorless states of hadrons are formed,
whereas at smaller distances one deals with the unphysical colored states -- quarks, gluons, 
and ghosts. A question, which may naturally be posed at this point, is whether the QCD Lagrangian,
operating with these colored states only, eventually yields the correct description of colorless
degrees of freedom as well. A reliable indication that this is really the case comes from the 
simulations of the QCD Lagrangian on a lattice (for recent reviews see e.g.~\cite{digiacomo, bali, latrev}),
which yield a reasonably good description of hadronic spectra. However, as of today, a systematic
{\it analytic} way of 
describing large-distance effects in QCD in terms of microscopic (colored) degrees of freedom 
is unfortunately lacking. The breakdown of the perturbative expansion at the large distances under 
discussion naturally
introduces the notion {\it nonperturbative} for these effects, as well as for the techniques 
which physicists attempt to invent for their description. 

It is actually not a great fault that the perturbative expansion is inapplicable at large distances,
since, being formulated in terms of colored states, it is anyway not suitable to keep gauge invariance 
under control. Indeed, an individual diagram is always gauge-dependent, merely because a certain gauge fixing
should be performed before one starts to compute diagrams.
Contrary to that, hadronic states, being colorless, are gauge-invariant. Moreover, since the QCD vacuum 
possesses an unbroken gauge symmetry,\footnote{\,We do not discuss in this 
review the modern line of 
research devoted to the so-called color 
superconductivity~\cite{colorsuper}. In the color superconducting phase 
of QCD, which takes place 
at high baryonic densities corresponding to the values of the quark chemical potential 
$>{\mathcal O}\left(400~{\rm MeV}\right)$, the gauge symmetry is spontaneously broken.\\[-2mm] ~} only gauge-invariant 
amplitudes, being averaged over the vacuum, yield a nonvanishing 
result.\footnote{\,To get a description of large-distance 
effects in terms of microscopic degrees of freedom, it looks natural 
to integrate over the 
high-momentum modes in the same way as is usually done in a derivation 
of renormalization-group (RG)
equations in statistical mechanics. For instance, this approach gives 
the correct RG
behavior of the running strong-coupling constant (at the one-loop 
level)~\cite{polbook}. However, it typically
suffers from the above-mentioned problem of violation of gauge 
invariance, as far as RG equations for 
Green functions are concerned. Recent progress in this direction of research 
is 
concentrated around the concept of the so-called exact RG 
flow (see e.g. Ref.~\cite{ergf} for a recent review).}
Although the gauge dependence disappears when 
one sums up contributions of individual diagrams to a certain gauge-invariant quantity, it would be 
more natural to have an expansion already operating with such quantities at any intermediate stage.
Such an expansion, which is the expansion in the inverse number of colors, 
has been proposed by 't~Hooft~\cite{thooft} and further developed in the framework of loop 
equations~\cite{loopeqs} 
(see e.g.~\cite{largen} for reviews). This is a classic example of a nonperturbative approach to QCD.

Another nonperturbative phenomenon in QCD, which is of the same fundamental importance as confinement, 
and whose complete analytic understanding is also still lacking, is the spontaneous chiral symmetry breaking 
(SCSB). This is the symmetry of the QCD Lagrangian with $N_f$ massless flavors 
under the global transformations, which are the U($N_f$)$\times$ 
U($N_f$) independent rotations of left- and right-handed quark fields. 
These are equivalent to the 
independent vector and axial U($N_f$) rotations of the full four-component 
Dirac spinors, under which the QCD Lagrangian remains invariant as well. 
At the same time, the axial transformations mix states with different 
$P$-parities. Therefore, were the chiral symmetry  
unbroken, one would observe parity degeneracy of all the states whose 
other quantum numbers are the same. The observed splittings between 
such states are, however, 
too large to be explained by the small quark masses. Namely, this splitting is of the order of hundreds of 
MeV, whereas the current masses of light $u$- and $d$-quarks are of the 
order of a few MeV.\footnote{\,The current mass of the $s$-quark, which  
is around 
150~MeV, is still smaller than the typical splitting 
values by at least a factor of three.\\[-2mm] ~} This observation tells us that 
the chiral symmetry of the QCD Lagrangian is broken down spontaneously. 
The phenomenon of SCSB
naturally leads to the appearance of light pseudoscalar Goldstone 
bosons, whose role is played by pions, which are indeed the lightest 
of all the hadrons. 
The order parameter of SCSB is the chiral quark condensate 
$\left<\bar\psi\psi\right>\simeq -(250{\,}{\rm MeV})^3$.
Its appearance is quite natural in QCD, where, due to the strong attraction between quarks and antiquarks,
it costs very little energy to create a massless quark-antiquark pair. Having zero total momentum and 
angular momentum, such pairs carry net chiral charge,  
hence is the notion ``chiral condensate''.

Since the early days of QCD, when it has been realized that its nonperturbative phenomena,
confinement and SCSB, are of fundamental importance, 
numerous nonperturbative approaches have been proposed in an attempt to describe one or other of these 
two phenomena in a controllable way.\footnote{\,Unfortunately, practically 
no approach exist, which 
describes both phenomena. Partially, that is because it is not yet fully 
accepted that these phenomena 
are related to each other microscopically, i.e. 
that the same vacuum configurations are responsible for both phenomena.
For instance, as has been shown in Ref.~\cite{konishi}, 
this is not so in the case of the so-called Faddeev-Niemi effective action \cite{fadnie}. An example of a  
nonperturbative approach, where the 
two phenomena are related to each other, is the stochastic vacuum model of QCD~\cite{revs}, which does not
refer to a particular microscopic vacuum configuration.\\[-2mm] ~} On the side of confinement, these approaches include
e.g. the already mentioned expansion in the large number of colors~\cite{thooft}, loop equations~\cite{loopeqs},
the stochastic vacuum model~\cite{revs}, method of Abelian 
projections \cite{AP}.\footnote{\,A comprehensive 
lattice analysis of the latter has recently been performed~\cite{ap} (see Ref.~\cite{ap1} for reviews).}
On the side of SCSB, the classic approach is the one based on the Nambu--Jona-Lasinio models~\cite{njl}, 
as well as 
on the related nonlinear chiral meson Lagrangians~\cite{cml} (the latter approach has further been 
developed in~\cite{cml1}; for a review see~\cite{cmlr}). There also exist microscopic models of SCSB in QCD, based on 
instantons~\cite{inst} or dyons~\cite{dy0}, which have been put forward in Refs.~\cite{sci} and~\cite{dy}, 
respectively.

Although a lot of physical effects have been captured by the above-mentioned methods, 
unfortunately none of them provides the full solution of QCD. 
The concept of ``full solution'' has several facets.
According to the conventional understanding, the solution of a certain field theory means 
a prescription of how to compute an 
arbitrary gauge-invariant vacuum amplitude. In QCD, the results should be in an agreement with the experimental and 
lattice data on the properties of hadrons. The gluonic and quark averages should then give a correct quantitative 
description of confinement and SCSB, respectively. The values of local averages (condensates) and relations between these 
quantities should also agree with those known from QCD sum rules~\cite{sumrules}. Furthermore,
the standard diagrammatic expansion
should be reproducible (giving, in particular, asymptotic freedom), 
and the nonperturbative dimensionful quantities should be 
made expressible in terms of the only dimensionful QCD parameter, $\Lambda_{\rm QCD}$.
Moreover, some constraints, which are accepted to be rigorous in QCD, should be obeyed. 
For instance, the above-mentioned expansion in the large number of 
colors should be correctly reproducible
and should respect the large-$N$ loop equation. 
The solution of QCD should also accommodate classical vacuum configurations of the gluodynamics 
action, such as instantons.

In this review, we are going to discuss the 3D Georgi--Glashow (GG) model, which is the QCD-related model possessing 
the property of confinement~\cite{pol}. 
Our primary goal will therefore be the study of confinement (and not of the SCSB), as well as of the 
deconfining phase transition at finite temperatures.
The advantage of the 3D GG model
with respect to QCD is that confinement in it takes place in the weak-coupling regime. 
It turns out that, already in this regime, the vacuum (i.e. the 
ground state) of this model is nonperturbative, being populated by 't~Hooft--Polyakov monopoles~\cite{mon},
which provide the permanent confinement of external fundamental charges. 
As a guiding principle of our analysis we will use the string picture of 
confinement, therefore let us briefly discuss it. 

In QCD, the linearly rising confining interquark potential is associated to a string-like configuration of the 
gluonic field between quarks, usually called the QCD string. Indeed, the energy of a string grows linearly 
with its length, $E(R)=\sigma R\,$.\footnote{\,In this review, we only briefly discuss 
the phenomenon of string breaking (see the end of subsection~4.2). 
String breaking always happens at a certain distance if dynamical matter fields, 
transforming by the same representation of the gauge group 
as the confined external ones, are present.} According to the Regge phenomenology, 
the string energy density $\sigma$, 
called the string tension, is approximately $(440{\,}{\rm MeV})^2$. The string can naturally be called confining
(the notion, which is always used in confining gauge theories other than QCD, such as the 3D GG model), 
since with the increase of the 
distance $R$, it stretches and prevents a quark and an antiquark from the separation to macroscopic distances.
As for any dimensionful quantity in QCD, the string tension should be proportional to the respective power of $\Lambda_{\rm QCD}$.
Namely, 
$$
\sigma\propto\Lambda_{\rm QCD}^2=a^{-2}\exp\Bigg[-\int\limits_{}^{g^2(a^{-2})}
\frac{d{g'}^2}{{g'}^2\beta({g'}^2)}\Bigg]\simeq$$
\vspace{-8mm}
\begin{equation}
\label{strtens}
\simeq a^{-2}\exp\bigg[-
\frac{16\pi^2}{\left(\frac{11}{3}N_c-\frac23N_f\right)g^2(a^{-2})}\bigg],
\end{equation}
where $a\to 0$ stands for the inverse UV cutoff (e.g. the lattice 
spacing). Furthermore, ``$\simeq$'' means ``at the one-loop level'', at which
$$
\beta(g^2)\simeq-\left(\frac{11}{3}N_c-\frac{2}{3}N_f\right)\frac{g^2}{16\pi^2}\ .
$$
As is seen explicitly from Eq.~(\ref{strtens}), all the coefficients in the expansion of 
$\sigma$ in (positive) powers of $g^2$ vanish, which means that the QCD string
is indeed an essentially nonperturbative object. The string should therefore be produced by some nonperturbative 
background fields. On top of these, however, one expects to have some quantum fluctuations of the gauge field, which give rise
to the string excitations.

The confining quark-antiquark potential corresponds to the so-called area law of 
the Wilson loop:\,\footnote{\,It should be compared with the 
perimeter law, 
$$\left<W(C)\right>\stackrel{{|C|\to\infty}}{\longrightarrow}
{\rm e}^{-{\rm const}{\,}\cdot{\,}|C|}\,,$$
which corresponds to the Coulomb potential and is found in non-confining theories, e.g. (non-compact) QED.}
$$
\left<W(C)\right>
\equiv\frac{1}{N_c}\left<{\rm tr}{\,}{\mathcal P}{\,}
\exp\left(ig\oint\limits_{C}^{}A_\mu^a T^adx_\mu\right)\right>
\stackrel{{|C|\to\infty}}{\longrightarrow}
{\rm e}^{-\sigma\left|\Sigma_{\rm min}(C)\right|}\ .
$$
Here, $\Sigma_{\rm min}(C)$ 
is the surface of the minimal area, bounded by the trajectory $C$ of the quark-antiquark pair, and $|\ldots|$ means
either a length or an area.
The confining string, which sweeps out the minimal surface, can naturally be viewed as a product of the above-mentioned 
strong background fields. Instead, quantum fluctuations around these enable the string to sweep out with a 
nonvanishing probability any other surface $\Sigma(C)$, different from $\Sigma_{\rm min}(C)$.
To derive the string representation of QCD would be to give sense to the formula 
\begin{equation}
\label{00}
\left<W(C)\right>\equiv\left<W(\Sigma_{\rm min}(C))\right>=\sum\limits_{\Sigma(C)}^{}
{\rm e}^{-S[\Sigma(C)]}\ .
\end{equation}
Here, $\sum\limits_{\Sigma(C)}^{}$ and 
$S[\Sigma(C)]$ stand for a certain sum over string world sheets
and a string effective action, both of which are yet unknown in QCD.

The first problem one encounters in trying to determine $S[\Sigma(C)]$ is that
fundamental strings~\cite{witten} can be quantized only in critical
dimensions: a consistent quantum theory describing 
strings out of the critical dimensions has not yet been found. The
simplest model, the Nambu--Goto string, can
be quantized only in space-time dimension $D = 26$ or $D \leq 1$ because of the
conformal anomaly. It is appropriate to describe 
an effective string picture for confinement in  QCD , but is inappropriate to
describe fundamental smooth strings
dual to QCD~\cite{poly1}, since large Euclidean world sheets are crumpled.
The picture of a fundamental string theory dual to QCD is strongly supported by a
recent lattice calculation by L\"uscher and Weisz~\cite{luscher}, where 
evidence of a  string behavior in the static quark-antiquark potential 
has been found down to distances of 0.5 fm.

In the rigid-string action~\cite{fi, more_rigid}, the marginal term proportional to
the square of the extrinsic curvature, introduced to avoid crumpling, turns
out to be infrared irrelevant and, thus, unable to provide smooth surfaces.

Recent progress in this field is based on a new type of action. In its
local formulation~\cite{tq,cs}, the string action is induced by an
antisymmetric tensor field. This action realizes explicitly
the necessary zig-zag invariance of confining strings~\cite{cs,zigzag}. It can be
derived without extra assumptions~\cite{theta} for the confining phase of compact U(1) gauge
theories~\cite{polbook}.
An alternative approach to the induced string action was originally proposed in~\cite{poly2}, 
and is based on a five-dimensional, curved space-time string action with
the quarks living on a four-dimensional horizon~\cite{alvarez}.
The formulation of the string theory in the five-dimensional curved space-time
is closely related to the AdS/CFT (Anti~de~Sitter/Conformal Field Theory) correspondence~\cite{kleb}. In fact, with a
special choice of the metric in the curved space, one recovers the ${\rm AdS}_5$
space, thereby providing a string theory description of a conformal gauge
theory~\cite{kleb}.

The main characteristic of the effective string action obtained by
integrating out the tensor field is a non-local interaction
with {\it negative stiffness},
that can be expressed as a
derivative expansion of the interaction between surface elements.
To perform an analytic analysis of the geometric properties of these
strings, this expansion must be truncated: this clearly makes the model
non-unitary, but in a spurious way. Moreover, since the stiffness is negative, a
stable truncation must, at least, include a sixth-order term in the derivatives.
The role of negative stiffness, as first pointed out in~\cite{theta,cris1}
is crucial. It is in fact the sixth-order term, forced by the negative stiffness,
that suppresses the formation of spikes on the surfaces and 
leads to a smooth surface in the large-D approximation.
In fact, in~\cite{cris1,cri} it has been shown that, in the large-D
approximation, this model has an infrared fixed point at zero stiffness, corresponding to a
tensionless smooth string whose world sheet has Hausdorff dimension 2, exactly
the desired properties to describe QCD flux tubes. As first
noticed in~\cite{stringa, stringa1}, the long-range orientational order in 
this model is due to an
antiferromagnetic interaction between normals to the surface, a mechanism
confirmed by numerical simulations~\cite{chern}.
The presence of the infrared fixed point does not depend on the
truncation~\cite{cri} and it is present for all ghost- and tachyon- free
truncations. Moreover, the effective theory
describing the infrared behavior of the confining string is
a conformal field theory with central charge $c=1$.

Another important feature of the negative-stiffness model is its
high-temperature behavior. Contrary to all previous string models for QCD, it is able to
reproduce the large-$N$ QCD behavior, found by Polchinski and Yang in~\cite{polc2},
in both {\it  sign and reality properties} \cite{Tfinita}.

It is remarkable that the role of negative stiffness,
while first discovered in the context of strings~\cite{theta, cris1} and membranes~\cite{dkt}, 
has been rediscovered, and actually experimentally tested, 
also in material science~\cite{lakes}. In fact, it has been found that
composites with negative stiffness inclusions have higher overall
stiffness than that of their
constituents. Such composites find applications in which high stiffness and
damping are needed, permitting extreme properties not previously anticipated.

Equation (\ref{00}) would clearly be a generalization of a path-integral representation 
for a propagator of a particle, for example boson, $\left<\phi(x)\phi(x')\right>=\sum\limits_{P_{xx'}}^{}
{\rm e}^{-S[P_{xx'}]}$, where $P_{xx'}$ is a path connecting the points $x$ and $x'$.    
In particular, within this analogy, the role of the classical trajectory of 
a particle would be played by $\Sigma_{\rm min}(C)$. For a point-like particle, the measure in the 
sum over paths is known and depends only on the dimensionality of the space-time. The world-line action
$S[P_{xx'}]$ can also be evaluated, either analytically [for certain potentials $V(\phi)$ or external
gauge fields if $\phi$ is charged], or using Feynman's variational method.
On the string side, a derivation of the measure in the sum over world sheets has been discussed in Ref.~\cite{emil}
for the case of the Abelian Higgs model in the London limit. As will be seen below, in a certain case the 
string effective action derivable in the 3D GG model is the same as that of the London limit 
of the dual Abelian Higgs model. The summation over string world sheets is, however, realized in these two
models in different ways. Namely, in the dual Abelian Higgs model it stems from the integration over 
singularities of the dual Higgs field (vortex cores), which is already present in the original partition function.
Instead, the 3D GG model does not contain any dual Abelian Higgs field, and the summation over 
world sheets in this model is realized by means of the resulting string
effective action itself~\cite{cs}. 
Numerous investigations of this action have further been performed (see e.g. Refs.~\cite{theta, stringa, stringa1}),
and some of these will be discussed in this review.
A separate section will be devoted to the  
finite-temperature properties of confining strings~\cite{Tfinita}.

For the purpose of the study of the so-called $k$-strings and merely to be closer to real QCD, we will
deal with the SU($N$)-generalization of the standard SU(2) GG model, which will be introduced 
in the next section. As for the $(N=2)$ 3D GG model, it is a
classic example~\cite{pol} of a theory which allows for an analytic
description of confinement. As has already been mentioned, confinement in this model
is due to the plasma of
point-like magnetic monopoles, which produce random magnetic fluxes
through the contour of the Wilson loop. In the weak-coupling regime of
the model, this plasma is dilute, and
the interaction between monopoles is Coulombic, being induced by the
dual-photon exchanges. Since the energy of a single monopole is a
quadratic function of its flux, it is energetically favorable for the
vacuum to support a configuration of two monopoles of unit charge (in
the units of the magnetic coupling constant, $g_m$), rather than a
single, doubly-charged monopole. Owing to this fact, monopoles of unit
charge dominate in the vacuum, whereas monopoles of higher charges
tend to dissociate to them.  Summing over the grand canonical
ensemble of monopoles of unit charge, interacting with each other by
the Coulomb law, one arrives at an effective low-energy theory, which
is a 3D sine-Gordon theory of a dual photon. The latter acquires a
mass (visible upon the expansion of the cosine potential) by means of
the Debye screening in the Coulomb plasma. The appearance of this
(exponentially small) mass and, hence, of a finite (albeit
exponentially large) vacuum correlation length is crucial for the
generation of the fundamental string tension, i.e. for the
confinement of an external fundamental matter.  It is worth noting
that a physically important interpretation of these ideas in terms of
spontaneous breaking of magnetic $Z_2$ symmetry has been presented in
reviews~\cite{Alik} and Refs. therein.

While the confining properties of the 3D GG model have been well known since  
Polyakov's pioneering paper~\cite{pol}, the finite-temperature properties of this model have been
addressed only recently, starting with the papers~\cite{az, dkkt}. It turns out~\cite{dkkt} that
charged matter fields of W-bosons play the crucial role for the dynamics of the phase transition.
Below we will review this issue and also discuss the influence of other matter fields on the 
finite-temperature properties
of the model. Such fields are either already present in the original Lagrangian (e.g. Higgs~\cite{higgs},
or dual photinos in the supersymmetric generalization of the model~\cite{susy}), or can be included in the framework of 
a certain extension of the model (e.g. massless fundamental quarks~\cite{quarks} 
or heavy fundamental bosonic matter~\cite{dkn}).

The outline of this review is the following. In the next section, we will introduce the 3D GG model in the general SU($N$) case.
In section~3, we will find 
the string tension of the fundamental Wilson loop defined at a flat contour (henceforce called for shortness 
``flat Wilson loop'').
In section~4,
we will develop a theory of confining strings
based on the Kalb--Ramond field, which enables one to deal with non-flat Wilson loops.
Using for concreteness the case of fundamental representation, we will present in subsection~4.1 
two methods by means of which the theory of confining strings can be derived.
In subsection~4.2, the case of the adjoint Wilson loop will be considered, and the corresponding theory of confining strings
will be constructed in the large-$N$ limit. In subsection~4.3, we will study the spectrum of $k$-strings, i.e. strings between
sources in (higher) representations with a  
nonvanishing $N$-ality. These sources  
carry a charge $k$ with respect
to the center of the gauge group $Z_N$ and can be seen as a superposition
of $k$ fundamental charges. Clearly, the spectrum of such strings is an important ingredient for the 
complete description of the confining dynamics of the 3D GG model.
In section~5,
the SU($N$)-theory of confining strings will be generalized to the SU($N$)-version 
of 4D compact QED (in the continuum limit)
with the field-theoretical $\theta$-term. As has been
found in Refs.~\cite{stringa1, tq}, for the usual compact QED, this term
leads to the appearance of the string $\theta$-term. The latter,
being proportional to the number of self-intersections of the world sheet, might help in
the solution of the problem of crumpling of large world sheets~\cite{fi, polbook}. 
The critical values of $\theta$, at which this happens, will be derived in the general SU($N$)-case for 
fundamental and adjoint representations, as well as for $k$-strings. 
In section~6, various geometric features of confining strings will be studied.
In section~7, the thermodynamics of confining strings 
will be discussed, and a derivation of the one-loop free energy of a string in the large-D limit 
will be presented. Again, we will show that it is the presence of negative stiffness that allows one 
to reproduce the large-$N$ behavior of high-temperature QCD. 

In section~8, we will pass from the thermodynamics of confining 
strings to the thermodynamics of the 3D GG model itself.
After an introduction to this subject in subsection~8.1, 
we will pay particular attention to the influence of various matter fields on the 
dynamics of the deconfining phase transition. In subsections~8.2 and 8.3, we will consider an
approximation where W-bosons are disregarded. Subsection~8.2 will be devoted to the influence of the 
Higgs field (when it is not infinitely heavy) on the RG flow, while in subsection~8.3 we will consider
the model in the presence of external dynamical fundamental quarks. In subsection~8.4, we will first 
discuss the crucial role of W-bosons in the dynamics of the phase transition in the finite-temperature 3D GG model
and then consider the supersymmetric generalization of the model. 
In section~9, the main points discussed in the review will 
be emphasized once again.

\setcounter{equation}{0}
\section{The SU(\boldmath{$N$}) 3D GG model}

The Euclidean action of the SU(2) 3D GG model reads~\cite{pol}
\begin{equation}
\label{SGG}
S=\int
d^3x\left[\frac{1}{4g^2}\left(F_{\mu\nu}^a\right)^2+\frac12
\left(D_\mu\Phi^a\right)^2+ 
\frac{\lambda}{4}\left(\left(\Phi^a\right)^2-\eta^2\right)^2\right],
\end{equation}
where the Higgs field $\Phi^a$ transforms by the adjoint
representation, i.e.
$D_\mu\Phi^a\equiv\partial_\mu\Phi^a+\varepsilon^{abc}A_\mu^b\Phi^c$.
The weak-coupling regime $g^2\ll m_W$, which will be assumed henceforth, parallels the requirement
that $\eta$ should be large enough to ensure spontaneous symmetry
breaking from SU(2) to U(1). At the perturbative level, the spectrum
of the model in the Higgs phase consists of a massless photon, two
heavy, charged $W$-bosons with mass $m_W=g\eta$, and a neutral Higgs
field with mass $m_H=\eta\sqrt{2\lambda}$.

What is, however, more important is the nonperturbative content of the
model, represented by the famous 't~Hooft--Polyakov
monopole~\cite{mon}. It is a solution to the
classical equations of motion, which has the following Higgs- and
vector-field parts:
\vspace{-1mm}
\begin{itemize}
\item
$\Phi^a=\delta^{a3}u(r)$, $u(0)=0$, 
$u(r)\stackrel{r\to\infty}{\longrightarrow}\eta-\exp(-m_Hr)/(gr)$;
\vspace{1mm}
\item
$A_\mu^{1,2}(\vec x)\stackrel{r\to\infty}{\longrightarrow}
{\mathcal O}\left({\rm e}^{-m_Wr}\right)$,
\vspace{-3mm}
$$H_\mu\equiv\varepsilon_{\mu\nu\lambda}\partial_\nu A_\lambda^3=
\frac{x_\mu}{r^3}-4\pi\delta(x_1)\delta(x_2)
\theta(x_3)\delta_{\mu 3};$$
\vspace{-6mm}
\item
as well as the following action $S_0=\frac{4\pi\epsilon}{\kappa}$. Here,
$\kappa\equiv g^2/m_W$ is the weak-coupling parameter, 
$\epsilon=\epsilon\left(m_H/m_W\right)$ is a certain monotonic, slowly
varying function, $\epsilon\ge 1$, $\epsilon(0)=1$ (BPS-limit)~\cite{bps},
\hfill \\
$\epsilon(\infty)\simeq 1.787$~\cite{kirk}.
\end{itemize}

The approximate saddle-point solution (which becomes exact
in the BPS-limit) was found in Ref.~\cite{pol} to be 
$$
S={\mathcal N}S_0+\frac{g_m^2}{8\pi}
\sum\limits_{{a,b=1\atop a\ne b}}^{\mathcal N}
\left(\frac{q_aq_b}{|\vec z_a-\vec z_b|}-
\frac{{\rm e}^{-m_H|\vec z_a-\vec z_b|}}{|\vec z_a-\vec z_b|}\right)+$$
\vspace{-7mm}
$$
~~~+\,{\mathcal O}\left(g_m^2m_H{\rm e}^{-2m_H|\vec z_a-\vec z_b|}\right)
+{\mathcal O}\left(\frac{1}{m_WR}\right),$$
where $m_W^{-1}\ll R\ll |\vec z_a-\vec z_b|$, $gg_m=4\pi$,
$[g_m]=[{\rm mass}]^{-1/2}$. Therefore, while at $m_H\to\infty$, the
usual compact-QED action is recovered, in the BPS-limit one has
$$S\simeq{\mathcal N}S_0+\frac{g_m^2}{8\pi}
\sum\limits_{{a,b=1\atop a\ne b}}^{\mathcal N}
\frac{q_aq_b-1}{|\vec z_a-\vec z_b|}\ ,$$ 
i.e. the interaction of two monopoles doubles for opposite and
vanishes for equal charges. 

When $m_H<\infty$, the summation over the grand canonical ensemble of
monopoles has been performed in Ref.~\cite{dietz} and reads
$$
{\mathcal Z}_{\rm mon}=1+\sum\limits_{{\mathcal N}=1}^{\infty}
\frac{\zeta^{\mathcal N}}{{\mathcal N}!}\prod\limits_{a=1}^{\mathcal N}\int d^3z_a
\sum\limits_{q_a=\pm 1}^{}{\rm e}^{-S}=\int {\mathcal D}\chi{\mathcal D}\psi\times
$$
\vspace{-4mm}
\begin{equation}
\label{ZMON}
\times\exp\left\{-
\int d^3x\left[\frac12(\partial_\mu\chi)^2+\frac12(\partial_\mu\psi)^2
+\frac{m_H^2}{2}\psi^2-2\zeta{\rm e}^{g_m\psi}\cos(g_m\chi)\right]\right\}.
\end{equation}
Here, $\chi$ is the dual-photon field and $\psi$ is the field
additional with respect to compact QED, which describes the Higgs
boson. Furthermore, the monopole fugacity (i.e. the statistical
weight of a single monopole), $\zeta$, has the following
form~\cite{pol}:
\begin{equation}
\label{0}
\zeta=\delta\,\frac{m_W^{7/2}}{g}\,
{\rm e}^{-S_0}\ .
\end{equation}
The function $\delta=\delta\left(m_H/m_W\right)$ is determined by the
loop corrections.  It is known~\cite{ks} that this function grows in
the vicinity of the origin (i.e. in the BPS limit). However, the
speed of this growth is such that it does not spoil the exponential
smallness of $\zeta$ in the weak-coupling regime under study.

Our next aim will be 
to construct the 
SU($N$)-generalization of the partition function~(\ref{ZMON}).
Let us first introduce the $(N-1)$-dimensional vector $\vec H$ of the 
mutually commuting diagonal generators of the group SU($N$).
Together with certain pairwise linear 
combinations of the off-diagonal generators which, in analogy with the SU(2)-group, are called step (rising and lowering)
generators $E_{\pm i}$, $i=1,\ldots,\frac{N(N-1)}{2}$, the diagonal generators form the following algebra:
$$
\left[\vec H, E_{\pm i}\right]=\vec q_{\pm i}E_{\pm i}\ , \quad
\left[\vec E_i, E_{-i}\right]=\vec q_i\vec H\ .$$
Vectors $\vec q_i$'s here are called root vectors of the group SU($N$).
The vector potential, $A_\mu^a$, $a=1,\ldots, N^2-1$, can be respectively decomposed into photons and W-bosons as 
$$
A_\mu^a=\sum\limits_{i}^{}\left[\left(W_\mu^{+}\right)^iE_{-i}+\left(W_\mu^{-}\right)^iE_i\right]+\vec A_\mu\vec H\ ,$$
where from now on $\sum\limits_{i}^{}\equiv\sum\limits_{i=1}^{N(N-1)/2}$.
Next, embeddings 
of the SU(2)-monopole into the maximal Abelian subgroup $U(1)^{N-1}$
of the group SU($N$) can be characterized by the space-time variations of the Higgs field $\Phi^a$. Outside the 
monopole core, $\Phi^a$ can be chosen along the $z$-axis: $\Phi^a(0,0,z) 
\stackrel{{r\to\infty}}{\longrightarrow}\vec\eta_a\vec H$. Here, $\vec\eta_a$ is the $(N-1)$-dimensional
vector of v.e.v.'s of the Higgs field $\Phi^a$. For an arbitrary space-time direction,
$$\Phi^a(r,\theta,\varphi)=\Omega(\theta,\varphi)\Phi^a(0,0,z)\Omega^{-1}(\theta,\varphi).$$ 
The matrix 
$\Omega$ can be chosen such that 
\begin{equation}
\label{bre}
\Phi^a(r,\theta,\varphi)=Y^a+X_i^a(r)\vec T_i\frac{\vec r}{r}.
\end{equation}
Here, the 3-component matrix-valued vector 

$$\vec T_i=\left(\frac{E_i+E_{-i}}{\sqrt{2}},\frac{E_i-E_{-i}}{\sqrt{2i}},\vec q_i\vec H\right)$$
(where ``$i$'' under the square root denotes $\sqrt{-1}$)
characterizes the embedding of the SU(2) Lie algebra into the root space of the group SU($N$), associated 
with the root $\vec q_i$. There exist therefore $\frac12N(N-1)$ embeddings, corresponding to the same number of monopoles.
The constant $Y^a=\left(\vec\eta_a-(\vec\eta_a\vec q_i)\vec q_i\right)\vec H$, which 
parametrically depends on $i$, $Y^a\equiv Y^a_{(i)}$, is the hypercharge associated with the $i$-th embedding. 
Since the vector $\vec\eta_a-(\vec\eta_a\vec q_i)\vec q_i$ belongs to the plane containing $\vec q_i$ 
and $\vec\eta_a$ (and is perpendicular to $\vec q_i$), the first term on the r.h.s. of Eq.~(\ref{bre})
breaks SU($N$) down to SU$(2)\times$U$(1)^{N-2}$, whereas the second term, with
$X_i^a\stackrel{r\to\infty}{\longrightarrow}\vec\eta_a\vec q_i$, further 
breaks SU(2) down to U(1). Again, as in the SU(2)-case, the only part of the monopole vector potential
which does not vanish exponentially at large distances is the diagonal (photonic) one. 
It is given by 
$$
A_\mu^i=\frac{1}{g}\varepsilon_{\mu\nu\lambda}T_\nu^i\frac{x_\lambda}{r},$$
that corresponds to the magnetic field
$$
H_\mu^i=g_m\frac{x_\mu}{4\pi r^3}\vec q_i\Omega(\theta,\varphi)\vec H\Omega^{-1}(\theta,\varphi),$$
where again $g_m=4\pi/g$. Therefore, in the SU($N$)-case, monopoles also interact by means of the long-ranged 
Coulomb forces.
This inter-monopole interaction, mediated by dual photons, results in
the following SU($N$)-analogue of the partition function~(\ref{ZMON}):
$$
{\mathcal Z}_{\rm mon}^N=
\int {\mathcal D}\vec\chi{\mathcal D}\psi\exp\left[-\int d^3x\times\right.
$$
\vspace{-8mm}
\begin{equation}
\label{zMON}
\left.\times\left(\frac12\left(\partial_\mu\vec\chi\right)^2
+\frac12\left(\partial_\mu\psi\right)^2+\frac{m_H^2}{2}\psi^2
-2\zeta{\rm e}^{g_m\psi}
\sum\limits_{i}^{}
\cos\left(g_m\vec q_i\vec\chi\right)\right)\right].
\end{equation}
Here, the dual-photon
field is described by the $(N\!-\!1)$-dimensional vector $\vec\chi$.

Averaging
in Eq.~(\ref{zMON}) over the Higgs field by means of the cumulant
expansion one gets in the second order of this expansion~\cite{mpla}:
$$
{\mathcal Z}_{\rm mon}^N\simeq
\int {\mathcal D}\vec\chi\exp\Bigg[-
\int d^3x\left(\frac12(\partial_\mu\vec\chi)^2-2\xi\sum\limits_{i}^{}
\cos\left(g_m\vec q_i\vec\chi\right)\right)
$$
\vspace{-7mm}
\begin{equation}
\label{morequart}
\hspace{1.5cm}+2\xi^2\int d^3xd^3y\sum\limits_{i,j}^{}
\cos\left(g_m\vec q_i\vec\chi(\vec x)\right){\mathcal K}(\vec x-\vec y)
\cos\left(g_m\vec q_j\vec\chi(\vec y)\right)\Bigg]\,.
\end{equation}
In this equation,
$\xi\equiv\zeta\exp\left[\frac{g_m^2}{2}D_{m_H}\left(m_W^{-1}\right)\right]$
is the modified fugacity (which can be shown to remain exponentially
small as long as the cumulant expansion is convergent), ${\mathcal
K}(\vec x)\equiv {\rm e}^{g_m^2D_{m_H}(\vec x)}-1$, and 
$D_m\left(\vec x\right)={\rm e}^{-m|\vec x|}/(4\pi|\vec x|)$ is the 3D Yukawa propagator.  The Debye mass
of the dual photon, stemming from Eq.~(\ref{morequart}) by virtue of the formula 
$\sum\limits_{i}^{}q_i^\alpha q_i^\beta=\frac{N}{2}\delta^{\alpha\beta}$, $\alpha,\beta=1,\ldots,N-1$,
reads
\begin{equation}
\label{eifi}
m_D=g_m\sqrt{N\xi}\left[1+\xi I\,\frac{N(N-1)}{2}\right],
\end{equation} 
where $I\equiv\int d^3x{\mathcal K}(\vec x)$.  At $m_H\sim m_W$, $I$ is given by~\cite{mpla}
\begin{equation}
\label{i}
I\simeq\frac{4\pi}{m_Hm_W^2}\exp\left(\frac{4\pi}{\kappa}{\rm e}^{-m_H/m_W}\right).
\end{equation}
The parameter of the cumulant expansion is ${\mathcal O}\left(\xi
IN^2\right)$.  By virtue of Eqs.~(\ref{0}) and~(\ref{i}), one can
readily see that the condition for this parameter to be
(exponentially) small reads
$N<\exp\left[\frac{2\pi}{\kappa}\left(\epsilon-\frac{1}{{\rm
e}}\right)\right]$.  Approximating $\epsilon$ by its value at
infinity, we find
$N<{\rm e}^{8.9/\kappa}$.
Therefore, when one takes into account the propagation of the heavy
Higgs boson, the necessary condition for the convergence of the
cumulant expansion is that the number of colors may grow not
arbitrarily fast, but should rather be bounded from above by some
parameter, which is nevertheless exponentially large.\footnote{\,As for the convergence of the 
cumulant expansion itself, it is a natural requirement, which should be 
obeyed by any field theory with a normal stochastic, rather than the coherent, vacuum.\\[-2mm] ~}
A similar
analysis can be performed in the BPS limit, $m_H\ll g^2$. There, one
readily finds $I\simeq\left(g_m/m_H\right)^2$, and
$$
\xi IN^2\propto N^2\exp\left[-\frac{4\pi}{\kappa}
\left(\epsilon-\frac12\right)\right].
$$ 
Approximating $\epsilon$ by its value at the origin, we see that the
upper bound for $N$ in this limit is smaller than in the vicinity of
the compact-QED limit and reads $N<{\rm e}^{\pi/\kappa}$.

\setcounter{equation}{0}
\section{String tension of the flat Wilson loop in the fundamental representation}

At zero temperature, the 3D GG model is a clear analogy of the 2D XY-model in its continuum limit
(see e.g. Refs.~\cite{pol, bmk}). The analogy is due to the fact that  
vortices of the 2D XY-model correspond to monopoles of the 3D GG model, whereas spin waves correspond
to free (non-dual) photons. Disorder in both theories is produced by topological defects, i.e. by
vortices or monopoles.\footnote{\,At finite temperatures, disorder is primarily generated by W-bosons~\cite{dkkt},
which, at zero temperature under discussion, are practically irrelevant due to their heaviness.
The quantitative discussion of the role of W-bosons at finite temperature will be presented in subsection~8.4.}
Instead, spin waves or free photons cannot disorder correlation 
functions at large distances and affect them only at the distances smaller than the vacuum correlation length.
As a result, monopoles lead to the area law of the Wilson loop, whereas free photons lead to its perimeter law.

Thus, since confinement and string tension, we are interested in, are 
generated by monopoles, photons will be omitted in this section.
The partition function of the grand canonical ensemble of monopoles in the Euclidean space-time reads
\begin{equation}
\label{1}
{\mathcal Z}_{\rm mon}^N=\sum\limits_{{\mathcal N}=0}^{\infty}\frac{\zeta^{\mathcal N}}{{\mathcal N}!}
\left<\exp\left[-\frac{g_m^2}{2}\int d^3x d^3y\vec\rho^{\mathcal N}(\vec x)D_0(\vec x-\vec y)
\vec\rho^{\mathcal N}(\vec y)\right]\right>_{\rm mon}.
\end{equation}
Here, $D_0(\vec x)=1/(4\pi|\vec x|)$
is the 3D Coulomb propagator, and  
the monopole density is defined as 
$\vec\rho^{\mathcal N}(\vec x)=\sum\limits_{k=1}^{\mathcal N}\vec q_{i_k}\delta(\vec x-\vec z_k)$ at ${\mathcal N}\ge 1$
and $\vec\rho^{{\mathcal N}=0}(\vec x)=0$. 
Furthermore, the average is defined as 
\begin{equation}
\label{monav}
\left<{\mathcal O}\right>_{\rm mon}=
\prod\limits_{n=0}^{\mathcal N}\int d^3z_n
\sum\limits_{i_n=\pm 1,\ldots,\pm\frac{N(N-1)}{2}}^{}{\mathcal O}\ .
\end{equation}
Upon explicit summation, the partition function~(\ref{1}) can be represented in the sine-Gordon--type form,
which is the large-$m_H$ limit of the partition function~(\ref{zMON}):
\begin{equation}
\label{2}
{\mathcal Z}_{\rm mon}^N=
\int {\mathcal D}\vec\chi\exp\left[-\int d^3x\left(\frac12\left(\partial_\mu\vec\chi\right)^2-
2\zeta\sum\limits_{i}^{}
\cos\left(g_m\vec q_i\vec\chi\right)\right)\right]\,.
\end{equation}
The Debye mass~(\ref{eifi}) becomes reduced to~\cite{mpla}: $m_D=g_m\sqrt{N\zeta}$.

The following comment is worth making at this point. The description of the grand canonical ensemble of 
monopoles in terms of the dual-photon field essentially implies the validity of the  
mean-field approximation. This approximation, which enables one to disregard fluctuations of fields 
of individual monopoles, is only valid if the number of monopoles contained in the Debye volume, $m_D^{-3}$, 
is large. Up to exponentially small corrections, the mean value of the monopole density, evaluated 
according to the formula $\rho_{\rm mean}=\frac{1}{V^{(3)}}\frac{\partial\ln{\mathcal Z}_{\rm mon}}{\partial
\ln\zeta}$, reads $\rho_{\rm mean}\simeq\zeta N(N-1)$, where $V^{(3)}$ is the three-volume occupied by the system.
Therefore, the number of monopoles contained in the Debye volume is 
$$
\rho_{\rm mean}m_D^{-3}\simeq\frac{N-1}{g_m^3\sqrt{N\zeta}}.$$
This is indeed an exponentially large quantity, even at $N\sim 1$\ . 
With the increase of $N$, the accuracy of the mean-field approximation is being further enhanced.

Next, one can introduce 
the monopole field-strength tensor $\vec F_{\mu\nu}^{\mathcal N}$ which violates the Bianchi 
identity as $\frac12\varepsilon_{\mu\nu\lambda}
\partial_\mu\vec F_{\nu\lambda}^{\mathcal N}=g_m\vec\rho^{\mathcal N}$. Recalling that photons are omitted
throughout this section, we obtain
\begin{equation}
\label{F}
\vec F_{\mu\nu}^{\mathcal N}\left(\vec x\right)=-g_m\varepsilon_{\mu\nu\lambda}\partial_\lambda\int d^3y 
D_0\left(\vec x-\vec y\right)
\vec\rho^{\mathcal N}\left(\vec y\right)\,.
\end{equation}
We will further use the formula 
$${\rm tr}{\,}\exp\left(i\vec {\mathcal O}\vec H\right)=
\sum\limits_{a=1}^{N}\exp\left(i\vec {\mathcal O}\vec\mu_a\right)\,,$$
where $\vec {\mathcal O}$ is an arbitrary $(N-1)$-component vector 
and $\vec\mu_a$'s are the weight vectors of the fundamental representation 
of the group SU($N$), $a=1,\ldots, N$. Together with the Stokes' theorem, this formula yields the following expression 
for the Wilson loop defined at the ${\mathcal N}$-monopole configuration:
\begin{equation}
\label{WcalN}
W(C)_{\rm mon}^{\mathcal N}=\frac{1}{N}{\,}{\rm tr}{\,}
\exp\left(\frac{ig}{2}\vec H\int d^3x\vec F_{\mu\nu}^{\mathcal N}\Sigma_{\mu\nu}\right)=
\frac{1}{N}\sum\limits_{a=1}^{N}W_a^{\mathcal N}\ .
\end{equation}
Here, 
\vspace{-6mm}
\begin{equation}
\label{Wa}
W_a^{\mathcal N}\equiv\exp\left(\frac{ig}{2}\vec\mu_a\int d^3x\vec F_{\mu\nu}^{\mathcal N}\Sigma_{\mu\nu}\right)
\end{equation}
and $\Sigma_{\mu\nu}\left(\vec x\right)=\int\limits_{\Sigma(C)}^{} 
d\sigma_{\mu\nu}\left(\vec x\left(\xi\right)\right)\delta\left(\vec x-\vec x(\xi)\right)$
is the vorticity tensor current defined at a certain surface 
$\Sigma(C)$ bounded by the contour $C$ and parametrized by the vector 
$\vec x(\xi)$ with $\xi=\left(\xi_0,\xi_1\right)$ standing for the 2D coordinate.
A straightforward calculation with the use of Eq.~(\ref{F}) yields
\begin{equation}
\label{W}
W_a^{\mathcal N}=\exp\left(i\vec\mu_a\int d^3x\vec\rho^{\mathcal N}\eta\right),
\end{equation}
where 
$\eta\left(\vec x; C\right)=\int\limits_{\Sigma(C)}^{} 
d\sigma_\mu\left(\vec x(\xi)\right)\partial_\mu\frac{1}{\left|\vec x-\vec x(\xi)\right|}$
is the solid angle under which the surface 
$\Sigma(C)$ is seen by an observer located at the point $\vec x$, $d\sigma_\mu=
\frac12\varepsilon_{\mu\nu\lambda}d\sigma_{\nu\lambda}$. (Note that, owing to
the Gauss' law, $\eta\equiv4\pi$ for a 
closed surface, i.e. when $C$ is shrunk to a point.) The ratio of two $W_a^{\mathcal N}$'s 
defined at different surfaces, $\Sigma_1$ and $\Sigma_2$, bounded by the contour $C$, reads
$$
\frac{W_a^{\mathcal N}\left(\Sigma_1\right)}{W_a^{\mathcal N}\left(\Sigma_2\right)}=\exp\Bigg(i\vec\mu_a
\int\limits_{\Sigma_1\cup\Sigma_2}^{}d\sigma_\mu\left(\vec x(\xi)\right)\int d^3x\vec\rho^{\mathcal N}
\left(\vec x{\,}\right)\partial_\mu\frac{1}{\left|\vec x-\vec x(\xi)\right|}\Bigg)=$$
\vspace{-6mm}
\begin{equation}
\label{fract}
=\prod\limits_{k=1}^{\mathcal N}\exp\Bigg(-i\vec\mu_a\vec q_{i_k}
\int\limits_{\Sigma_1\cup\Sigma_2}^{}d\sigma_\mu\left(\vec x(\xi)\right)\partial_\mu^{\vec x(\xi)}
\frac{1}{\left|\vec z_k-\vec x(\xi)\right|}\Bigg).
\end{equation}
Due to the Gauss' law, the last integral in this equation is equal either to $-4\pi$ or to $0$, depending on whether
the point $\vec z_k$ lies inside or outside the volume bounded by the surface $\Sigma_1\cup\Sigma_2$. Since the product
$\vec\mu_a\vec q_{i_k}$ is equal either to $\pm\frac12$ or to $0$, we conclude that 
$\frac{W_a^{\mathcal N}\left(\Sigma_1\right)}{W_a^{\mathcal N}\left(\Sigma_2\right)}=1$. This fact proves the 
independence of $W_a^{\mathcal N}$ of the choice of the surface $\Sigma$ in the definition~(\ref{Wa}).
Clearly, this is the consequence of the quantization condition $gg_m=4\pi$ used in the derivation of Eq.~(\ref{W}).

A representation of the partition function~(\ref{1}), alternative to Eq.~(\ref{2}) and more appropriate for the 
investigation of the Wilson loop, is the one in terms of the 
dynamical monopole densities~\cite{mpla}.
It can be obtained by multiplying Eq.~(\ref{1}) by the following unity:
$$
\int {\mathcal D}\vec\rho\delta\left(\vec\rho-\vec\rho^{\mathcal N}\right)=
\int {\mathcal D}\vec\rho{\mathcal D}\vec\chi\exp\left[ig_m\int d^3x\vec\chi\left(\vec\rho
-\vec\rho^{\mathcal N}\right)\right],$$
so that the field $\vec\chi$ plays the role of the Lagrange multiplier. We obtain for the partition function:
$$
{\mathcal Z}_{\rm mon}^N=\int{\mathcal D}
\vec\rho{\mathcal D}\vec\chi\exp\left[-\frac{g_m^2}{2}\int d^3x d^3y\vec\rho(\vec x)D_0(\vec x-\vec y)
\vec\rho(\vec y)+ig_m\int d^3x\vec\chi\vec\rho\right]\times$$
\vspace{-6mm}
\begin{equation}
\label{II}
\times\sum\limits_{{\mathcal N}=0}^{\infty}\frac{\zeta^{\mathcal N}}{{\mathcal N}!}
\left<\exp\left(-ig_m\int d^3x\vec\chi\vec\rho^{\mathcal N}\right)\right>_{\rm mon},
\end{equation}
where the last sum is equal to 
\begin{equation}
\label{III}
\exp\left[2\zeta\int d^3x
\sum\limits_{i}^{}
\cos\left(g_m\vec q_i\vec\chi\right)\right].
\end{equation}
Accordingly, Eq.~(\ref{W}) becomes replaced by
$$
W_a^{\mathcal N}\to W_a=\frac{1}{{\mathcal Z}_{\rm mon}^N}
\int{\mathcal D}\vec\rho{\mathcal D}\vec\chi\exp\Bigg\{-\frac{g_m^2}{2}
\int d^3x d^3y\vec\rho(\vec x)D_0(\vec x-\vec y)
\vec\rho(\vec y)+$$
\vspace{-6mm}
$$+\int d^3x\Bigg[ig_m\vec\chi\vec\rho+
2\zeta
\sum\limits_{i}^{}
\cos\left(g_m\vec q_i\vec\chi\right)+
i\vec\mu_a\vec\rho\eta\Bigg]\Bigg\}\,,$$
and the full expression for the monopole contribution to the Wilson loop [instead of Eq.~(\ref{WcalN})]
reads $W(C)_{\rm mon}=\frac{1}{N}\sum\limits_{a=1}^{N}W_a$.

Let us next introduce the magnetic field according to the formulae 
$\partial_\mu\vec B_\mu=\vec\rho$, 
$\varepsilon_{\mu\nu\lambda}\partial_\nu\vec B_\lambda=0$. 
This yields

$$
W_a=\frac{1}{{\mathcal Z}_{\rm mon}^N}
\int{\mathcal D}\vec B_\mu\delta\left(\varepsilon_{\mu\nu\lambda}\partial_\nu\vec B_\lambda\right)
\int {\mathcal D}\vec\chi
\exp\Bigg\{\int d^3x\Big[-\frac{g_m^2}{2}\vec B_\mu^2+$$
\vspace{-5mm}
\begin{equation}
\label{WWW}
+ig_m\vec\chi\partial_\mu\vec B_\mu+
2\zeta
\sum\limits_{i}^{}
\cos\left(g_m\vec q_i\vec\chi\right)\Big]+
4\pi i\vec\mu_a\int\limits_{\Sigma(C)}^{}d\sigma_\mu\vec B_\mu\Bigg\},
\end{equation}
where, as has been discussed, the surface $\Sigma(C)$ is arbitrary. 
The magnetic and the dual-photon fields 
can be integrated out of Eq.~(\ref{WWW}) by solving the respective saddle-point equations. 
From now on in this section, we will 
consider the contour $C$ located in the $(x,y)$-plane. This naturally leads to the following Ans\"atze for the  
fields to be inserted into the saddle-point equations: 
$\vec B_\mu=\delta_{\mu 3}\vec B(z)$, $\vec\chi=\vec\chi(z)$. 
For the points 
$(x,y)$ lying inside the contour $C$, these equations then read:
\begin{equation}
\label{sp1}
ig_m\vec\chi'+g_m^2\vec B-4\pi i\vec\mu_a\delta(z)=0,
\end{equation}
\vspace{-8mm}
\begin{equation}
\label{sp2}
i\vec B'-2\zeta\sum\limits_{i}^{}\vec q_i\sin\left(g_m\vec q_i\vec\chi\right)=0,
\end{equation}
where $'\equiv d/dz$.
[Note that, differentiating Eq.~(\ref{sp1}) and substituting the result into Eq.~(\ref{sp2}), we obtain the equation
$$\vec\chi''-2g_m\zeta\sum\limits_{i}^{}\vec q_i\sin\left(g_m\vec q_i\vec\chi\right)=g\vec\mu_a
\delta'(z),$$
which is the SU($N$)-generalization of the saddle-point equation obtained in Ref.~\cite{pol}.]
These equations can be solved by using one more natural Ansatz for the saddle points of the fields, 
$\vec B(z)=\vec\mu_a B(z)$, $\vec\chi(z)=\vec\mu_a\chi(z)$, that, due to the formula 
\begin{equation}
\label{norm}
\vec\mu_a\vec\mu_b=\frac12\left(\delta_{ab}-\frac{1}{N}\right),
\end{equation}
makes $W_a$ $a$-independent. Let us next multiply Eq.~(\ref{sp2}) by $\vec\mu_a$, taking into account that, 
for any $a$, $(N-1)$ positive roots yield the scalar product with $\vec\mu_a$ equal to $1/2$, while the others are 
orthogonal to $\vec\mu_a$. Equations~(\ref{sp1}), (\ref{sp2}) then go over to
\begin{equation}
\label{SP}
2i\phi'+g_m^2B=4\pi i\delta(z),~~ B'+2i\zeta N\sin\phi=0,
\end{equation}
where $\phi\equiv g_m\chi/2$. One can readily 
check that the solution to this system of equations has the form

\begin{equation}
\label{solution}
B(z)=i\frac{8m_D}{g_m^2}\frac{{\rm e}^{-m_D|z|}}{1+{\rm e}^{-2m_D|z|}},~~
\phi(z)=4{\,}{\rm sgn}{\,}z\cdot\arctan\left({\rm e}^{-m_D|z|}\right).
\end{equation}
Taking the value $B(0)$ for the evaluation of the Wilson loop, we obtain for the string tension:
\begin{equation}
\label{Sig}
\sigma=4\pi\cdot\frac{N-1}{2N}\frac{4m_D}{g_m^2}=
\frac{8\pi}{g_m}\frac{N-1}{\sqrt{N}}\sqrt{\zeta}.
\end{equation}
This result demonstrates explicitly the nonanalytic dependence of $\sigma$ on $g$, but 
does not yield the correct overall numerical factor. That is because $B(z)$, necessary for this 
calculation, is defined ambiguously. 
The ambiguity originates from the exponentially
large thickness of the string: $|z|<d\equiv m_D^{-1}$. It is therefore unclear which value of $B(z)$
we should take: either $B(0)$, as was done, 
or the one averaged over some range of $z$. It turns out that, in the weak-field (low-density) limit, 
the problem can be solved for an arbitrarily-shaped surface, by 
using the representation in terms of the Kalb-Ramond field, which will
be described in the next section.

\setcounter{equation}{0}
\section{SU(\boldmath{$N$}) confining strings}

\subsection{Fundamental representation}

The dual photon can alternatively be described in terms of the Kalb--Ramond field. 
One of the ways to introduce this field is to consider it as the field-strength 
tensor corresponding to the field 
$\vec B_\mu$, namely $\vec B_\mu=\frac{1}{2g_m}\varepsilon_{\mu\nu\lambda}\vec h_{\nu\lambda}$.
The Wilson loop~(\ref{WWW}) then takes the form\,\footnote{\,Note the following correspondence between the 
Coulomb interaction of monopole densities and the actions of the magnetic- and Kalb--Ramond fields:
$$
\frac14\int d^3x\vec h_{\mu\nu}^2=\frac{g_m^2}{2}\int d^3x\vec B_\mu^2=
\frac{g_m^2}{2}\int d^3x d^3y\vec\rho(\vec x)D_0(\vec x-\vec y)
\vec\rho(\vec y)\;.$$}
$$
\hspace{-0.8cm} W_a=
\frac{1}{{\mathcal Z}_{\rm mon}^N}
\int{\mathcal D}\vec h_{\mu\nu}\delta\left(\partial_\mu\vec h_{\mu\nu}\right)\int {\mathcal D}\vec\chi
\exp\Bigg\{\int d^3x\Bigg[-\frac14\vec h_{\mu\nu}^2+$$
\vspace{-6mm}
\begin{equation}
\label{WwW}
\hspace{1.8cm}+\,\frac{i}{2}\,\vec\chi\,\varepsilon_{\mu\nu\lambda}\partial_\mu\vec h_{\nu\lambda}+
2\zeta
\sum\limits_{i}^{}
\cos\left(g_m\vec q_i\vec\chi\right)\Bigg]+
\frac{ig}{2}\,\vec\mu_a\!\!\!\int\limits_{\Sigma(C)}^{}\!d\sigma_{\mu\nu}\vec h_{\mu\nu}\Bigg\}\,.
\end{equation}
The field $\vec\chi$ can further be integrated out by solving the saddle-point equation of the form~(\ref{sp2}),
where $\vec B'$ is replaced by  
$\frac{1}{2g_m}\varepsilon_{\mu\nu\lambda}
\partial_\mu\vec h_{\nu\lambda}$ (and $\vec\chi$ depends on all three coordinates). 
Using the Ansatz $\vec h_{\mu\nu}=\vec\mu_a 
h_{\mu\nu}$, we arrive at the following substitution in Eq.~(\ref{WwW}):
$$
\int d^3x\Big[\frac{i}{2}\,\vec\chi\,\varepsilon_{\mu\nu\lambda}\partial_\mu\vec h_{\nu\lambda}+
2\zeta
\sum\limits_{i}^{}
\cos\left(g_m\vec q_i\vec\chi\right)\Big]\Longrightarrow$$
\vspace{-6mm}
\begin{equation}
\label{Vh}
\Longrightarrow -V\left[\vec h_{\mu\nu}\right]\equiv -2(N-1)\zeta\int d^3x \left[H_a{\rm arcsinh}{\,} H_a
-\sqrt{1+H_a^2}\right].
\end{equation}
Here, 
\begin{equation}
\label{Ha}
H_a\equiv \frac{1}{2g_m\zeta (N-1)}\,\vec\mu_a\varepsilon_{\mu\nu\lambda}\partial_\mu\vec h_{\nu\lambda}\ ,
\end{equation} 
and $V\big[\vec h_{\mu\nu}\big]$ is the 
multi-valued potential of the Kalb--Ramond field (or of monopole densities).
Similarly to the case of compact QED~\cite{cs}, it is the summation over branches of this potential, which yields the 
summation over world sheets $\Sigma(C)$ in Eq.~(\ref{WwW}). 

Another way to discuss the connection of the Kalb--Ramond field with the string world sheets is based on the 
semi-classical analysis of the saddle-point equations stemming from Eq.~(\ref{WwW}). These are 
analogous to Eqs.~(\ref{SP}) and read:
\vspace{-3mm}
\begin{equation}
\label{SSP1}
2i\partial_\mu\phi+\frac{g_m}{2}\varepsilon_{\mu\nu\lambda}h_{\nu\lambda}=4\pi i\Sigma_\mu\ ,
\end{equation}
\vspace{-6mm}
\begin{equation}
\label{SSP2}
\frac{1}{2g_m}\varepsilon_{\mu\nu\lambda}\partial_\mu h_{\nu\lambda}+2i\zeta N\sin\phi=0\ .
\end{equation}
Here, $\Sigma_\mu\equiv\frac12\varepsilon_{\mu\nu\lambda}\Sigma_{\nu\lambda}$,
and the field $\phi$
is defined by the relation $\vec\chi=\frac{2}{g_m}\vec\mu_a\phi$. It is this auxiliary field $\phi$, which establishes 
the correspondence
between the Kalb--Ramond field and the stationary surface in the present approach.
The summation over branches of the potential $V\big[\vec h_{\mu\nu}\big]$ is now replaced by the following procedure.
One should restrict oneself to the domain $|\phi|\le\pi$ and solve Eqs.~(\ref{SSP1}), (\ref{SSP2}) with the conditions
\begin{equation}
\label{cond1}
\lim\limits_{|\vec x|\to\infty}^{}\phi(\vec x)=0,
\end{equation}
\vspace{-6mm}
\begin{equation}
\label{cond2}
\lim\limits_{\varepsilon\to 0}^{}\left\{\phi\left[\vec x(\xi)+\varepsilon\vec n(\xi)\right]-
\phi\left[\vec x(\xi)-\varepsilon\vec n(\xi)\right]\right\}=2\pi,
\end{equation}
where $\vec n(\xi)$ is the normal vector to $\Sigma$ at an arbitrary point $\vec x(\xi)$. After that, one should 
get rid of the so-appearing $\Sigma$-dependence of $W_a$ by extremizing the latter with respect to $\vec x(\xi)$.
Owing to the formula 
$$
\delta\int d\sigma_{\mu\nu}h_{\mu\nu}\left[\vec x(\xi)\right]=
\int d\sigma_{\mu\nu}H_{\mu\nu\lambda}\left[\vec x(\xi)\right]\delta x_\lambda(\xi)\ ,
$$
such an extremization is equivalent to the definition of the extremal surface through the equation 
$H_{\mu\nu\lambda}\left[\vec x(\xi)\right]=0$. Due to Eqs.~(\ref{SSP2}), (\ref{cond2}), $H_{\mu\nu\lambda}$ 
indeed vanishes on both sides of the stationary surface. The reason for that is the following 
distinguished property of the extremal surface:
$\phi$ is equal to $\pi$ on one of its sides and to $-\pi$ on the other, whereas for any other surface, 
$|\phi|$ is smaller than $\pi$ on one side and larger than $\pi$ on the other side. Therefore, 
according to Eq.~(\ref{SSP2}), on both
sides of the extremal surface, $\varepsilon_{\mu\nu\lambda}\partial_\mu h_{\nu\lambda}=0$. This is 
equivalent to the 
equation $\varepsilon_{\mu\nu\lambda}H_{\mu\nu\lambda}=0$ and hence (upon the multiplication of 
both sides by $\varepsilon_{\mu'\nu'\lambda'}$)
to $H_{\mu\nu\lambda}=0$.
[In particular, for a flat contour the extremal surface is the flat surface inside this contour.
This can explicitly be seen from the solutions~(\ref{solution}) of Eqs.~(\ref{SP}). Namely, 
at any $z$ such that $|z|\ll d$,
$\phi(z)\simeq\pi\cdot {\rm sgn} z$, $B'(z)\simeq 0$, where both equalities hold with the exponential 
accuracy. The equation $B'(z)\simeq 0$
is equivalent to the condition of vanishing of $H_{\mu\nu\lambda}$ on both sides of the flat surface.]

Next, the general form of an antisymmetric rank-2 tensor field $\vec h_{\mu\nu}$ is 
\begin{equation}
\label{ttf}
\vec h_{\mu\nu}=
\partial_\mu\vec A_\nu-\partial_\nu\vec A_\mu+\varepsilon_{\mu\nu\lambda}\partial_\lambda\vec C\ .
\end{equation}
The constraint $\partial_\mu\vec h_{\mu\nu}=0$, 
imposed in Eq.~(\ref{WwW}), is equivalent to setting $\vec A_\mu$ to zero.
From now on, we will promote 
$\vec h_{\mu\nu}$ to include also the fields of free photons, $\vec A_\mu$, by abolishing this 
constraint. Let us further go into the weak-field limit, $|H_a|\ll 1$. Using the Cauchy inequality, we have 
\begin{equation}
\label{wf}
\left|H_a\right|\le\frac{\left|\vec\mu_a\right|\left|\vec\rho
\right|}{\zeta(N-1)}=
\frac{|\vec\rho|}{\zeta\sqrt{2N(N-1)}}\ ,
\end{equation}
so that the weak-field limit is equivalent to the following low-density approximation: $|\vec\rho|\ll\zeta\sqrt{2N(N-1)}$.
To understand in which sense this 
inequality implies the low-density approximation, recall the mean monopole density we had from Eq.~(\ref{2}): 
\begin{equation}
\label{me}
\left|\vec\rho\right|_{\rm mean}\simeq\zeta N(N-1)\ .
\end{equation}
Therefore, at large-$N$, the low-density approximation implies that $|\vec\rho|$ 
should be of the order $N$ times smaller than its mean value.

In the weak-field limit and with the constraint 
$\partial_\mu\vec h_{\mu\nu}=0$ removed, 
we now have the following expression for the total Wilson loop:
$$
W\left(C,\Sigma\right)_{\rm weak-field}^{{\rm tot},~ a}=\frac{1}{{\mathcal Z}^{\rm tot}}
\int{\mathcal D}\vec h_{\mu\nu}\times
$$
\vspace{-6mm}
\begin{equation}
\label{Wwf}
\times\exp\left\{-\int d^3x\left[\frac{1}{12m_D^2}\vec H_{\mu\nu\lambda}^2
+\frac14\vec h_{\mu\nu}^2-\frac{ig}{2}\vec\mu_a\vec h_{\mu\nu}\Sigma_{\mu\nu}\right]\right\}\,.
\end{equation}
Here, ${\mathcal Z}^{\rm tot}$ is given by the same integral over 
$\vec h_{\mu\nu}$, but with $\Sigma_{\mu\nu}$ set to zero, and 
$\vec H_{\mu\nu\lambda}=\partial_\mu\vec h_{\nu\lambda}+\partial_\lambda\vec h_{\mu\nu}+
\partial_\nu\vec h_{\lambda\mu}$ is the Kalb--Ramond field-strength tensor. 
[The apparent $\Sigma$-dependence of the r.h.s. of Eq.~(\ref{Wwf})
is due to the fact that in course of the $\vec h_{\mu\nu}$-expansion 
of $V\big[\vec h_{\mu\nu}\big]$, only one branch of this potential
has been taken into account. As discussed above, the 
$\Sigma$-dependence disappears upon the summation over all the branches.]
Owing to Eq.~(\ref{ttf}),
Eq.~(\ref{Wwf}) is obviously factorized as 
$W\left(C,\Sigma\right)_{\rm weak-field}^{{\rm tot},~ a}=W_a W_a^{\rm phot}$, where the free 
photonic contribution to the Wilson loop reads 
\begin{equation}
\label{Fp}
W_a^{\rm phot}=\exp\Bigg[-g^2\frac{N-1}{4N}\oint\limits_{C}^{}dx_\mu
\oint\limits_{C}^{}dx'_\mu D_0\left(\vec x-\vec x'\right)\Bigg].
\end{equation}
Doing the integration over $\vec h_{\mu\nu}$ in Eq.~(\ref{Wwf}), we obtain
$$\hspace{-0.8cm}W\left(C,\Sigma\right)_{\rm weak-field}^{{\rm tot},~ a}=
\exp\Bigg\{-g^2\frac{N-1}{4N}\Bigg[\oint\limits_C^{}dx_\mu\oint\limits_C^{}dx'_\mu
D_{m_D}(\vec x-\vec x')+$$
\vspace{-6mm}
\begin{equation}
\label{WCSig}
+\frac{m_D^2}{2}\int\limits_{\Sigma(C)}^{}d\sigma_{\mu\nu}(\vec x(\xi))
\int\limits_{\Sigma(C)}^{}d\sigma_{\mu\nu}(\vec x(\xi'))D_{m_D}\left(\vec x(\xi)-\vec x(\xi')\right)
\Bigg]\Bigg\}\,.
\end{equation}
Note that the free photonic contribution completely cancels out of this expression, i.e. it is only the dual photon 
(of the mass $m_D$) which mediates the $C\times C$- and $\Sigma\times\Sigma$-interactions.

One can further expand the nonlocal string effective action 
\begin{equation}
\label{snonloc}
S_{\rm str}=(gm_D)^2\frac{N-1}{8N}\int\limits_{\Sigma(C)}^{}d\sigma_{\mu\nu}(\vec x(\xi))
\int\limits_{\Sigma(C)}^{}d\sigma_{\mu\nu}(\vec x(\xi'))D_{m_D}\left(\vec x(\xi)-\vec x(\xi')\right)
\end{equation}
in powers of derivatives with respect to the world-sheet 
coordinates $\xi_a$'s. Note that the actual parameter of this expansion 
is $1/(m_DR)^2$, where $R\sim\sqrt{{\rm Area}(\Sigma)}$ is the size of $\Sigma$ 
(see the discussion below). The resulting quasi-local action is
(cf. Refs.~\cite{theta, dva}):
$$
S_{\rm str}=\sigma\int d^2\xi\sqrt{\sf g}+
\alpha^{-1}\int d^2\xi\sqrt{\sf g}{\sf g}^{ab}
(\partial_at_{\mu\nu})(\partial_bt_{\mu\nu})+
$$
\vspace{-6mm}
\begin{equation}
\label{sstr}
+\kappa
\int d^2\xi\sqrt{\sf g}{\mathcal R}+{\mathcal O}\left(\frac{\sigma}{m_D^4R^2}\right).
\end{equation}
Here, 
$\partial_a\equiv\partial/\partial\xi^a$, and the following quantities characterize $\Sigma$: ${\sf g}_{ab}(\xi)=
(\partial_a x_\mu(\xi))(\partial_b x_\mu(\xi))$ is the induced-metric tensor, ${\sf g}=\det\| {\sf g}^{ab}\|$,
$t_{\mu\nu}(\xi)=\varepsilon^{ab}(\partial_a 
x_\mu(\xi))(\partial_b x_\nu(\xi))/\sqrt{\sf g}$ is the extrinsic-curvature tensor,
and ${\mathcal R}=\left(\partial^a\partial_a\ln
\sqrt{\sf g}\right)/\sqrt{\sf g}$ is the expression for the scalar curvature 
in the conformal gauge ${\sf g}_{ab}=\sqrt{{\sf g}}\delta_{ab}$.

The third 
term on the r.h.s. of Eq.~(\ref{sstr})
is known to be a full derivative, and therefore it does not actually contribute to the string effective action, while 
the second 
term describes the so-called rigidity (or stiffness) of the string~\cite{fi, more_rigid}.
The reason for the notation $\alpha^{-1}$, introduced in Ref.~\cite{fi}, is that, as has been 
shown in that paper, it is $\alpha$ 
which is asymptotically free. This asymptotic freedom then indicates that the rigidity term can only be infrared 
relevant if the respective $\beta$-function
has a zero in the infrared region. However, such a zero has not been
found. Due to this fact, the rigidity term is not a good candidate to solve the old-standing 
problem of crumpling of large world sheets in Euclidean space-time. This necessitates seeking 
other possible solutions of this problem.
One such solution, based on the string $\theta$-term, has been proposed in Ref.~\cite{fi}. 
A possible derivation of such a term, within the SU($N$)-analogue of the 4D compact QED will be presented below.

The string coupling constants in Eq.~(\ref{sstr}) read
\begin{equation}
\label{tef}
\sigma=2\pi^2\,\frac{N-1}{\sqrt{N}}\frac{\sqrt{\zeta}}{g_m}\ ,
\end{equation}
\vspace{-7mm}
\begin{equation}
\label{alf}
\alpha^{-1}=-3\kappa/2=-\frac{\pi^2(N-1)}{4g_m^3N^{3/2}\sqrt{\zeta}}\ .
\end{equation} 
A comment is in order regarding the negative sign of $\alpha$. Up to a total derivative, the 
rigidity term reads
$$\int d^2\xi\sqrt{\sf g}{\sf g}^{ab}
(\partial_at_{\mu\nu})(\partial_bt_{\mu\nu})=
\int d^2\xi\sqrt{{\sf g}}(\Delta x_\mu)^2\ ,$$
where the Laplacian associated to the metric ${\sf g}^{ab}$ acts on $x_\mu(\xi)$ as 
$\Delta x_\mu=\frac{1}{\sqrt{{\sf g}}}\partial_a\left(\sqrt{{\sf g}}{\sf g}^{ab}\partial_b x_\mu\right)$.
In the conformal gauge, one thus readily gets for the rigidity term $\int d^2\xi
\frac{1}{\sqrt{{\sf g}}}\left(\partial^2x_\mu\right)^2$. For a nearly flat surface, ${\sf g}\simeq{\,}
{\rm const}{\,}\equiv c^2$, and we have for the string propagator (in 2D Euclidean space):
$$
\left<x_\mu(\xi)x_\nu(0)\right>=\frac{\delta_{\mu\nu}}{\sigma}\int d^2p\;\frac{{\rm e}^{ip\xi}}{p^2\left(1+\frac{1}{\alpha c
\sigma}p^2\right)}\ .$$
For $\alpha<0$, this propagator has a tachyonic pole, whereas an unphysical mass pole shows up otherwise.
We therefore conclude that the negative sign of $\alpha$ is important for the stability of strings.

In the weak-field (or low-density) approximation, we can thus fix the 
proportionality factor $2\pi^2$ at $\sigma$, that is close to the factor $8\pi$ of 
Eq.~(\ref{Sig}) which, however, could have not 
been fixed within the method of section~3. Another important fact is that 
(in the weak-field approximation) the string tension and higher string coupling constants 
are the same for all large enough surfaces $\Sigma(C)$. 
[The words ``large enough'' here mean the validity of the inequality 
$R\gg d$, i.e. the string length $R$ should 
significantly exceed the exponentially large string thickness $d$. Indeed, 
the general $n$-th term of the derivative (or curvature) 
expansion of $S_{\rm str}$ is of the order of $\sigma m_D^2R^4(R/d)^{-2n}$.]
It is, however, worth noting that some of the terms in the expansion~(\ref{sstr}) vanish at the surface 
of the minimal area corresponding to
a given contour $C$ (e.g. in the case of a flat surface 
considered in section~3). This concerns, for instance, the rigidity 
term.\footnote{\,Indeed, one can prove the equality 
${\sf g}^{ab}(\partial_at_{\mu\nu})(\partial_bt_{\mu\nu})=({\mathcal D}^a{\mathcal D}_ax_\mu)({\mathcal D}^b{\mathcal D}_bx_\mu)$, where 
${\mathcal D}_a{\mathcal D}_bx_\mu=\partial_a\partial_bx_\mu-\Gamma_{ab}^c\partial_cx_\mu$, 
and $\Gamma_{ab}^c$ is the Christoffel symbol corresponding to 
the metric ${\sf g}^{ab}$. On the other hand, the surface of the 
minimal area is defined by the equation ${\mathcal D}^a{\mathcal D}_ax_\mu(\xi)=0$ together 
with the respective boundary condition at the contour $C$.}

Let us finally demonstrate how to introduce 
the Kalb-Ramond field in a way alternative to its definition via the $\vec B_\mu$-field 
and incorporating automatically the free-photonic contribution to the Wilson loop. It is based on the 
direct combination of Eqs.~(\ref{1}) and (\ref{W}), that yields
$$
W_a=\sum\limits_{{\mathcal N}=0}^{\infty}\frac{\zeta^{\mathcal N}}{{\mathcal N}!}
\left<\exp\left[-\frac{g_m^2}{2}\int d^3x d^3y\vec\rho^{\mathcal N}(\vec x)D_0(\vec x-\vec y)
\vec\rho^{\mathcal N}(\vec y)+\right.\right.$$
\vspace{-6mm}
$$\left.\left.+i\vec\mu_a\int d^3x\vec\rho^{\mathcal N}\eta\right]\right>_{\rm mon}
=\frac{1}{{\mathcal Z}_{\rm mon}^N}\int {\mathcal D}\vec\varphi\times$$
\vspace{-6mm}
\begin{equation}
\label{alphaa}
\times\exp\Biggl\{-\int d^3x
\Biggl[\frac12\left(\partial_\mu\vec\varphi-\frac{\vec\mu_a}{g_m}
\partial_\mu\eta\right)^2-2\zeta\sum\limits_{i}^{}
\cos\left(g_m\vec q_i\vec\varphi\right)\Biggr]\Biggr\}\  ,
\end{equation}
where $\vec\varphi=\vec\chi+\frac{\vec\mu_a}{g_m}\eta$. One can further use the following identity, which 
shows explicitly how the Kalb--Ramond field unifies the monopole and the free-photonic contributions
to the Wilson loop:
$$\exp\left[-\frac12\int d^3x\left(\partial_\mu\vec\varphi-\frac{\vec\mu_a}{g_m}
\partial_\mu\eta\right)^2-g^2\frac{N-1}{4N}
\oint\limits_C^{}dx_\mu\oint\limits_C^{}dx'_\mu
D_0(\vec x-\vec x')\right]=$$
\vspace{-6mm}
\begin{equation}
\label{unify}
=\int {\mathcal D}\vec h_{\mu\nu}\exp\left[-\int d^3x\left(
\frac14\vec h_{\mu\nu}^2+\frac{i}{2}\vec\varphi\varepsilon_{\mu\nu\lambda}\partial_\mu
\vec h_{\nu\lambda}-\frac{ig}{2}\vec\mu_a\vec h_{\mu\nu}\Sigma_{\mu\nu}\right)\right].
\end{equation}
One can prove this by showing that both sides of this formula are equal 
to\,\footnote{\,Obviously, in the noncompact case, when monopoles are 
disregarded and $\vec h_{\mu\nu}=\partial_\mu\vec A_\nu-
\partial_\nu\vec A_\mu$, the r.h.s. of Eq.~(\ref{unify}) yields the free-photonic contribution to the Wilson loop, 
Eq.~(\ref{Fp}).} 
$$\exp\bigg[-\frac12\int d^3x\left(\partial_\mu\vec\varphi+g\vec\mu_a\Sigma_\mu\right)^2\bigg].$$
Inserting further Eq.~(\ref{unify}) into Eq.~(\ref{alphaa}), we obtain
$$W_a\to W(C)_a^{\rm tot}
=\frac{1}{{\mathcal Z}^{\rm tot}}
\int{\mathcal D}\vec h_{\mu\nu}{\mathcal D}\vec\varphi\exp\Big[-\int d^3x\Big(
\frac14\vec h_{\mu\nu}^2+$$
\vspace{-5mm}
\begin{equation}
\label{fffi}
+\,\frac{i}{2}\,\vec\varphi\,\varepsilon_{\mu\nu\lambda}\partial_\mu
\vec h_{\nu\lambda}-2\zeta\sum\limits_{i}^{}\cos\left(g_m\vec q_i\vec\varphi\right)
-\frac{ig}{2}\vec\mu_a\vec h_{\mu\nu}\Sigma_{\mu\nu}\Big)\Big]\ .
\end{equation}
Denoting $\vec\chi\equiv-\vec\varphi$, 
we arrive back to Eq.~(\ref{WwW}) with ${\mathcal Z}_{\rm mon}^N\to {\mathcal Z}^{\rm tot}$ and 
the constraint $\partial_\mu\vec h_{\mu\nu}=0$ removed, as it should be.

\vspace{-2mm}

\subsection{Adjoint representation}

Let us now extend the ideas of the previous subsection to the case of the Wilson loop in the adjoint representation. 
The charges of quarks in this representation are distributed along the 
roots, so that in the formulae for the Wilson loop, Eqs.~(\ref{Wa}) and (\ref{W}), 
$\vec\mu_a$ should be replaced by $\vec q_i$. Noting
that any root is a difference of two weights, we will henceforth in this subsection label roots by two indices
running from $1$ to $N$, e.g. $\vec q_{ab}=\vec\mu_a-\vec\mu_b\,$.\,\footnote{\,In particular, this makes explicit 
the number of positive roots, $\frac{N^2-N}{2}$.} Equation~(\ref{norm}) then leads to the following formula:
\begin{equation}
\label{qq}
\vec q_{ab}\vec q_{cd}=\frac12\left(\delta_{ac}+\delta_{bd}-\delta_{ad}-\delta_{bc}\right),
\end{equation}
according to which the product $\vec q_{ab}\vec q_{cd}$ may take the values $0, \pm \frac12, \pm 1$. Owing to this fact,
the ratio~(\ref{fract}) is again equal to 1, i.e. for adjoint quarks, whose charge obeys the quantization condition 
$gg_m=4\pi$, the Wilson loop is as surface independent, as it is for fundamental quarks.

The adjoint-case version of Eq.~(\ref{WwW}) with the constraint $\partial_\mu\vec h_{\mu\nu}=0$ abolished 
[or of Eq.~(\ref{fffi})] has the form:
$$
W(C)_{ab}^{\rm tot}=
\frac{1}{{\mathcal Z}^{\rm tot}}
\int{\mathcal D}\vec h_{\mu\nu}{\mathcal D}\vec\chi
\exp\Bigg\{\int d^3x\Big[-\frac14\vec h_{\mu\nu}^2+$$
\vspace{-6mm}
\begin{equation}
\label{wab}
+\,\frac{i}{2}\,\vec\chi\,\varepsilon_{\mu\nu\lambda}\partial_\mu\vec h_{\nu\lambda}+
\zeta\sum\limits_{c,d=1}^{N}
\cos\left(g_m\vec q_{cd}\vec\chi\right)\Big]+
\frac{ig}{2}\vec q_{ab}\int\limits_{\Sigma(C)}^{}d\sigma_{\mu\nu}\vec h_{\mu\nu}\Bigg\}\,.
\end{equation}
The saddle-point equation, emerging from the above formula in the course of integration over $\vec\chi\,$,
\begin{equation}
\label{neweq}
\frac{i}{2}\varepsilon_{\mu\nu\lambda}\partial_\mu\vec h_{\nu\lambda}=
g_m\zeta\sum\limits_{c,d=1}^{N}\vec q_{cd}
\sin\left(g_m\vec q_{cd}\vec\chi\right),
\end{equation}
is to be solved in a way similar to the one of the previous subsection, namely by using the Ansatz
$\vec h_{\mu\nu}=\vec q_{ab}h_{\mu\nu}$, $\vec\chi=\vec q_{ab}\chi$. Multiplying both sides of Eq.~(\ref{neweq})
by $\vec q_{ab}$, we obtain:
\begin{equation}
\label{rhs}
\frac{i}{2g_m\zeta}\varepsilon_{\mu\nu\lambda}\partial_\mu h_{\nu\lambda}=
\sum\limits_{c,d=1}^{N}\vec q_{ab}\vec q_{cd}\sin\left(g_m\vec q_{ab}\vec q_{cd}\chi\right).
\end{equation}
Let us now compute the sum on the r.h.s. of this equation by virtue of Eq.~(\ref{qq}).
To this end, imagine the antisymmetric $N\times N$-matrix of roots $\vec q_{cd}$'s.
Obviously, for a fixed root $\vec q_{ab}$, there is one root $\vec q_{cd}$ (equal to $\vec q_{ab}$) and 
a negative symmetric to it, whose
scalar products with $\vec q_{ab}$ are equal to $1$ and $-1$. Our aim is to calculate the number $n$ of roots,
whose scalar product with $\vec q_{ab}$ is equal to 
$\pm\frac12\,$.\footnote{\,For all other $N^2-N-2-n$ roots, the scalar product
vanishes, and so does the r.h.s. of Eq.~(\ref{rhs}).} According to Eq.~(\ref{qq}), these roots belong either to the rows
$c=a$, $c=b$, or to the columns $d=a$, $d=b$, which in total contain $4N-4$ elements. Among these, two are diagonal and 
other two yield the scalar product $\pm 1$. The remaining $n=4(N-2)$ roots are precisely those which 
yield the scalar product $\pm\frac12$, so that this product equals $\frac12$ for $2(N-2)$ of them and $-\frac12$ 
for the other $2(N-2)$. The sum on the r.h.s. of Eq.~(\ref{rhs}) thus takes the form
$$
\sin(2\phi)-\sin(-2\phi)+2(N-2)\left[\frac12\sin\phi-\frac12\sin(-\phi)\right]=$$
\vspace{-6mm}
$$
=2\left[\sin(2\phi)+(N-2)\sin\phi\right]\,,
$$
where the field $\phi$ has been 
defined after Eq.~(\ref{SP}). Let us now consider the limit $N\gg 1$, in which Eq.~(\ref{rhs})
takes the form $\sin\phi=iH_{ab}$, where [cf. Eq.~(\ref{Ha})]
\vspace{-2mm}
$$H_{ab}\equiv\frac{1}{4Ng_m\zeta}\,\vec q_{ab}\,\varepsilon_{\mu\nu\lambda}\partial_\mu\vec h_{\nu\lambda}\ .$$
The analogue of the potential~(\ref{Vh}) in the same limit then reads
$$
V\left[\vec h_{\mu\nu}\right]=4N\zeta\int d^3x \left[H_{ab}{\,}{\rm arcsinh}{\,} H_{ab}
-\sqrt{1+H_{ab}^2}\right].$$
The string representation of the adjoint Wilson loop in 
the large-$N$ limit is therefore given by Eq.~(\ref{wab}) with the 
substitution 
$$
\int d^3x\left[\frac{i}{2}\vec\chi\varepsilon_{\mu\nu\lambda}\partial_\mu\vec h_{\nu\lambda}+
\zeta\sum\limits_{c,d=1}^{N}
\cos\left(g_m\vec q_{cd}\vec\chi\right)\right]\Longrightarrow -V\left[\vec h_{\mu\nu}\right]$$
(and the symbol ``${\mathcal D}\vec\chi$'' removed).

We are finally interested in the adjoint-case counterparts of Eqs.~(\ref{tef}), (\ref{alf}), one can obtain in the 
weak-field limit $\left|H_{ab}\right|\ll 1\,$.\footnote{\,Note that, 
using the formula $H_{ab}=\frac{1}{2N\zeta}\vec q_{ab}\vec\rho$
and the Cauchy inequality, one gets the following analogue of Eq.~(\ref{wf}): 
$\left|H_{ab}\right|\le\frac{\left|\vec\rho\right|}{2N\zeta}$. This clearly leads to the same definition of the 
weak-field limit in terms of the low-density approximation, as in the fundamental case. Namely, the weak-field limit
corresponds to densities 
$\left|\vec\rho\right|$, which are of the order $N$ times smaller than the mean one~(\ref{me}).\\[-2mm] ~}
In this limit, formula~(\ref{Wwf}) is reproduced, with the substitution 
$\vec\mu_a\to\vec q_{ab}$. As a consequence, the ratios of adjoint-case
values of string couplings $\sigma_{\rm adj}$, $\alpha_{\rm adj}^{-1}$, and $\kappa_{\rm adj}$ 
to the respective fundamental-case values, Eqs.~(\ref{tef}) and (\ref{alf}),
are equal to $\frac{2N}{N-1}\simeq2$. In particular, for the string tensions this ratio coincides with the known 
leading large-$N$ QCD-result (see e.g. Ref.~\cite{mak}). However, unlike the fundamental string, the adjoint one 
is unstable at large distances due to the production of $W^{+}W^{-}$-pairs. The string-breaking distance, $R_c$, 
can be estimated from the balance of the string free energy, $\sigma_{\rm adj}R_c$, and the mass of the 
produced pair, $2m_W$~\cite{Alik}. By virtue of Eq.~(\ref{tef}), one obtains $R_c\propto d/\kappa$.
Therefore, the breaking length of the adjoint string is in a factor $\kappa^{-1}$ larger 
than its thickness. However, since the very existence of the string implies that its length is much larger than 
its thickness [cf. the discussion in the paragraph preceding Eq.~(\ref{alphaa})], the question of 
(non-)existence of the adjoint string becomes purely numerical.

\vspace{-3mm}

\subsection{$k$-strings}

The large-$N$ ideas discussed in the Introduction have recently found a novel realization  
in the studies of the spectrum of $k$-strings in SU($N$) gauge
theories. A $k$-string is defined as the confining flux tube between
sources in higher representations, carrying a charge $k$ with respect
to the center of the gauge group, $Z_N$, i.e. representations with
nonvanishing $N$-ality. These sources can be seen as the superposition
of $k$ fundamental charges, and charge conjugation exchanges $k$- and
$(N-k)$-strings, so that non-trivial $k$-strings exist only for
$N>3\,$;\,\footnote{\,We will not consider here high-dimensional
representations that are screened by gluons and do not yield a genuine
asymptotic string tension.} their string tensions $\sigma_k$ can, 
and should, be used to constrain mechanisms of
confinement~\cite{DelDebbio:2002yp,Greensite:2002yn}. Results for the
values of $\sigma_k$ can be obtained by various approaches. Early
results suggested the
so-called ``Casimir scaling'' hypothesis for the ratio of string
tensions~\cite{Ambjorn:1984mb}:
\begin{equation}
\label{r}
R(k,N)\equiv\frac{\sigma_k}{\sigma_1} = \frac{k(N-k)}{N-1} \equiv C(k,N),
\end{equation}
where $\sigma_1$ is the fundamental string tension. These are based on the arguments of the 
Parisi-Sourlas dimensional reduction of the 4D QCD to the 2D one, which is due to stochastic 
vacuum fields, and the further use of the fact that, in 2D, confinement is produced by the one-gluon exchange.
Recent studies in
supersymmetric Yang--Mills theories and M-theory suggest instead a ``Sine
scaling'' formula:
\vspace{-3mm}
\begin{equation}
R(k,N)=\frac{\sin\left(k \pi/N\right)}{\sin\left(\pi/N\right)}\ .
\end{equation}
Corrections are expected to both formulae, but the form of such
corrections is unknown for the physically relevant case of a 4D, 
nonsupersymmetric, SU($N$) gauge theory.

In the large-$N$ limit, where the interactions between flux tubes are
suppressed by powers of $1/N$, the lowest-energy state of the system
should be made of $k$ fundamental flux tubes connecting the sources,
hence:
\begin{equation}
R(k,N) \stackrel{{k-{\rm fixed}}\atop{N\to \infty}}{\longrightarrow} k\ .
\end{equation}
Both the Casimir and the Sine scaling formulae satisfy this
constraint; they also remain invariant under the replacement $k \leftrightarrow
(N-k)$, which corresponds to the exchange of quarks with
antiquarks. However, it has been argued in Refs.~\cite{Armoni:2003ji,
Armoni:2003nz} that the correction to the large-$N$ behavior should
occur as a power series in $1/N^2$ rather than $1/N\,$.\,\footnote{\,One
should study with some care whether the arguments presented in
Refs.~\cite{Armoni:2003ji, Armoni:2003nz} hold independently of the
space-time dimensionality.} Clearly, such a behavior would exclude
Casimir scaling as an {\it exact}\ description of the $k$-string
spectrum.

Recent lattice calculations have provided new results for the spectrum
of $k$-strings both in three and four
dimensions~\cite{Lucini:2000qp,Lucini:2001ej,Lucini:2001nv,
DelDebbio:2001kz,DelDebbio:2001sj,DelDebbio:2003tk}. They all confirm
that Casimir scaling is a {\it good approximation}\ to the Yang--Mills
results. To be more quantitative, one could say that all lattice
results are within 10\% of the Casimir-scaling prediction, and that
deviations from it are larger in four than they are in three
dimensions, in agreement with strong-coupling
predictions~\cite{DelDebbio:2001sj}. The taming of systematic errors
is a crucial matter for such lattice calculations, and it can only be
achieved by an intensive numerical analysis. In four dimensions, the
higher statistics simulations presented in
Ref.~\cite{DelDebbio:2001sj} show that corrections to the Casimir-scaling 
formula are statistically significant, and actually favor the
Sine scaling. Finally, it has been pointed out in
Ref.~\cite{DelDebbio:2003tk} that higher-dimensional representations
with common $N$-ality do yield the same string tension, as expected
because of gluon screening.

These numerical results trigger a few comments on Casimir scaling. The
original argument~\cite{Ambjorn:1984mb} was based on the idea that a
4D gauge theory in a random magnetic field could be described by a
2D theory without such a field. Besides the numerical results, there
is little support for such an argument in QCD; moreover it is not
clear that the same hypothesis could explain the approximate Casimir
scaling observed in three dimensions. On the other hand, Casimir
scaling appears ``naturally'' as the lowest-order result, both at
strong-coupling in the case of $k$-strings in the Hamiltonian
formulation of gauge theories, and in the case of the spectrum of
bound states in chiral models. Corrections can be computed in the
strong-coupling formulation and they turn out to be $\propto (D-2)/N$
-- see e.g. Ref.~\cite{DelDebbio:2001sj} for a summary of results and
references. While strong-coupling calculations are
not directly relevant to describe the physics of the continuum theory, 
it is nonetheless instructive to have some quantitative
analytic control within that framework. Last but not least, Casimir
scaling also appears at the lowest order in the stochastic model of
the QCD vacuum~\cite{xxx}. In view of these considerations, it is fair
to say that approximate Casimir scaling should be a prerequisite for
any model of confinement, that corrections should be expected, and
that these corrections are liable to yield further information about
the nonperturbative dynamics of strong interactions. Moreover, it
would be very interesting to improve our understanding of some other
aspects of the $k$-string spectrum, like e.g. the origin of the Sine
scaling for nonsupersymmetric theories, or the structure of the
corrections to this scaling form.

We will prove below that Casimir scaling occurs to a 
high accuracy in the low-density approximation 
of the SU($N$) 3D GG model~\cite{luidim}.
The $k$-string tension is defined by means of the $k$-th power of the
fundamental Wilson loop. The surface-dependent part of the latter 
can be written, in terms of the dual-photon field, as follows:
\vspace{-1mm}
\begin{equation}
\label{299}
\hspace{0.1cm}
\left\langle W_k({\mathcal C})\right\rangle_{\rm mon}=\!\!\!\!\!\sum\limits_{a_1,\ldots,
a_k=1}^{N}\!\!\left\langle\exp\left[-ig\left(
\sum\limits_{i=1}^{k}\vec\mu_{a_i}\right)\!\! \int\!\! d^3x\,\Sigma\left(\vec
x\right) \vec\chi\left(\vec x\right) \right]\right\rangle_{\rm mon}\!\!.
\end{equation}
\vspace{-2mm}
In this equation,
$$\Sigma\left(\vec x\right)\equiv\int\limits_{\Sigma({\mathcal
C})}^{}d\sigma_\mu\left(\vec
x(\xi)\right)\partial_\mu^x\delta\left(\vec x- \vec x(\xi)\right),$$
where again $\Sigma({\mathcal C})$ is an arbitrary surface bounded by the
contour ${\mathcal C}$ and parametrized by the vector $\vec x(\xi)$.

The independence of Eq.~(\ref{299}) of the choice of $\Sigma({\mathcal C})$
can readily be seen in the same way as for the $(k=1)$-case. 
The $\Sigma$-dependence rather appears in the weak-field, or low-density,
approximation, which is equivalent to keeping only the quadratic term in the
expansion of the cosine in Eq.~(\ref{2}). As has been demonstrated in subsection~4.1 
[cf. the discussion around Eqs.~(\ref{wf}), (\ref{me})],
the notion ``low-density'' implies that the
typical monopole density is related to the mean one, $\rho_{\rm
mean}=\zeta N(N-1)$, by the following sequence of inequalities:
\vspace{-4mm}
\begin{equation}
\label{rhorho} 
\rho_{\rm typical}\ll\zeta\cdot{\mathcal O}(N)\ll\rho_{\rm
mean}=\zeta\cdot{\mathcal O}\left(N^2\right).
\end{equation}
We will discuss in some more detail the
correspondence between the low-density and the large-$N$
approximations after Eq.~(\ref{quart}).

Denoting for brevity $\vec a\equiv-g\int d^3x\Sigma\left(\vec x\right)
\vec\chi\left(\vec x\right)$, we can rewrite Eq.~(\ref{299}) as
\vspace{-4mm}
\begin{equation}
\label{rep}
\left\langle W_k({\mathcal C})\right\rangle_{\rm mon}=\hspace{-4mm}\sum\limits_{{a_1,\ldots,
a_k=1}\choose{{\rm with}~ {\rm possible}~ {\rm
coincidences}}}^{N}\left\langle {\rm e}^{i\vec
a\left(\vec\mu_{a_1}+\cdots+\vec\mu_{a_k}\right)}\right\rangle,
\end{equation}
where in the low-density approximation the average is defined with
respect to the action
\begin{equation}
\label{quadr}
\int d^3x\left[
\frac12\left(\partial_\mu\vec\chi\right)^2+\frac{m_D^2}{2}\vec\chi^2\right].
\end{equation} 
Similarly to the fundamental representation, in this approximation the string tension for a given $k$
is the same for all surfaces $\Sigma({\mathcal C})$, which are large
enough in the sense $\sqrt{S}\gg d$, where $S$ is the area of
$\Sigma({\mathcal C})$. In
particular, the fundamental string tension found above reads
$\sigma_1=\frac{N-1}{2N}\bar\sigma$, where $\bar\sigma\equiv 4 \pi^2
\frac{\sqrt{\zeta N}}{g_m}$, and the factor $\frac{N-1}{2N}$ is the
square of a weight vector.

To evaluate Eq.~(\ref{rep}) for $k>1$, we should calculate the
expressions of the form
\begin{equation}
\label{25}
\Bigg(n\vec\mu_{a_i}+\sum\limits_{j=1}^{k-n}\vec\mu_{a_j}\Bigg)^2,
\end{equation}
where $(k-n)$ weight vectors $\vec\mu_{a_j}$'s are mutually different
and also different from the vector $\vec\mu_{a_i}$. By virtue of the
formula
$\vec\mu_a\vec\mu_b=\frac12\left(\delta_{ab}-\frac{1}{N}\right)$, we
obtain for Eq.~(\ref{25}):
\begin{equation}
\frac{N-1}{2N}\left(n^2\!+\!k\!-\!n\right)-\frac{1}{2N}\left[2n(k\!-\!n)+2\!\!\sum
\limits_{l=1}^{k-n-1}l\right]   
=\frac{k(N\!-\!k)}{2N}+\frac12\left(n^2\!-\!n\right).
\label{27}
\end{equation}

We should further calculate the number of times a term with a given
$n$ appears in the sum~(\ref{25}). In what follows, we will consider
the case $k<N$, although $k$ may be of the order of $N$. Then,
$C_k^n\equiv\frac{k!}{n!(k-n)!}$ possibilities exist to choose out of
$k$ weight vectors $n$ coinciding ones, whose index may acquire any
values from $1$ to $N$.  The index of any weight vector out of the remaining 
$(k-n)$ ones may then acquire only $(N-1)$ values, and so on. Finally,
the index of the last weight vector may acquire $(N-k+n)$
values. Therefore, the desired number of times a term with a given
$n$ appears in the sum~(\ref{25}) is
$$
C_k^nN\cdot(k-n)(N-1)\cdot(k-n-1)(N-2)\cdots1(N-k+n)=
$$
\vspace{-7mm}
\begin{equation}
\label{29}
=C_k^nA_N^{k-n+1}(k-n)!=\frac{k!N!}{n!(n+N-k-1)!}\ ,
\end{equation}
where $A_N^{k-n+1}\equiv\frac{N!}{(N-k+n-1)!}$. Equations~(\ref{27})
and (\ref{29}) together yield for the monopole contribution to the
Wilson loop, Eq.~(\ref{rep}):
\vspace{-2mm}
\begin{equation}
\label{2991}
\left\langle W_k({\mathcal C})\right\rangle_{\rm mon}=k!N!{\rm e}^{-C\bar\sigma
  S}\cdot\sum\limits_{n=1}^{k} 
\frac{1}{n!(n+N-k-1)!}{\rm e}^{-\frac{n^2-n}{2}\bar\sigma S}\ ,
\end{equation}
where $C\equiv\frac{k(N-k)}{2N}$ is proportional to the Casimir of the
rank-$k$ antisymmetric representation of SU($N$). We have thus arrived
at a Feynman--Kac--type formula, where, in the asymptotic regime of
interest, $S\to\infty$, only the first term in the sum is
essential. The $k$-string tension is therefore 
$\sigma_k=C\bar\sigma$, yielding the Casimir-scaling law~(\ref{r}).
It is interesting to note that the Casimir of the original unbroken
SU($N$) group is recovered. This is a consequence of the Dirac
quantization condition~\cite{roots}, which distributes the quark
charges along the weights of the fundamental representation and the
monopole ones along the roots. The orthonormality of the roots then
yields the action (\ref{quadr}), which is diagonal in the dual
magnetic variables; the sum of the weights squared is responsible for
the Casimir factor, since
$C=\left(\vec\mu_{a_1}+\cdots+\vec\mu_{a_k}\right)^2$, where all $k$
weight vectors are different from each other. Therefore, terms where
all $k$ weight vectors are mutually different 
yield the dominant contribution to the sum~(\ref{rep}).  Their number
in the sum is equal to $\frac{k!N!}{(N-k)!}$, that corresponds to the
$(n=1)$-term in Eq.~(\ref{2991}).

Let us further address the leading correction to Casimir scaling, which
originates from the non-diluteness of plasma.  Expanding the cosine up to the quartic term and using the formula
$$\sum\limits_{i=1}^{N(N-1)/2}q_i^\alpha q_i^\beta q_i^\gamma
q_i^\delta=\frac{N}{2(N+1)}
\left(\delta^{\alpha\beta}\delta^{\gamma\delta}+\delta^{\alpha\gamma}
\delta^{\beta\delta}+\delta^{\alpha\delta}\delta^{\beta\gamma} 
\right),$$ 
which stems from the orthonormality of roots, we obtain the action 
\begin{equation}
\label{quart}
\int d^3x\left[\frac12
\left(\partial_\mu\vec\chi\right)^2+\frac{m_D^2}{2}
\left(\vec\chi^2-\frac{g_m^2}{12(N+1)}\vec\chi^4\right)\right].
\end{equation}
By virtue of this formula, one can analyze the correspondence between
the $1/N$-expansion and corrections to the low-density approximation.
The natural choice for defining the behavior of the electric coupling
constant in the large-$N$ limit is the
QCD-inspired one, $g={\mathcal O}(N^{-1/2})$. To make some estimates, 
let us use an obvious argument that the $\vec\chi$-field
configuration dominating in the partition function is the one where
every term in the action~(\ref{quart}) is of the order of unity. When
applied to the kinetic term, this demand tells us that the
characteristic wavelength $l$ of the field $\vec\chi$ is related to
the amplitude of this field as $l\sim|\vec\chi|^{-2}$. Substituting
further this estimate into the condition $l^3m_D^2|\vec\chi|^2\sim 1$,
we get $|\vec\chi|^2\sim m_D$. The ratio of the quartic and mass terms,
being of the order of $|\vec\chi|^2g_m^2/N$, can then be estimated as
$\frac{m_Dg_m^2}{N}=g_m^3\sqrt{\frac{\zeta}{N}}\sim N\sqrt{\zeta}$. With
the exponentially high accuracy, this ratio is small, provided
$N\lesssim{\mathcal O}\left({\rm e}^{S_0/2}\right)$.\footnote{\,This constraint
is similar to those which were derived at the end of sect.~2 as the necessary 
conditions for the stochasticity of the Higgs vacuum. We therefore conclude that 
the conditions of stochasticity of the Higgs vacuum and of the validity of the dilute-plasma 
approximation parallel each other.}  
Therefore, the non-diluteness corrections are suppressed only for 
$N$'s bounded from above by a certain parameter. However, due to the exponential largeness of this parameter,
the obtained constraint leaves enough space for $N$ to be sufficiently large 
to provide the validity of the inequalities~(\ref{rhorho}).

To proceed with the study of the non-diluteness correction, one needs
to solve iteratively the saddle-point equation, which corresponds to the
average~(\ref{rep}) taken with respect to the approximate
action~(\ref{quart}). Since it has been demonstrated above that the
string tension is defined by the averages $\left\langle{\rm e}^{i\vec
a\left(\vec\mu_{a_1}+\cdots+\vec\mu_{a_k}\right)}\right\rangle$, where
all $k$ weight vectors are mutually different, let us restrict
ourselves to such terms in the sum~(\ref{rep}) only.  Solving then the
saddle-point equation with the Ansatz
$\vec\chi=\vec\chi_0+\vec\chi_1$, where $|\vec\chi_1|\ll|\vec\chi_0|$,
we obtain for such a term:
\begin{equation}
\label{31}
-\ln\left\langle{\rm e}^{i\vec a
\left(\vec\mu_{a_1}+\cdots+\vec\mu_{a_k}\right)}\right\rangle=
\frac{g^2}{2}\,C\!\!\int\! d^3xd^3y\Sigma\left(\vec x\right)D_{m_D}
\left(\vec x-\vec y\right)
\Sigma\left(\vec y\right)+\Delta {\mathcal S}\ .
\end{equation}
The first term on the r.h.s. of Eq.~(\ref{31})
yields the string tension $\sigma_k=C\bar\sigma$, while the second
term yields the desired correction. This term reads
$$
\Delta {\mathcal S}=\frac{2\pi^2}{3}\frac{(gm_DC)^2}{N+1}\int d^3x
\prod\limits_{l=1}^{4}\left[\int d^3x_lD_{m_D}
\left(\vec x-\vec x_l\right)\Sigma\left(\vec x_l\right)\right]=$$
\vspace{-5mm}
\begin{equation}
\label{6}
=\frac{\pi^2}{6}\frac{(gm_DC)^2}{N+1}
\!\int\! d\sigma_{\mu\nu}\left(\vec x_1\right)d\sigma_{\mu\nu}\left(\vec x_2\right)
d\sigma_{\lambda\rho}\left(\vec x_3\right)d\sigma_{\lambda\rho}
\left(\vec x_4\right)
\partial_\alpha^{x_1}\partial_\alpha^{x_2}\partial_\beta^{x_3}
\partial_\beta^{x_4}I\ ,
\end{equation}
where $I\equiv\int d^3x\prod\limits_{l=1}^{4}D_{m_D}\left(\vec x-\vec
x_l\right)$.  The action~(\ref{6}) can be represented in the form
\vspace{-3mm}
\begin{eqnarray}
\label{7}
\Delta {\mathcal S}=\frac{(gC)^2}{(N+1)m_D^5} \int d\sigma_{\mu\nu}(\vec x_1) 
d\sigma_{\mu\nu}(\vec x_2) D(\vec x_1 - \vec x_2) \; \times
\nonumber \\ 
\times \int d\sigma_{\lambda\rho}(\vec x_3) 
d\sigma_{\lambda\rho}(\vec x_4) D(\vec x_3 - \vec x_4) \;\times\; 
G(\vec x_1 - \vec x_3)\,.
\end{eqnarray}
Here, $D$ and $G$ are some positive functions, which depend on $m_D|\vec
x_i-\vec x_j|$ and vanish exponentially at the distances $\gtrsim
d$. They can be represented as $D(\vec x)=m_D^4{\mathcal D}(m_D|\vec
x|)$, $G(\vec x)=m_D^4{\mathcal G}(m_D|\vec x|)$, where the functions ${\mathcal
D}$ and ${\mathcal G}$ are dimensionless.\footnote{\,Our investigations can
readily be translated to the stochastic vacuum model of QCD~\cite{xxx}
for the evaluation of a correction to the string tension, produced by
the four-point irreducible average of field strengths. In that case,
the functions $D$ and $G$ would be proportional to the gluonic
condensate.}

The derivative expansion yields the Nambu--Goto
action as the leading term:
\begin{equation}
\int d\sigma_{\mu\nu}(\vec x_1) 
d\sigma_{\mu\nu}(\vec x_2) D(\vec x_1 - \vec x_2) =
\sigma_D\int d^2\xi\sqrt{{\sf g}(\vec
x_1)}+{\mathcal O} \left(\frac{\sigma_D}{m_D^2}\right).
\end{equation}
Here, $\sigma_D=2m_D^2\int d^2z{\mathcal D}(|z|)$ (with $z$ being
dimensionless).
Recalling that $d\sigma_{\alpha\beta}(\vec x) =\sqrt{{\sf g}(\vec x)} t_{\alpha\beta}(\vec
x)d^2\xi$, we may further take into
account that we are interested in the {\it leading} term of the
derivative expansion of the action $\Delta {\mathcal S}$, which corresponds to
the so short distance $\left|\vec x_1-\vec x_3\right|$, that
$t_{\alpha\beta}\left(\vec x_1\right)t_{\alpha\beta}\left(\vec x_3\right) \simeq
2$. (Higher terms of the derivative expansion contain derivatives of
$t_{\alpha\beta}$ and do not contribute to the string tension.) This yields
for the integral in Eq.~(\ref{7}):
\vspace{-5mm}
$$
\sigma_D^2\int d^2\xi d^2\xi'\sqrt{{\sf g}(\vec x_1){\sf g}(\vec x_3)}\,G(\vec x_1-\vec x_3)\simeq$$
\vspace{-7mm}
$$\simeq\frac{\sigma_D^2}{2}
\!\int\! d\sigma_{\alpha\beta}\left(\vec x_1\right)d\sigma_{\alpha\beta}
\left(\vec x_3\right)
G\left(\vec x_1-\vec x_3\right)=
\frac{\sigma_D^2}{2} \left[\sigma_G\!\int\! d^2\xi\sqrt{\sf g}+{\mathcal O}
\left(\frac{\sigma_G}{m_D^2}\right)\right],$$ 
where $\sigma_G=2m_D^2\int d^2z{\mathcal G}(|z|)$. We finally obtain from
Eq.~(\ref{7}) that 

$$
\Delta\sigma_k\simeq\frac{(gC)^2\sigma_D^2\sigma_G}{2(N+1)m_D^5}=
\frac{\alpha}{4}
\frac{(gC)^2m_D}{N+1}=\alpha\frac{\bar\sigma C^2}{N+1},$$
where $\alpha$ is some dimensionless positive constant. This yields
$$
\sigma_k+\Delta\sigma_k=\bar\sigma C\left(1+\frac{\alpha C}{N+1}\right).
$$ 
In the limit $k\sim N\gg 1$ of interest, the obtained correction is
${\mathcal O}(1)$, while for $k={\mathcal O}(1)$ it is ${\mathcal
O}(1/N)$.  The latter fact enables one to write down the following
final result for the leading correction to Eq.~(\ref{r}) due to the
non-diluteness of monopole plasma:
$$
R(k,N)+\Delta R(k,N)\equiv
\frac{\sigma_k+\Delta\sigma_k}{\sigma_1+\Delta\sigma_1}=
C(k,N)\left[1+\alpha\frac{(k-1)(N-k-1)}{2N(N+1)}\right].
$$
This expression is invariant under the replacement $k\leftrightarrow
(N-k)$ just as the expression~(\ref{r}), which does not account for
non-diluteness. The fact that, at $k\sim N\gg 1$, the obtained
correction to the Casimir-scaling law is ${\mathcal O}(1)$, indicates that
non-diluteness effects can
significantly distort the Casimir-scaling behavior.

\setcounter{equation}{0}
\section{Generalization to the SU(\boldmath{$N$})-analogue of 4D compact QED with the $\theta$-term}

In this section, we will  
consider the 4D-case and introduce the field-theoretical 
$\theta$-term.\footnote{\,In the case when the ensemble of 
Abelian-projected monopoles is modeled by the magnetically-charged dual Higgs field, the effects produced by 
this term on the respective 
string effective action have been studied in Ref.~\cite{term}.}
As was first found for 
compact QED in Refs.~\cite{theta, tq} by means of the derivative expansion of the resulting 
nonlocal string effective action,
this term generates the string $\theta$-term. Being proportional to the number of self-intersections of the world sheet,
the latter might be important for the solution of the problem of crumpling of large world sheets~\cite{polbook, fi}.
In this section, we will 
perform the respective analysis for the general SU($N$)-case under study, in the fundamental and 
in the adjoint representations, as well as for $k$-strings.
It is worth noting that, in the lattice 4D compact QED, confinement holds only at strong coupling. On the other hand,
the continuum counterpart~\cite{theta, tq} [of the SU($N$)-version] of this model, which we are going to explore,
possesses confinement at arbitrary values of coupling. However, we will see that the solution to the problem of crumpling
due to the $\theta$-term is only possible in the strong-coupling regime, implied in a certain sense.
Note also that the continuum
sine-Gordon theory of the dual-photon field 
possesses an ultraviolet cutoff, $\Lambda$, which appears in course of the path-integral average over the shapes of
monopole loops.
However, the $\theta$-parameter is dimensionless, and consequently its values,
at which one may expect the disappearance of crumpling, will be cutoff-independent.

The full partition function, including the $\theta$-term and the average over free photons, reads [cf. Eq.~(\ref{1})]
\begin{equation}
\label{Z4D}
{\mathcal Z}^N=\int {\mathcal D}\vec A_\mu{\rm e}^{-\frac14\int d^4x\vec F_{\mu\nu}^2}\sum\limits_{{\mathcal N}=0}^{\infty}
\frac{\zeta^{\mathcal N}}{{\mathcal N}!}\left<\exp\left\{-{\mathcal S}\left[\vec j_\mu^{\mathcal N}, 
\vec A_\mu\right]\right\}\right>_{\rm mon},
\end{equation}
where
\begin{equation}
\label{Sini}
{\mathcal S}\left[\vec j_\mu^{\mathcal N}, \vec A_\mu\right]=
\frac{1}{2}\int d^4xd^4y\vec j_\mu^{\mathcal N}(x)
D_0(x-y)\vec j_\mu^{\mathcal N}(y)+
\frac{i\theta g^2}{8\pi^2}
\int d^4x\vec A_\mu\vec j_\mu^{\mathcal N},
\end{equation}
and $\vec F_{\mu\nu}=\partial_\mu\vec A_\nu-\partial_\nu\vec A_\mu$.
For ${\mathcal N}=0$, the monopole current $j_\mu^{\mathcal N}$ is equal to zero, whereas for
${\mathcal N}\ge 1$, it is defined as
$\vec j_\mu^{\mathcal N}=
g_m\sum\limits_{k=1}^{\mathcal N}\vec q_{i_k}
\oint dz_\mu^k(\tau)\delta\left(x-x^k(\tau)\right)$. The couplings 
$g$ and $g_m$ are now dimensionless and obey the same condition $gg_m=4\pi$
as in the 3D-case. We have also parametrized the trajectory of the
$k$-th monopole by the vector $x_\mu^k(\tau)=y_\mu^k+z_\mu^k(\tau)$,
where $y_\mu^k=\int\limits_{0}^{1}d\tau x_\mu^k(\tau)$ is the
position of the trajectory, whereas the vector $z_\mu^k(\tau)$
describes its shape, both of which should be averaged over.
The fugacity of a single-monopole loop, $\zeta$, entering Eq.~(\ref{Z4D}), has the dimensionality $[{\rm mass}]^4$,
$\zeta\propto {\rm e}^{-S_{\rm mon}}$, where
the action of a single $k$-th loop, obeying the
estimate $S_{\rm mon}\propto\frac{1}{g^2}\int\limits_{0}^{1}d\tau
\sqrt{\left(\dot z^k\right)^2}$, is assumed to be of the same order of magnitude
for all loops. Finally, in Eq.~(\ref{Z4D}), 
$D_0(x)=1/(4\pi^2x^2)$ is the 4D Coulomb propagator, and the average over monopole loops
is defined similarly to Eq.~(\ref{monav}) as follows:
$$\left<{\mathcal O}\right>_{\rm mon}=
\prod\limits_{n=0}^{\mathcal N}\int d^4y_n
\sum\limits_{i_n=\pm 1,\ldots,\pm\frac{N(N-1)}{2}}^{}\left<{\mathcal O}\right>_{z_n(\tau)}\ .$$
The particular form of the path-integral average over the shapes of 
the loops, $z_n(\tau)$'s, here is immaterial for the final (ultraviolet-cutoff dependent)
expression for the partition function 
[see e.g. Ref.~\cite{jh} for a similar situation in the plasma of closed dual strings in the Abelian--Higgs--type
models]. The only thing which matters is the normalization $\left<1\right>_{z_n(\tau)}=1$, that will be implied
henceforth. The analogue of the partition function~(\ref{II}), (\ref{III}) then reads
$$
{\mathcal Z}_{\rm mon}^N\left[\vec A_\mu\right]=\int {\mathcal D}\vec j_\mu\exp\left\{-{\mathcal S}
\left[\vec j_\mu, \vec A_\mu\right]\right\}\times$$
\vspace{-8mm}
\begin{equation}
\label{interm}
\times\int{\mathcal D}\vec\chi_\mu
\exp\Bigg\{\int d^4x\Bigg[2\zeta\sum\limits_{i}^{}
\cos\left(\frac{\vec q_i\left|\vec\chi_\mu\right|}{\Lambda}\right)+i\vec\chi_\mu\vec j_\mu\Bigg]\Bigg\}\,,
\end{equation}
with the action ${\mathcal S}$ given by Eq.~(\ref{Sini}),
while the full partition function is
$${\mathcal Z}^N=\int {\mathcal D}\vec A_\mu{\rm e}^{-\frac14\int d^4x\vec F_{\mu\nu}^2}{\mathcal Z}_{\rm mon}^N
\left[\vec A_\mu\right].$$
In Eq.~(\ref{interm}),
\vspace{-4mm}
\begin{equation}
\label{absv}
\left|\vec\chi_\mu\right|\equiv
\left(\sqrt{\chi_\mu^1\chi_\mu^1},\ldots,\sqrt{\chi_\mu^{N-1}\chi_\mu^{N-1}}\right),
\end{equation}
and $\Lambda\sim\left|y^k\right|/\left(z^k\right)^2
\gg \left|z^k\right|^{-1}$ is the ultraviolet cutoff.
Clearly, unlike the 3D-case without the $\theta$-term, the $\vec A_\mu$-field is now coupled to $\vec j_\mu$, making
${\mathcal Z}_{\rm mon}^N$ $\vec A_\mu$-dependent. 
This eventually leads to the change of the mass of the dual photon $\vec\chi_\mu$, as well
as to the appearance of the string $\theta$-term.

Let us address the fundamental case first.
After the saddle-point integration over $\vec\chi_\mu\,$,\footnote{\,Unlike the superrenormalizable 
3D-case, where loop corrections to the saddle point are rapidly converging, this is not necessarily the 
case in four dimensions. This fact is, however, clearly unimportant in the weak-field limit 
[Eq.~(\ref{STAR}) below], where the cosine in Eq.~(\ref{interm}) is approximated by the quadratic
term only, and the saddle-point integration over $\vec\chi_\mu$ becomes Gaussian. As we have seen in the previous section, 
this limit is already enough for the discussion of the string representation. One of the authors (D.A.) is grateful to H.~Gies and 
E.~Vicari for drawing his attention to this issue.}
the analogue of Eqs.~(\ref{WwW}), (\ref{Vh}) takes the form
\begin{equation}
\label{danuovo}
W_a=
\frac{1}{{\mathcal Z}^N}
\int{\mathcal D}\vec h_{\mu\nu}
\exp\Bigg\{-S\left[\vec h_{\mu\nu}\right]
+\frac{ig}{2}\vec\mu_a\int\limits_{\Sigma(C)}^{}d\sigma_{\mu\nu}\vec h_{\mu\nu}\Bigg\}\,,
\end{equation}
where the constraint $\partial_\mu\vec h_{\mu\nu}=0$ was already abolished. Here, the Kalb--Ramond action reads
\begin{equation}
\label{fourkalb}
S\left[\vec h_{\mu\nu}\right]=\int d^4x\Biggl(\frac{1}{4}\vec h_{\mu\nu}^2-\frac{i\theta g^2}{32\pi^2}
\vec h_{\mu\nu}\tilde{\vec h}_{\mu\nu}\Biggr)+V\left[\vec h_{\mu\nu}\right].
\end{equation}
The potential $V$ in this equation is given by Eq.~(\ref{Vh}) with the symbol $\int d^3x$ replaced by $\int d^4x$, and

$$H_a=\frac{g\Lambda}{\zeta(N-1)}\vec\mu_a\left|\partial_\mu\tilde{\vec h}_{\mu\nu}\right|.$$
In these formulae,
$\tilde{\mathcal O}_{\mu\nu}\equiv\frac12\varepsilon_{\mu\nu\lambda\rho}
{\mathcal O}_{\lambda\rho}$,
and the absolute value is defined in the same way as in Eq.~(\ref{absv}), 
i.e. again with respect to the Lorentz indices only.
Note that the form to which the $\theta$-term has been transformed is quite natural, since
the respective initial expression of Eq.~(\ref{Sini}) can be rewritten modulo full derivatives as
$$
\frac{i\theta g^2}{8\pi^2}
\int d^4x\vec A_\mu\vec j_\mu^{\mathcal N}=-\frac{i\theta g^2}{32\pi^2}\int d^4x
\left(\vec F_{\mu\nu}+\vec F_{\mu\nu}^{\mathcal N}\right)\left(
\tilde{\vec F}_{\mu\nu}+\tilde{\vec F}_{\mu\nu}^{\mathcal N}\right),$$
where [cf. Eq.~(\ref{F})]
$$
\vec F_{\mu\nu}^{\mathcal N}(x)=-\varepsilon_{\mu\nu\lambda\rho}\partial_\lambda\int 
d^4yD_0(x-y)\vec j_\rho^{\mathcal N}(y),~~
\partial_\mu\tilde{\vec F}_{\mu\nu}^{\mathcal N}=\vec j_\nu^{\mathcal N}.$$
Next, the mass of the Kalb--Ramond field, equal to
the Debye mass of the dual photon,
which follows from the action~(\ref{fourkalb}), 
reads $m_D=\frac{g\eta}{4\pi}\sqrt{\left(\frac{4\pi}{g^2}\right)^2+\left(\frac{\theta}{2\pi}\right)^2}$,
where $\eta\equiv\sqrt{N\zeta}/\Lambda$. In the extreme strong-coupling limit, 
$g\to\infty$, this expression demonstrates the important
difference of the case $\theta=0$ from the case $\theta\ne 0$. 
Namely, since $\eta(g)\propto {\rm e}^{-{\rm const}/g^2}\to 1$,
$m_D\to 0$ at $\theta=0$, whereas $m_D\to\infty$ at $\theta\ne 0$. 
In another words, in the extreme strong-coupling limit, the correlation
length of the vacuum, equal to $d$, goes large (small) at $\theta=0$ ($\theta\ne 0$).

In the weak-field limit, $\left|H_a\right|\ll 1$, Eq.~(\ref{danuovo}) yields [cf. Eq.~(\ref{Wwf})]:
$$
W\left(C,\Sigma\right)_{\rm weak-field}^{{\rm tot},~ a}=\frac{1}{{\mathcal Z}^{\rm tot}}
\int {\mathcal D}\vec h_{\mu\nu}\times$$
\vspace{-6mm}
\begin{equation}
\label{STAR}
\times\exp\Bigg\{-\int d^4x\Bigg[\frac{g^2}{12\eta^2}\vec H_{\mu\nu\lambda}^2+
\frac{1}{4}\vec h_{\mu\nu}^2-\frac{i\theta g^2}{32\pi^2}
\vec h_{\mu\nu}\tilde{\vec h}_{\mu\nu}-\frac{ig}{2}\vec\mu_a
\vec h_{\mu\nu}
\Sigma_{\mu\nu}\Bigg]\Bigg\}\,,
\end{equation}
where again ${\mathcal Z}^{\rm tot}$ is the same integral over $\vec h_{\mu\nu}$, but with $\Sigma_{\mu\nu}$ set to zero.
Integrating over $\vec h_{\mu\nu}$, we then obtain:
$$W\left(C,\Sigma\right)_{\rm weak-field}^{{\rm tot},~ a}
=\exp\Bigg\{-\frac{N-1}{4N}\Bigg[g^2\oint\limits_{C}^{}dx_\mu
\oint\limits_{C}^{}dx'_\mu D_{m_D}(x-x')+$$
\vspace{-5mm}
\begin{equation}
\label{w4d}
+\frac{\eta^2}{2}\int d^4xd^4x'
D_{m_D}(x-x')\left(\Sigma_{\mu\nu}(x)\Sigma_{\mu\nu}(x')+
\frac{i\theta g^2}{8\pi^2}\Sigma_{\mu\nu}(x)\tilde\Sigma_{\mu\nu}(x')
\right)\Bigg]\Bigg\}\,,
\end{equation}
where $D_m(x)\equiv mK_1\left(m|x|\right)/(4\pi^2|x|)$ is the 4D Yukawa propagator with $K_1$ denoting the
modified Bessel function. [Note that, at $\theta=0$, Eq.~(\ref{w4d}) takes the form of Eq.~(\ref{WCSig}).]
Further curvature expansion of the
$\Sigma_{\mu\nu}\times\tilde\Sigma_{\mu\nu}$ interaction
[analogous to the expansion of the $\Sigma_{\mu\nu}\times\Sigma_{\mu\nu}$ 
interaction in the action~(\ref{snonloc})] yields the
string $\theta$-term equal to $ic_{\rm fund}\nu$. Here,
$\nu\equiv\frac{1}{2\pi}
\int d^2\xi\sqrt{{\sf g}}{\sf g}^{ab}(\partial_a t_{\mu\nu})(\partial_b
\tilde t_{\mu\nu})$ is the number of self-intersections of the
world sheet, and the coupling constant $c_{\rm fund}$ reads
\begin{equation}
\label{ccrit}
c_{\,\rm fund}=\frac{(N-1)\theta}{8N\Big[
\left(\frac{4\pi}{g^2}\right)^2+\left(\frac{\theta}{2\pi}\right)^2\Big]}\ .
\end{equation}
As has already been discussed,
$c_{\,\rm fund}$ is $\Lambda$-independent, since (similarly to the rigidity coupling constant $\alpha$) it is
dimensionless. We therefore see that, at 
\vspace{-3mm}
\begin{equation}
\label{thcr}
\theta_{\pm}^{\rm fund}=\frac{\pi}{2}\left[\frac{N-1}{2N}\pm\sqrt{\left(\frac{N-1}{2N}\right)^2-\left(\frac{16\pi}{g^2}
\right)^2}\right],
\end{equation}
$c_{\rm fund}$ becomes equal to $\pi$, and 
self-intersections are weighted in the string partition function with the factor $(-1)^\nu$, that might
cure the problem of crumpling. This is only possible at $g\ge g^{\rm fund}\equiv 
4\sqrt{\frac{2\pi N}{N-1}}$,\,\footnote{\,Clearly, 
at $g=g^{\rm fund}$,\; $\theta_{+}^{\rm fund}=\theta_{-}^{\rm fund}=
\pi\,\frac{N-1}{4N}\,$.} paralleling the strong-coupling regime, which 
should hold
in the lattice version of the model (cf. the discussion in the beginning 
of this section).
In the extreme strong-coupling limit, understood in 
the sense $g\gg g^{\rm fund}$,
only one critical value, $\theta_{+}^{\rm fund}$, 
survives, $\theta_{+}^{\rm fund}\to\pi\frac{N-1}{2N}$, whereas
$\theta_{-}^{\rm fund}\to 0$, i.e. $\theta_{-}^{\rm fund}$ 
becomes a spurious solution, since $\left.c_{\rm fund}\right|_{\theta=0}=0$.

With the use of the results of subsection~4.2, 
the adjoint-case large-$N$ counterpart of Eq.~(\ref{thcr}) can readily 
be found.
Indeed, Eq.~(\ref{ccrit}) in that case becomes replaced by
\vspace{-3mm}
$$
c_{\rm adj}=\frac{\theta}{4\Big[\left(\frac{4\pi}{g^2}\right)^2+\left(\frac{\theta}{2\pi}\right)^2\Big]},
$$
so that $c_{\rm adj}$ is equal to $\pi$ at
\vspace{-3mm}
$$
\theta_{\pm}^{\,\rm adj}=\frac{\pi}{2}\left[1\pm\sqrt{1-\left(\frac{16\pi}{g^2}
\right)^2}\right].$$
Note that the respective lower bound for the critical value of $g$, $g^{\rm adj}_{N\gg 1}=4\sqrt{\pi}\,$,\,\footnote{\,As in the 
previous footnote, at the particular value of $g$, $g=g^{\rm adj}_{N\gg 1}$, $\theta_{+}^{\rm adj}=\theta_{-}^{\rm adj}=
\frac{\pi}{2}$.\\[-2mm] ~} is only slightly different 
from the value $g^{\rm fund}_{N\gg 1}=4\sqrt{2\pi}$. Similarly to the fundamental case, at $g\gg
g^{\rm adj}_{N\gg 1}$, $\theta_{-}^{\rm adj}$ becomes a spurious solution, whereas $\theta_{+}^{\rm adj}\to\pi$.
Thus, at $N\gg 1$ and in the strong-coupling limit, 
understood in the sense $g\gg g^{\rm fund}_{N\gg 1}$, 
the critical fundamental- and adjoint-case values of $\theta$, at which 
the problem of crumpling might be solved, are given by the following simple formula:
$\theta_{+}^{\rm fund}=\frac12\theta_{+}^{\rm adj}=\frac{\pi}{2}$.

Finally, for $k$-strings, the ratio $c_k/c_{\rm fund}$ is the same Casimir one, $C(k,N)$,
as the ratio of string tensions, as long as $N\lesssim{\mathcal O}\left({\rm e}^{S_{\rm mon}/2}\right)$.\,\footnote{\,In 4D, the ratio of string tensions 
indeed equals $C(k,N)$ as in 3D, as long as 
the dual-photon field can be treated as a free massive field, since  
the rest of the derivation of 
Eq.~(\ref{2991}) is based on the properties of fundamental weights only.}
This enables one to readily find $\theta_{\pm}^k$ as well. Namely, 
$$\theta_{\pm}^k=\frac{\pi}{2}\left[\frac{N-1}{2NC(k,N)}
\pm\sqrt{\left(\frac{N-1}{2NC(k,N)}
\right)^2-\left(\frac{16\pi}{g^2}
\right)^2}\right],$$
that is only possible at $g>g_k\equiv g^{\rm fund}\sqrt{C(k,N)}\ge g^{\rm fund}$.

\setcounter{equation}{0}
\section{Geometric aspects of confining strings: the physics of negative stiffness}

The role of antisymmetric tensor field theories for confinement has been
studied in detail in Ref.~\cite{tq}. By analyzing compact
antisymmetric tensor field theories of rank $h-1$ Quevedo and
Trugenberger have shown that, starting from a ``Coulomb'' phase, the condensation
of $(d-h-1)$-branes  (where $D = d +1$ is the space-time dimension) leads to a
generalized confinement phase for $(h-1)$-branes. Each phase in the model has two
dual descriptions in terms of antisymmetric tensor of different ranks, massless
for the Coulomb phase and massive for the confinement phase.
Upon integration over the massive antisymmetric tensor field (in the case of an even 
number of dimensions -- when the $\theta$-term is absent), one obtains the string effective action~(\ref{snonloc}).
The derivative expansion of this nonlocal string effective action, up to the term next to the rigidity one, 
produces the following confining-string action:
\begin{equation}
S = \int d^2{\xi } \sqrt{{\sf g}} \ {\sf g}^{ab}
{\mathcal D}_a x_{\mu } \left( T - s{\mathcal D}^2 +
{1\over M^2} {\mathcal D}^4 \right)
 {\mathcal D}_b x_{\mu }.
\label{eq:one}
\end{equation}
Here, ${\mathcal D}_a$ is the covariant derivative with
respect to the induced metric ${\sf g}_{ab}=(\partial_a x_{\mu })(\partial _bx_{\mu })$
on the surface parametrized by $\vec x\left( \xi _0, \xi_1\right)$.

In (\ref{eq:one}), the first term provides a bare surface tension
$2T$, while the second accounts for rigidity 
with stiffness parameter $s$.
The last term can be written (up to surface terms)
as a combination of the fourth
power and the square of the gradient of the extrinsic curvature matrices,
with $M$ being a new mass scale. It thus suppresses world-sheet
configurations with rapidly changing extrinsic curvature; due to its
presence, the stiffness $s$ may not necessarily be negative, as in Eq.~(\ref{alf}) above, but also positive.
We analyze the model (\ref{eq:one}) in the
large-$D$ approximation. To this end, we introduce a
Lagrange multiplier matrix $L ^{ab}$
to impose the constraint ${\sf g}_{ab}=(\partial _ax_{\mu })(\partial_bx_{\mu })$,
extending the action (\ref{eq:one}) to
\vspace{-3mm}
\begin{equation}
 S+ \int d^2\xi \sqrt{{\sf g}} \ \ L
^{ab} \left( \partial _a x_{\mu } \partial _bx_{\mu } - {\sf g}_{ab} \right) \ .
\label{eq:two}
\end{equation}
Then we parametrize the world sheet in a Gauss' map by
$x_{\mu } (\xi ) = \left( \xi _0, \xi _1, \phi ^i (\xi )
\right) $, $(i=2, \dots , D-1)$,
where $-\beta /2\le \xi_0 \le \beta /2$,
$-R/2 \le \xi _1 \le R/2$, and
$\phi ^i(\xi )$ describe the $D-2$
transverse fluctuations.
With the usual homogeneity and isotropy Ansatz
${\sf g}_{ab}=\rho \delta_{ab} $,  $L ^{ab} =
L  {\sf g}^{ab}$
of infinite surfaces
($\beta ,R \to \infty $) at the saddle
point, we obtain
\begin{eqnarray}
\hspace{-4mm}S &&=2\int d^2\xi  \left[T +L
(1-\rho ) \right]
+ \int d^2\xi  \ \partial_a\phi ^i
V\left( T, s, M, L, {\mathcal D}^2
\right)  \partial_a \phi ^i\ ,
\label{eq:three}
\end{eqnarray}
where
\vspace{-3mm}
\begin{eqnarray}
V\left(T, s, M, L, {\mathcal D}^2 \right) = T+L - s {\mathcal D}^2 +
{1\over M^2} {\mathcal D}^4.
\label{eq:four}
\end{eqnarray}
Integrating over the transverse fluctuations, in the infinite-area 
limit, we get the effective action
$$
S^{\rm eff} = 2A_{\rm ext}  \left[T+L
(1-\rho ) \right]+$$
\vspace{-6mm}
\begin{equation}
+ A_{\rm ext} {{D-2}\over 8\pi^2 }\rho
\int d^2p\ {\rm ln}
\left[ p^2 V\left( T, s, M, L, p \right)
\right] \ ,
\label{eq:five}
\end{equation}
where $A_{\rm ext}=\beta R$ is the
extrinsic, physical, space-time area.
For large
$D$, the fluctuations of $L$ and $\rho $ are suppressed and these
variables take their ``classical values", determined by the two saddle-point
equations
\begin{equation}
0 = f\left( T, s, M, L \right) \ ,\quad 
\rho = {1\over f'\left( T, s, M , L \right) }\ ,
\label{eq:six}
\end{equation}
where the prime denotes a derivative with respect to $L$ and
the ``saddle-function" $f$ is defined by
\begin{eqnarray}
f\left( T, s, M, L \right) \equiv L - {{D-2}\over {8\pi }}
\int dp \ p\  {\rm ln}
\left[ p^2 V
\left( T, s, M, L, p \right) \right] .
\label{eq:additional1}
\end{eqnarray}
Using (\ref{eq:six}) in (\ref{eq:five}) we get
$S^{\rm eff}= 2\left( T+L \right)
A_{\rm ext}$
showing that ${\mathcal T} =2\left( T+L \right)$
is the physical string tension.

The stability condition for the Euclidean surfaces is that
$V\left( T, s, M, L, p \right)$ be positive for all $p^2 \ge 0$.
However, we will require the same condition also for $p^2\le 0$,
so that strings propagating in Minkowski space-time are
not affected by the
propagating states of negative norm which plague rigid strings.
The stability condition becomes thus
$\sqrt{T+L} \ge |sM/2|$, which
allows us to introduce the real variables $R$ and $I$ defined by
\begin{equation}
R^2 \equiv {M\over 2} \sqrt{T+L}
+ {s M^2\over 4}\ ,\quad 
I^2 \equiv {M\over 2} \sqrt{T+L}
- {s M^2\over 4}\ .
\label{eq:eight}
\end{equation}
In terms of these, the kernel $V$ can be written as
\begin{equation}
M^2 V\left( T, s, M, L, p \right) =
\left( R^2+I^2 \right) ^2 + 2 \left( R^2-I^2 \right) p^2
+ p^4 \ .
\label{eq:newone}
\end{equation}

In order to analyze the geometric properties of the string model (\ref{eq:one}) we
will study two correlation functions. First, we
consider the orientational correlation function
$$g_{ab}(\xi -\xi ') \equiv \langle \partial_a \phi ^i (\xi )
\ \partial_b \phi ^i (\xi ') \rangle $$
 for the normal components of
tangent vectors to the world sheet.

Secondly, we compute the scaling law of
the distance $d_E$ in embedding space
between two points on the
world sheet  when changing its projection $d$
on the reference plane. The exact relation between
the two lengths is
\begin{equation}
d_E^2=d^2 + \sum_i \langle |\phi ^i (\xi) - \phi ^i (\xi ')
|^2\rangle \ .
\label{eq:newtwo}
\end{equation}
When $d_E^2 \propto d^2$ this implies that the Hausdorff dimension of the surfaces,
defined as $d_E^2 = (d^2)^{2/D_H}$, is equal to 2, and the surface is smooth.

In order to  establish the properties
of our model, we analyze
the saddle-point function $f\left( T, s, M, L \right)$
in (\ref{eq:additional1}). To this end, we must
prescribe a
regularization for the ultraviolet divergent
integral. We use dimensional regularization,
computing
the integral in $(2-\epsilon)$ dimensions.
For small $\epsilon$, this leads to
\begin{equation}
f\left(T, s, M, L\right)\! =\! L\! +\! {1\over 16 \pi ^3} \left(R^2\!-\!I^2
\right) {\rm ln}\, {R^2\!+\!I^2\over \Lambda ^2}
\!-\! {1\over 8 \pi ^3} RI\! \left( {\pi \over 2}\! +\! {\rm arctan}
\,{I^2\!-\!R^2 \over 2RI} \right) \, ,
\label{eq:newfour}
\end{equation}
where $\Lambda \equiv \mu \ {\rm exp} (2/\epsilon)$ and $\mu $ is a
reference scale which must be introduced for dimensional reasons.
The scale $\Lambda $ plays the role of an
ultraviolet cutoff,
diverging for $\epsilon \to 0$.

The saddle-point function above
is best studied by introducing the dimensionless
couplings $t\equiv T/\Lambda ^2$, $m\equiv M/\Lambda $ and
$l\equiv L/\Lambda ^2$.
We will study in detail
the case $s=0$, in which
the saddle-point equations can be solved analytically
since $R=I$. This choice is not too restrictive since, as we will show, $s=0$
is the infrared fixed point. We get
\begin{eqnarray}
l &&= {m^2 c^2\over 2}\ \left(
1+\sqrt{1+{4t\over m^2 c^2}} \right) \ ,\\
\rho &&= \left( 1-{m c\over 2\sqrt{t+l}} \right) ^{-1}\, ,
\label{eq:newfive}
\end{eqnarray}
where $c\equiv 1/(32 \pi ^2)$. This shows that the point ($t^*=0$,
$s^*=0$, $m^*=0$) constitutes
an infrared-stable fixed point with vanishing
physical string tension ${\mathcal T}$. This point is characterized by
long-range correlations
$g(d) = 2\pi ^2/a$,
with a constant $a$, and by the scaling law
\begin{equation}
d_E^2 = {\pi ^2\over a} \rho ^*  d^2 \ , \quad
\rho ^* \equiv \left( 1-{1\over 2a} \right) ^{-1} \ ,
\label{eq:newseven}
\end{equation}
which shows that the Hausdorff dimension of world sheets is
$D_H=2$. For $s=0$, the constant $a$ can be computed analytically:
\begin{equation}
a^2 = \lim _{{t\to 0}\atop {m\to 0}}
{{1+(2t/m^2 c^2) + \sqrt{1+ 4t/m^2 c^2 }}\over 2}\ ,
\label{eq:neweight}
\end{equation}
from which we recognize that $1\le \rho ^* \le 2$.

At the infrared fixed point we can remove the cutoff. The
renormalization of the model is easily obtained by noting that
the effective action for transverse fluctuations to quadratic
order decouples from other modes and is identical with the
second term in (\ref{eq:three}) with ${\mathcal D}^2=\partial^2/\rho$
and $\rho $ taking its saddle-point value.
From here we identify the physical tension, stiffness and mass as
\vspace{-3mm}
\begin{equation}
{\mathcal T} = \Lambda ^2 (t+l)\ ,\quad
{\mathcal S} = {s\over \rho}\ ,\quad
{\mathcal M} = \Lambda  m \rho\ .
\label{eq:newnine}
\end{equation}
For $s=0$, we can compute analytically the corresponding
$\gamma$-functions:
\begin{eqnarray}
\gamma_t &&\equiv -\Lambda {d\over d\Lambda} {\rm ln}\ t = 2+
O\left( {t\over m^2} \right) \ ,\\
\gamma _m  &&\equiv -\Lambda {d\over d\Lambda} {\rm ln}\ m = 1+
O\left( {t^2\over m^4} \right) \ .
\label{newten}
\end{eqnarray}

The vicinity of the infrared fixed point defines a new
theory of smooth strings for which the range of the orientational
correlations in embedding space
is always of the same order or bigger than the
length scale $1/\sqrt{{\mathcal T}}$
 associated with the tension. The
naively irrelevant term ${\mathcal D}^4/M^2$ in
(\ref{eq:one}) becomes relevant in the
large-$D$ approximation since it generates a
string tension proportional to $M^2$ which
takes over the control of the fluctuations
after the orientational
correlations die off.
Note moreover, that it is exactly this new quartic
term which guarantees that the spectrum $p^2 V\left(
T, s, M, L, p\right)$ has no other pole than $p=0$,
contrary to the rigid string, which necessarily has
a ghost pole at $p^2= -T/s$.

By studying the finite-size scaling~\cite{cardy} of the Euclidean model
(\ref{eq:one}) on a cylinder of (spatial)
circumference $R$  it is possible to
determine the universality class of confining
strings.
In the limit of large $R$, the effective action on the cylinder 
takes the form~\cite{cri}
\begin{equation}
\lim_{\beta \to \infty} {S^{\rm eff}\over \beta} =
{\mathcal T} R-{\pi c (D-2)\over 6R} + \dots \ ,
\label{eq:mone}
\end{equation}
for $(D-2)$ transverse degrees of freedom,
the universality class being encoded in the pure number $c$.
This suggests that the effective theory
describing the infrared behavior is
a conformal field theory with central charge $c$.
In this case the number $c$ also fixes the L\"uscher term~\cite{luscher1} 
in the quark-antiquark potential:
\vspace{-3mm}
\begin{equation}
V(R) = {\mathcal T} R -{\pi c (D-2)\over 24R}+\dots \ .
\label{eq:zero}
\end{equation}

In~\cite{cri} it has been shown that confining strings are characterized
by $c=1$. Although they share the same value of $c$, confining strings
are clearly different ($c=1$)-theories than Nambu--Goto or rigid strings.
Indeed, the former are smooth strings on any scale, while the latter
crumple and fill the ambient space, at least in the infrared region.
Our result $c=1$ is in agreement with recent precision numerical
determinations of L\"uscher and Weisz~\cite{luscher} of this constant.

Having established that the model (\ref{eq:one}) describes smooth strings
with $c=1$, the question arises as to how much these results depend on
the truncation of the original non-local action after the
${\mathcal D}^4$-term. In~\cite{cri} it has been proved that the answer to this
question is no: the value of $c$ and the smooth geometric properties
are independent of an infinite set of truncations, provided
that a solution for the polynomial ``gap'' equation exists. These
properties are presumably common to a large class of non-local
world-sheet interactions.

\setcounter{equation}{0}
\section{High-temperature behavior of confining strings}

The high-temperature behavior of
large-$N$ QCD has been studied by Polchinski  in~\cite{polc3}, where he shows
that the deconfining transition in QCD is due to the condensation of Wilson
lines, and the
partition function of QCD flux tubes can be continued above the deconfining
transition; this high-temperature continuation can be evaluated
perturbatively. So, any string theory that is equivalent to QCD {\it must} reproduce
this behavior. However, the Nambu--Goto action has the wrong temperature dependence,
while the rigid string parametrically has the correct high-temperature behavior but with the 
wrong sign and an imaginary part signaling a world-sheet
instability \cite{polc2}.

The high-temperature behavior of the confining-string model proposed in~\cite{cris1} 
has been studied in~\cite{Tfinita}, where it has been shown that
 this model has a high-temperature behavior that agrees  in temperature
dependence,  {\it  sign and reality properties}
with the large-$N$ QCD result~\cite{polc3}.
The starting point will be again~(\ref{eq:two}). In the Gauss' map, the
value of the periodic coordinate $\xi_0$ is
$-\beta/2 \leq \xi_0 \leq \beta/2$ with $\beta = 1/T$ and $T$  the
temperature.
Note that,  at high temperatures ($\beta \ll 1$), the scale $M^2$ can be temperature-dependent.
This is not unusual in closed string theory as  has been shown by Atick
and Witten~\cite{witten1}.
The value of $\xi_1$  is
$-R/2 \leq \xi_1 \leq R/2$; $\phi^i(\xi)$'s describe the $D-2$ transverse
fluctuations. We look for a saddle-point solution with a diagonal metric $g_{ab}
= {\rm diag}\ (\rho_0, \rho_1)$, and a Lagrange multiplier of the form
$\lambda^{ab} = {\rm diag}\ (\lambda_0/\rho_0, \lambda_1/\rho_1)$.

After integration over transverse fluctuations we obtain, in the limit $R \to
\infty$,  four gap equations:
\begin{eqnarray}
&&{ 1 -\rho_0 \over \rho_0} =  0\ , \label{eq:gap1}\\[1mm]
&&{1 \over \rho_1} = 1 - {D - 2\over 2} { 1 \over 4 \beta }
{\sqrt{2 M} \over (\lambda_1 + t )^{3/4} }\ , \label{eq:gap2} 
\\[1mm]
&&{1\over 2}(t - \lambda_1) + {1 \over 2\rho_1}(\lambda_1 + t ) - t -
\lambda_0 + {D - 2\over 2} { \pi \over 2 \beta^2 } = 0 \ ,\label{eq:gap3} \\[1mm]
&&(t - \lambda_1) - {1 \over \rho_1}(\lambda_1 + t ) +
{D - 2\over 2} { 1 \over  \beta } \left[ \sqrt{2 M} \left( \lambda_1
+ t \right)^{1/4} - {\pi \over \beta } \right] = 0 \label{eq:gap4} \ ,
\end{eqnarray}
and a simplified form of the effective action:
\begin{equation}
S^{\rm eff} = A_{\rm ext}  {\mathcal T} \sqrt{1 \over \rho_1}
\label{eq:effac} \ ,
\end{equation}
with ${\mathcal D} =  2 (\lambda_1 + t )$  representing  the physical string
tension.
By inserting (\ref{eq:gap2}) into (\ref{eq:gap4}), we obtain an equation for
$(\lambda_1 + t )$ alone:
\begin{eqnarray}
(\lambda_1 + t ) -  {D - 2\over 2} {5\over 8 \beta}\sqrt{2 M}
\left( \lambda_1 + t \right)^{1/4} +{D - 2\over 2} { \pi \over  2 \beta^2 } - t = 0 \ .
\label{eq:quart}
\end{eqnarray}

Without loss of generality we  set
\begin{equation}
\left( \lambda_1 + t \right)^{1/4} = {\sqrt{2 M} \over \gamma} \ ,
\label{eq:condiz}
\end{equation}
where $\gamma $ is a dimensionless parameter.
It is possible to show that, at high temperatures, when
\begin{equation}
t \beta^2 \ll {D - 2\over 2} \ ,
\label{eq:cont}
\end{equation}
we can completely neglect $t$ in (\ref{eq:quart}). Indeed, as we now show, $\lambda_1$ is proportional
to $(D-2)^2/\beta^2$.
We can thus rewrite (\ref{eq:quart}) as
\begin{equation}
\lambda_1 -  {D - 2\over 2}\, {5\over 8 \beta}\,\gamma
\,\lambda_1^{1/2} + {D - 2\over 2}\, { \pi \over 2 \beta^2 }  = 0\ .
\label{eq:lamgap}
\end{equation}
We now restrict ourselves to the regime
\begin{equation}
G( \gamma, D) = {25 \over 64}\, \gamma^2 \left({D- 2 \over 2}\right)^2 - 2
\pi \, {D- 2 \over 2} > 0  \ ,\label{eq:posdis}
\end{equation}
for which (\ref{eq:lamgap}) admits two real solutions:
\begin{eqnarray}
(\lambda_1^1)^{1/2} &&= {5 \over 16 \beta}\ \gamma\ {D- 2 \over 2}+\ {1\over 2 \beta} \sqrt{G( \gamma, D)} \
,\label{eq:possol}
\\[1mm]
(\lambda_1^2)^{1/2} &&= {5 \over 16 \beta}\ \gamma\ {D- 2 \over 2}-\ {1\over 2 \beta} \sqrt{G( \gamma, D)}\ . \label{eq:negsol}
\end{eqnarray}
In both cases, $\lambda_1$ is proportional to $(D-2)/ \beta^2$, which justifies neglecting $t$ in
(\ref{eq:quart}) and implies
that the scale $M^2$ must be chosen  proportional to $1/\beta^2$.
Moreover, since the physical string tension is real we are guaranteed that $M^2
> 0$, as required by the stability of our model. Any complex solutions for $
{\mathcal D}$ would have
been incompatible with the stability of the truncation.

Let us start by analyzing the first solution
(\ref{eq:possol}).
By inserting (\ref{eq:possol}) into (\ref{eq:gap3}), we obtain the following
equation for $\rho_1\,$:
\begin{equation}
{1 \over \rho_1} = 1 - {4 \over 5  + \sqrt{25 - {128 \pi \over \gamma^2 {D-2 \over 2}}}}\ .
\label{eq:rhopos}
\end{equation}
Owing to the condition (\ref{eq:posdis}),
$1/\rho_1$ is positive
and, since $\lambda_1^2$ is real, the squared free energy is also positive:
\begin{equation}
F^2(\beta) \equiv {S^2_{\rm eff} \over R^2} ={1\over \beta^2}
\left({5 \over 16 }\
\gamma\ {D\!-\! 2 \over 2}
\!-\!{1\over 2 } \sqrt{
 G( \gamma , D)}
\right)^4  \left(1 \!-\! {4 \over 5 \! +\! \sqrt{25\! -\! {128 \pi \over \gamma^2 {D-2 \over 2}}}} \right) \, .
\label{eq:freew}
\end{equation}
In this case the high-temperature behavior is the same as in QCD, but the sign is wrong,
exactly as for the rigid string.
There is, however, a crucial difference: (\ref{eq:freew}) is real, while the
squared free energy for the rigid string is imaginary, signaling an instability in the model.

If we now look at the behavior of $\rho_1$ at low temperatures, below the deconfining transition
\cite{cri}, we see that $1/\rho_1$ is positive. The deconfining transition is indeed
determined by the
vanishing of $1/\rho_1$ at $\beta = \beta_{\rm dec}$. In the case of
(\ref{eq:possol}) this means that   $1/\rho_1$ is positive below the Hagedorn transition, touches
zero at $\beta_{\rm dec}$ and remains positive above it. Exactly the same will happen also for
$F^2$, which is positive below $\beta_{\rm dec}$ , touches
zero at $\beta_{\rm dec}$ and remains positive above it.
This solution thus describes an unphysical ``mirror'' of the low-temperature behavior of the
confining string, without a real deconfining Hagedorn transition. For this reason we discard it.

Let us now study the solution~(\ref{eq:negsol}).
Again, by inserting (\ref{eq:negsol}) into (\ref{eq:gap3}), we obtain 
\begin{equation}
{1 \over \rho_1} = 1 - {4 \over 5  - \sqrt{25 - {128 \pi \over \gamma^2 {D-2 \over 2}}}}\ .
\label{eq:rhoneg}
\end{equation}
In this case, when
\begin{equation}
\gamma > 4 \sqrt{{\pi \over 3}} \left({D- 2 \over 2}\right)^{-1/2}\ ,
\label{eq:congam}
\end{equation}
$1/\rho_1$ becomes negative.
The condition (\ref{eq:congam}) is consistent with (\ref{eq:posdis}) and will
be taken to  fix the values of the range of parameter $\gamma$ that enters 
(\ref{eq:condiz}).
We will restrict to those that satisfy (\ref{eq:congam}).
Since $\rho_0 = 1$ and $\lambda_1$ is real and proportional to $1/\beta^2$,
we obtain the following form of the squared free energy:
\begin{equation} 
F^2(\beta) = -{1\over \beta^2}\left( {5 \over 16 }\ \gamma\ {D- 2 \over 2}
-{1\over 2 } \sqrt{G(\gamma , D)}
\right)^4 \left({4 \over 5  - \sqrt{25 - {128 \pi \over \gamma^2 {D-2 \over 2}}}} -1\right) \ .
\label{eq:freec}
\end{equation}
In the range defined by (\ref{eq:congam}) this is {\it negative}. For this solution, thus,
both $1/\rho_1$ and $F^2$ pass from positive values at low temperatures to negative values at high
temperatures, exactly as one would expect for a string model undergoing the Hagedorn transition at an
intermediate temperature. In fact, this is also what happens in the rigid-string case, 
but there, above the Hagedorn transition, there is a second transition above which, at
high temperature, $\lambda_1$ becomes large and essentially imaginary, giving a positive squared free
energy. This second transition is absent in our model.

Let us now  compare the result (\ref{eq:freec}) with the corresponding one for large-$N$ QCD \cite{polc3}:
\begin{equation}
F^2(\beta)_{\rm QCD} = - {2 g^2(\beta) N \over \pi^2 \beta^2}\ ,\label{eq:qcd}
\end{equation}
where $g^2(\beta)$ is the QCD coupling constant.
First of all let us simplify our result by choosing large values of $\gamma$:
$$
\gamma \gg \sqrt{128 \pi \over 25} \left({D-2 \over 2}\right)^{-1/2} \ .
$$
In this case (\ref{eq:freec}) reduces to
\begin{equation}
F^2(\beta) = - {1\over \beta^2 }\, {8 \pi^3 \over 125}\, {D-2\over \gamma^2}\ .
\label{eq:bingo}
\end{equation}
This corresponds {\it exactly} to the QCD result (\ref{eq:qcd}) with the
identifications
\begin{eqnarray}
g^2 &&\propto {1\over \gamma^2} \ , \nonumber \\[1mm]
N && \propto D - 2 \ . \nonumber
\end{eqnarray}
The weak $\beta$-dependence of the QCD coupling $g^2(\beta)$ can be accommodated in the parameter
$\gamma$. Note that our result is valid at large values of $\gamma$, i.e. small values of $g^2$, as it
should be for QCD at high temperatures \cite{wilczek}. Note also the interesting identification
between the order of the gauge group and the number of transverse space-time dimensions. Moreover,
since the sign of $\lambda_1$ does not change at high temperatures, the field $x_\mu$ is not
unstable. The opposite happens in the rigid-string case \cite{polc2}, where the change of
sign of $\lambda_1$ gives rise to a world-sheet instability.

\setcounter{equation}{0}
\section{The influence of matter fields on the 
deconfinement phase transition in the 3D GG model at finite temperature}

\subsection{Introduction}

The phase structure of the 3D GG model at finite temperature
has for the first time been addressed in Ref.~\cite{az}, where it
has been shown that, in the absence of W-bosons, the
weakly coupled monopole plasma 
undergoes the Berezinsky-Kosterlitz-Thouless (BKT)~\cite{bkt}
phase transition into the molecular phase
at the temperature $T_{\rm BKT}=g^2/ (2\pi)$. Then, in
Ref.~\cite{dkkt}, it has been shown that the true phase transition, which 
takes into account W-bosons, occurs at 
approximately half this temperature.
Let us first discuss in some more detail the
above-mentioned BKT phase transition, which takes place in the absence of 
W-bosons (i.e. in the continuum limit of the 3D lattice compact QED extended by the Higgs field).

At finite temperature $T\equiv1/\beta$, equations of motion of the fields $\chi$ and $\psi$,
entering the partition function~(\ref{ZMON}), 
$${\mathcal Z}_{\rm mon}\equiv \int {\mathcal D}\chi{\mathcal D}\psi \exp\left\{-
\int d^3x{\mathcal L}[\chi,\psi|g_m,\zeta]\right\},$$
should be supplemented by the periodic boundary conditions in the temporal direction, with the period equal to
$\beta$. Because of that, the lines of magnetic field emitted by a monopole cannot cross
the boundary of the one-period region. Consequently, at the distances larger than $\beta$ in the direction 
perpendicular to the temporal one, magnetic-field lines approaching the boundary   
should run almost parallel to it. 
Therefore, monopoles separated by such distances
interact via the 2D Coulomb potential, rather than the 3D one. Since the average distance
between monopoles in the plasma is of the order of $\zeta^{-1/3}$, we see that, at $T\gtrsim\zeta^{1/3}$,
the monopole ensemble becomes two-dimensional. Owing to the fact that $\zeta$ is exponentially 
small in the weak-coupling regime under discussion, the idea of dimensional reduction is perfectly applicable
at the temperatures of the order of the above-mentioned critical temperature $T_{\rm BKT}$.

Note that, due to the $T$-dependence of the strength of the 
monopole-antimonopole interaction, which is a consequence of the dimensional reduction, the BKT 
phase transition in the 3D Georgi-Glashow model is inverse with respect to the standard one of the 2D 
XY model. Namely, monopoles exist in the plasma phase at the temperatures below $T_{\rm BKT}$
and in the molecular phase above this temperature.
As has already been discussed, the analogy with the 2D XY-model
established in Ref.~\cite{bmk} is that spin waves of that model correspond
to free photons of the 3D GG model at zero temperature, while vortices correspond to magnetic
monopoles. At finite temperature, 
disorder is rather produced by the thermally excited W-bosons~\cite{dkkt}, whereas 
monopoles order the system, binding W-bosons into pairs. However, the analogy is still true
in the case of the continuum version of 3D compact QED (with the Higgs field) under discussion.

Let us therefore briefly discuss the BKT phase transition, occurring in the 2D XY-model at 
a certain critical temperature $T=T_{\rm BKT}^{\rm XY}$.
At $T<T_{\rm BKT}^{\rm XY}$, the spectrum of the model
is dominated by massless spin waves, and the periodicity of the angular
variable is unimportant in this phase. The spin waves are unable to disorder
the spin-spin correlation functions, and these decrease at large distances by
some power law. On the contrary, at $T>T_{\rm BKT}^{\rm XY}$,
the periodicity of the angular variable becomes important. This leads to the
appearance of topological singularities (vortices) of the angular variable,
which, contrary to spin waves, have nonvanishing winding numbers.
Such vortices condense and disorder the spin-spin correlation
functions, so that those start decreasing exponentially with the distance.
Thus, the nature of the BKT phase transition is the condensation
of vortices at $T>T_{\rm BKT}^{\rm XY}$. In another words, at $T>T_{\rm BKT}^{\rm XY}$, free vortices do exist
and mix in the ground state (vortex condensate) of an indefinite
global vorticity. Contrary to that, at $T<T_{\rm BKT}^{\rm XY}$, free vortices cannot exist,
and they rather mutually couple into bound states of vortex-antivortex pairs.
Such vortex-antivortex molecules are small-sized and short-living (virtual)
objects. Their dipole-type fields are short-ranged and therefore cannot
disorder significantly the spin-spin correlation functions. However, when
the temperature starts rising, the size of these molecules grows,
until at $T=T_{\rm BKT}^{\rm XY}$ it diverges, that corresponds to the dissociation of the molecules
into pairs. Therefore,
one of the methods to determine the critical
temperature of the BKT phase transition is to evaluate the mean squared separation
in the molecule and to find the temperature at which it
starts diverging.

Let us now 
return to the continuum limit of the finite-temperature 3D compact QED, extended by the Higgs field, 
and determine there the mean squared separation in the monopole-antimonopole molecule.
One can then see that,
up to exponentially small corrections, the respective critical temperature is 
unaffected by the finiteness of the Higgs-boson mass. Indeed, the mean squared 
separation reads\,\footnote{\,In this section, 3-vectors are denoted as $\vec a$, whereas
2-vectors are denoted as ${\bf a}\,$.}
\begin{equation}
\label{lsquared}
\left<L^2\right>=\frac{\int\limits_{|{\bf x}|>m_W^{-1}}^{} d^2{\bf x}|{\bf x}|^{2-\frac{8\pi T}{g^2}}\exp
\left[\frac{4\pi T}{g^2}K_0\left(m_H|{\bf x}|
\right)\right]}{\int\limits_{|{\bf x}|>m_W^{-1}}^{} 
d^2{\bf x}|{\bf x}|^{-\frac{8\pi T}{g^2}}\exp\left[\frac{4\pi T}{g^2}K_0\left(m_H|{\bf x}|
\right)\right]}\ ,
\end{equation}
where $K_0$ denotes the modified Bessel function.
Disregarding the exponential factors in the numerator and denominator of this equation, we obtain 
$\left<L^2\right>\simeq\frac{4\pi T-g^2}{2m_W^2\left(2\pi T-g^2\right)}$,
that yields the above-mentioned value of the BKT critical temperature $T_{\rm BKT}=g^2/(2\pi)$. 
Besides that, we see that, as long as $T$ does not tend to $T_{\rm BKT}$, and in the weak-coupling 
regime under study, the value of $\sqrt{\left<L^2\right>}$ is exponentially smaller than the characteristic distance in the 
monopole plasma, $\zeta^{-1/3}$, i.e. molecules are very small-sized with respect to that distance.

The factor $\beta$ at the 
action of the dimensionally-reduced theory, $S_{{\rm d.-r.}}=\beta\int d^2x{\mathcal L}[\chi,\psi|g_m,\zeta]$,
can be removed [and this action can be cast to 
the original form of Eq.~(\ref{ZMON}) with the substitution $d^3x\to d^2x$]
by the obvious rescaling: 
\vspace{-3mm}
\begin{equation}
\label{sDR}
S_{{\rm d.-r.}}=\int d^2x{\mathcal L}\left[\chi^{\rm new},\psi^{\rm new}|\sqrt{K},\beta\zeta\right].
\end{equation}
Here, $K\equiv g_m^2T$, 
$\chi^{\rm new}=\sqrt{\beta}\chi$, $\psi^{\rm new}=\sqrt{\beta}\psi$, and in what follows
we will denote for brevity $\chi^{\rm new}$ and $\psi^{\rm new}$ simply as $\chi$ and $\psi$, respectively.
Averaging then over the field $\psi$ with the use of the cumulant expansion we arrive at the 
following action:
\vspace{-3mm}
$$
S_{{\rm d.-r.}}\simeq\int d^2x\left[\frac12(\partial_\mu\chi)^2-2\xi\cos\left(\sqrt{K}\chi\right)\right]-
$$
\vspace{-8mm}
\begin{equation}
\label{2t}
-2\xi^2\int d^2xd^2y\cos\left(\sqrt{K}\chi({\bf x})\right){\mathcal K}^{(2)}({\bf x}-{\bf y})
\cos\left(\sqrt{K}\chi({\bf y})\right).
\end{equation}
In this expression, similarly to Eq.~(\ref{morequart}),
we have disregarded all the cumulants higher 
than the quadratic one, and the limits of applicability of 
this bilocal approximation will be discussed below. 
Further, in Eq.~(\ref{2t}), ${\mathcal K}^{(2)}({\bf x})\equiv{\rm e}^{KD_{m_H}^{(2)}({\bf x})}-1$, where  
$D_{m_H}^{(2)}({\bf x})\equiv K_0(m_H|{\bf x}|)/(2\pi)$ is the 2D Yukawa propagator, and 
$\xi\equiv\beta\zeta\exp\left[\frac{K}{2}D_{m_H}^{(2)}\left(m_W^{-1}\right)\right]$
denotes the monopole fugacity modified by the interaction of monopoles via the Higgs field.
The exponential factor entering $\xi$ reads
$$\xi\propto\exp\left[-\frac{4\pi}{g^2}\left(m_W\epsilon+T\ln\left(\frac{{\rm e}^{\gamma}}{2}c\right)
\right)\right].$$
Here, we have introduced the notation $c\equiv m_H/m_W$, $c<1$, and $\gamma\simeq 0.577$ is the 
Euler constant, so that $\frac{{\rm e}^{\gamma}}{2}\simeq 0.89<1$. We see that 
the modified fugacity remains exponentially small, provided that 
\begin{equation}
\label{3t}
T<-\frac{m_W\epsilon}{\ln\left(\frac{{\rm e}^{\gamma}}{2}c\right)}.
\end{equation}

This constraint should be updated by another one, which would provide the convergence
of the cumulant expansion. Analogously to the zero-temperature case, the divergence of 
the cumulant expansion 
would indicate that the Higgs vacuum loses its normal stochastic property and 
becomes a coherent one.
In order to get the 
new constraint, notice that the parameter of the cumulant expansion reads 
$\xi I^{(2)}$, where $I^{(2)}\equiv\int d^2x{\mathcal K}^{(2)}({\bf x})$.
Evaluation of the integral $I^{(2)}$ yields~\cite{higgs}:

\begin{equation}
\label{I}
I^{(2)}\simeq\frac{2\pi}{m_H^2}\left[\frac12\left(c^2-1+\left(\frac{2}{{\rm e}^{\gamma}}
\right)^{\frac{8\pi T}{g^2}}\frac{1-c^{2-
\frac{8\pi T}{g^2}}}{1-\frac{4\pi T}{g^2}}\right)
+{\rm e}^{\frac{a}{{\rm e}}}-1+\frac{a}{{\rm e}}\right].
\end{equation}
(Note that, at $T\to\frac{g^2}{4\pi}$, $\frac{1-c^{2-\frac{8\pi T}{g^2}}}{1-\frac{4\pi T}{g^2}}\to
-2\ln c$, i.e. $I^{(2)}$ remains finite.)
In the derivation of this expression, 
the parameter $a\equiv4\pi\sqrt{2\pi}T/g^2$ was assumed to be of the order of unity.
That is because the temperatures we are working at are of the order of $T_{\rm BKT}$. 
Due to the exponential term in Eq.~(\ref{I}),
the violation of the cumulant expansion may occur at high enough temperatures [that parallels the above-obtained 
constraint~(\ref{3t})].
The most essential, exponential, part of the parameter
of the cumulant expansion thus reads
$$\xi I^{(2)}\propto\exp\left\{-\frac{4\pi}{g^2}\left[m_W\epsilon+T\left[
\ln\left(\frac{{\rm e}^{\gamma}}{2}c\right)-\frac{\sqrt{2\pi}}{{\rm e}}\right]\right]\right\}.$$
Therefore, the cumulant expansion converges at the temperatures obeying the inequality
$$T<\frac{m_W\epsilon}{\frac{\sqrt{2\pi}}{{\rm e}}-\ln\left(\frac{{\rm e}^{\gamma}}{2}c\right)}\ ,$$
which strengthens the inequality~(\ref{3t}). On the other hand, since we are working in the plasma phase, i.e. 
$T\le T_{\rm BKT}$, it is enough to impose the following upper 
bound on the parameter of the weak-coupling approximation, $\kappa$:
$$\kappa\le\frac{2\pi\epsilon}{\frac{\sqrt{2\pi}}{{\rm e}}-
\ln\left(\frac{{\rm e}^{\gamma}}{2}c\right)}.$$
Note that, although this inequality is satisfied automatically at $\frac{{\rm e}^\gamma}{2}c\sim 1$
(or $c\sim 1$), since  
it then takes the form $\kappa\le\sqrt{2\pi}{\rm e}\epsilon$, this is not so automatic 
in the BPS limit, $c\ll 1$. Indeed, in this case,
we have $\kappa\ln\left(\frac{2}{c{\rm e}^{\gamma}}\right)\le 2\pi\epsilon$. 
It is, however, quite feasible to obey this inequality, since the logarithm is a weak function.

\subsection{Higgs-inspired corrections to the RG flow in the absence of W-bosons}

Although we have seen in the previous subsection that the propagating Higgs field does not change the 
value of the critical temperature $T_{\rm BKT}$, it is instructive to derive the corrections it produces to the 
RG flow. Such a derivation can also be extended to the SU($N$)-case~(\ref{zMON}).
In that case~\cite{suNmp}, it is qualitatively clear that $T_{\rm BKT}$ should remain the same, since it only 
differs from that of the SU(2)-case by the factor $\vec q_i{\,}^2$, which is equal to unity. However, some 
peculiarities of the SU($N$)-case at $N>2$ become clear only upon a derivation of the RG equations, which 
will be given below.  

Let us thus start with the SU(2)-case~(\ref{ZMON}), whose action after the dimensional reduction 
has the form~(\ref{sDR}). 
In what follows, we will use the usual RG strategy based on the integration over the
high-frequency modes. Note that this procedure will be applied to {\it all} the fields, i.e. 
not only to $\chi$, but also to $\psi$.
Splitting the momenta into two ranges, $0<p<\Lambda'$ and $\Lambda'<p<\Lambda$, one can define the high-frequency
modes as $h=\chi_\Lambda-\chi_{\Lambda'}$, $\phi=\psi_\Lambda-\psi_{\Lambda'}$, where 
${\mathcal O}_{\Lambda'}({\bf x})=
\int\limits_{0<p<\Lambda'}^{}\frac{d^2p}{(2\pi)^2}{\rm e}^{i{\bf p}{\bf x}}{\mathcal O}({\bf p})$ and
consequently, e.g.
$h({\bf x})=\int\limits_{\Lambda'<p<\Lambda}^{}\frac{d^2p}{(2\pi)^2}{\rm e}^{i{\bf p}{\bf x}}\chi({\bf p})$.
The partition function,
$${\mathcal Z}_\Lambda=\int\limits_{0<p<\Lambda}^{}{\mathcal D}\chi({\bf p}){\mathcal D}\psi({\bf p})
\exp\left\{-S_{\rm d.-r.}\left[\chi_\Lambda,\psi_\Lambda\right]\right\},$$
can be rewritten as follows:

$$
{\mathcal Z}_\Lambda=\!\!\!\int\limits_{0<p<\Lambda'}^{}\!\!\!{\mathcal D}\chi({\bf p}){\mathcal D}\psi({\bf p})
\exp\left\{\frac12\int d^2x\left[\chi_{\Lambda'}\partial^2\chi_{\Lambda'}+
\psi_{\Lambda'}\left(\partial^2-m_H^2\right)\psi_{\Lambda'}\right]\right\}{\mathcal Z}',
$$
where
$$
{\mathcal Z}'=\int\limits_{\Lambda'<p<\Lambda}^{}{\mathcal D}\chi({\bf p}){\mathcal D}\psi({\bf p})
\exp\left\{\int d^2x\left[\frac12h\partial^2h+\frac12\phi\left(\partial^2-m_H^2\right)\phi+\right.\right.$$
\vspace{-6mm}
$$\left.\left.+2\xi{\rm e}^{\sqrt{K}\left(\psi_{\Lambda'}+\phi\right)}\cos\left(\sqrt{K}\left(\chi_{\Lambda'}+h\right)
\right)\right]\right\},$$
and $\xi\equiv\beta\zeta$.
Owing to the exponential smallness of the fugacity, ${\mathcal Z}'$ can further be expanded as
$${\mathcal Z}'\simeq 1+2\xi\int d^2x\left<{\rm e}^{\sqrt{K}\left(\psi_{\Lambda'}+\phi\right)}\right>_{\phi}
\left<\cos\left(\sqrt{K}\left(\chi_{\Lambda'}+h\right)\right)\right>_h
+2\xi^2\int d^2x d^2y\times$$
\vspace{-5mm}
$$\times\left[
\left<{\rm e}^{\sqrt{K}\left(\psi_{\Lambda'}
({\bf x})+\phi({\bf x})\right)}{\rm e}^{\sqrt{K}\left(\psi_{\Lambda'}({\bf y})+\phi({\bf y})\right)}
\right>_{\phi}\left<\cos\left(\sqrt{K}\left(\chi_{\Lambda'}({\bf x})+h({\bf x})\right)\right)\times
\right.\right.$$
\vspace{-5mm}
$$\left.\times
\cos\left(\sqrt{K}\left(\chi_{\Lambda'}({\bf y})+h({\bf y})\right)\right)\right>_h-
\left<{\rm e}^{\sqrt{K}\left(\psi_{\Lambda'}({\bf x})+\phi({\bf x})\right)}\right>_{\phi}
\left<{\rm e}^{\sqrt{K}\left(\psi_{\Lambda'}({\bf y})+\phi({\bf y})\right)}
\right>_{\phi}\times$$
\vspace{-5mm}
$$\left.\times\left<\cos\left(\sqrt{K}\left(\chi_{\Lambda'}({\bf x})+h({\bf x})\right)\right)\right>_h
\left<\cos\left(\sqrt{K}\left(\chi_{\Lambda'}({\bf y})+h({\bf y})\right)\right)\right>_h\right],$$
where
$$
\left<{\mathcal O}\right>_h\equiv\frac{\int\limits_{\Lambda'<p<\Lambda}^{}{\mathcal D}\chi({\bf p})
\exp\left(\frac12\int d^2xh\partial^2h\right){\mathcal O}}{\int\limits_{\Lambda'<p<\Lambda}^{}{\mathcal D}\chi({\bf p})
\exp\left(\frac12\int d^2xh\partial^2h\right)},$$
\vspace{-5mm}
$$
\left<{\mathcal O}\right>_{\phi}\equiv\frac{\int\limits_{\Lambda'<p<\Lambda}^{}{\mathcal D}\psi({\bf p})
\exp\left[\frac12\int d^2x\phi\left(\partial^2-m_H^2\right)\phi\right]{\mathcal O}}
{\int\limits_{\Lambda'<p<\Lambda}^{}{\mathcal D}\psi({\bf p})
\exp\left[\frac12\int d^2x\phi\left(\partial^2-m_H^2\right)\phi\right]}.$$
Carrying out the averages we arrive at the following expression for ${\mathcal Z}'$:
$${\mathcal Z}'\simeq 1+2\xi A(0)B(0)\int d^2x{\rm e}^{\sqrt{K}\psi_{\Lambda'}}\cos\left(\sqrt{K}\chi_{\Lambda'}\right)+
\left(\xi A(0)B(0)\right)^2\int d^2xd^2y\times$$
\vspace{-6mm}
$$
\times{\rm e}^{\sqrt{K}\left(\psi_{\Lambda'}({\bf x})+\psi_{\Lambda'}({\bf y})\right)}
\sum\limits_{k=\pm 1}^{}
\left[A^{2k}({\bf x}-{\bf y})B^2({\bf x}-{\bf y})-1\right]
\times$$
\vspace{-5mm}
\begin{equation}
\label{zR}
\times\cos\left[\sqrt{K}\left(\chi_{\Lambda'}({\bf x})+k\chi_{\Lambda'}({\bf y})\right)\right],
\end{equation}
where $A({\bf x})\equiv{\rm e}^{-KG_h({\bf x})/2}$,
$B({\bf x})\equiv{\rm e}^{KG_{\phi}({\bf x})/2}$,

$$G_h({\bf x})=\!\!\!\int\limits_{\Lambda'<p<\Lambda}^{}\!\!\!\frac{d^2p}{(2\pi)^2}
\frac{{\rm e}^{i{\bf p}{\bf x}}}{p^2}\,,\qquad
G_{\phi}({\bf x})=\!\!\!\int\limits_{\Lambda'<p<\Lambda}^{}\!\!\!\frac{d^2p}{(2\pi)^2}
\frac{{\rm e}^{i{\bf p}{\bf x}}}{p^2+m_H^2}\ .$$
Since in what follows we will take $\Lambda'=\Lambda-d\Lambda$,
the factors $\left[A^{2k}({\bf x}-{\bf y})B^2({\bf x}-{\bf y})-1\right]$, $k=\pm 1$, are small. 
Owing to this fact, it is convenient
to introduce the coordinates ${\bf r}\equiv{\bf x}-{\bf y}$ and ${\bf R}\equiv\frac12({\bf x}+{\bf y})$, and
Taylor expand Eq.~(\ref{zR}) in powers of ${\bf r}$. Clearly, this expansion should be performed up to the
induced-interaction term $\sim\xi^2\int d^2R{\rm e}^{2\sqrt{K}\psi_{\Lambda'}({\bf R})}\cos\left(2\sqrt{K}
\chi_{\Lambda'}({\bf R})\right)$, that itself should
already be disregarded. 
As a result, we obtain the following expression for ${\mathcal Z}_{\Lambda}$ (For the
sake of uniformity, we replace $d^2R$ by $d^2x$.):
$$
{\mathcal Z}_\Lambda=\int\limits_{0<p<\Lambda'}^{}{\mathcal D}\chi({\bf p}){\mathcal D}\psi({\bf p})\exp\left\{
\int d^2x\left\{
\frac12\psi_{\Lambda'}\left(\partial^2-m_H^2\right)\psi_{\Lambda'}+\right.\right.
$$
\vspace{-5mm}
$$
+a_2(\xi A(0)B(0))^2
{\rm e}^{2\sqrt{K}\psi_{\Lambda'}}-
\frac12\left[1+a_1(\xi A(0)B(0))^2\frac{K}{2}{\rm e}^{2\sqrt{K}\psi_{\Lambda'}}\right]
\left(\partial_\mu\chi_{\Lambda'}
\right)^2+$$
\vspace{-5mm}
\begin{equation}
\label{renR}
\left.\left.+2\xi A(0)B(0){\rm e}^{\sqrt{K}\psi_{\Lambda'}}\cos\left(\sqrt{K}\chi_{\Lambda'}\right)
\right\}\right\},
\end{equation}
where we have introduced the notations 
\begin{equation}
\label{aaR}
a_1\equiv\int d^2rr^2\left[A^{-2}({\bf r})B^2({\bf r})-1\right],~
a_2\equiv\int d^2r\left[A^{-2}({\bf r})B^2({\bf r})-1\right].
\end{equation}
Taking into account that $\Lambda'=\Lambda-d\Lambda$ it is straightforward to get
$$
a_1=\alpha_1 K\frac{d\Lambda}{\Lambda^5}\left(1+\frac{\Lambda^2}{m_H^2}\right),
\qquad
a_2=\alpha_2 K\frac{d\Lambda}{\Lambda^3}\left(1+\frac{\Lambda^2}{m_H^2}\right).
$$
Here, $\alpha_{1,2}$ stand for some momentum-space-slicing dependent positive constants,
whose concrete values will turn out to be unimportant for the final expressions describing the
RG flow.

Next, since
$a_{1,2}$ occur to be infinitesimal (being proportional to $d\Lambda$), the
terms containing these constants on the r.h.s. of Eq.~(\ref{renR}) can be treated
in the leading-order approximation of the cumulant expansion that we will apply for the average over $\psi$. In fact,
we have
$$
\int\limits_{0<p<\Lambda'}^{}{\mathcal D}\psi({\bf p})\exp\left[
\frac12\int d^2x\psi_{\Lambda'}\left(\partial^2-m_H^2\right)\psi_{\Lambda'}\right]\times$$
\vspace{-5mm}
$$
\times
\exp\left(b\int d^2x {\rm e}^{2\sqrt{K}\psi_{\Lambda'}}f({\bf x})\right)\simeq
\exp\left[b{\rm e}^{2KG(0)}\int d^2x f({\bf x})+\right.$$
\vspace{-5mm}
$$
\left.+\,
\frac{b^2}{2}{\rm e}^{4KG(0)}\int d^2xd^2y\left({\rm e}^{4KG({\bf x}-{\bf y})}-1\right)f({\bf x})f({\bf y})\right],$$
where $f$ is equal either to unity or to $(\partial_\mu\chi_{\Lambda'})^2$, $b\sim a_{1,2}(\xi A(0)B(0))^2$, and
$$G({\bf x})\equiv\int\limits_{0<p<\Lambda'}^{}\frac{d^2p}{(2\pi)^2}
\frac{{\rm e}^{i{\bf p}{\bf x}}}{p^2+m_H^2},~
G(0)=\frac{1}{4\pi}\ln\left(1+\frac{{\Lambda'}^2}{{m_H}^2}\right).$$
In order to estimate the parameter of the cumulant expansion,
$\kappa\equiv b{\rm e}^{2KG(0)}\int d^2x\left({\rm e}^{4KG({\bf x})}-1\right)$,
note that we are working in the phase where
monopoles form the plasma, i.e. below $T_{\rm BKT}$. Because of this fact,
$4K|G({\bf x})|\le 32\pi(\Lambda'/m_H)^2$, 
that, due to the factor $(\Lambda'/m_H)^2$, is generally much smaller than unity.
Owing to that, we get
$$\kappa\simeq b\left(1+\frac{{\Lambda'}^2}{{m_H}^2}\right)^{\frac{K}{2\pi}}
\!\!\!\!\!\cdot 4K\!\!\int\! d^2x G({\bf x})\simeq
\frac{2bK}{\pi m_H^2}\left[1+\frac{K}{2\pi}\left(\frac{\Lambda'}{m_H}\right)^2\right].$$
Choosing for concreteness $b=a_2(\xi A(0)B(0))^2$ and taking into account that
\begin{equation}
\label{abRA}
A(0)\simeq 1-\frac{K}{2}G_h(0)=1-\frac{K}{4\pi}\frac{d\Lambda}{\Lambda}\ ,
\end{equation}
\vspace{-6mm}
\begin{equation}
\label{abRB}
B(0)\simeq 1+\frac{K}{2}G_{\phi}(0)\simeq 1+\frac{K}{4\pi}\left(\frac{\Lambda}{m_H}\right)^2
\frac{d\Lambda}{\Lambda}\ ,
\end{equation}
we obtain to the leading order:
$\kappa\simeq 2a_2K\xi^2/(\pi m_H^2)$. This quantity possesses double smallness --
firstly, because $a_2$ is infinitesimal and, secondly, due to the exponential smallness of $\xi$.

Such an extremely rapid convergence of the cumulant expansion enables one to replace
${\rm e}^{2\sqrt{K}\psi_{\Lambda'}}$ in the terms proportional to $a_{1,2}$ on the r.h.s. of Eq.~(\ref{renR})
by the average value of this exponent equal to $\left(1+\frac{{\Lambda'}^2}{{m_H}^2}\right)^{\frac{K}{2\pi}}$.
Comparing the so-obtained expression with the initial one, we arrive at the following renormalizations
of fields and parameters of the Lagrangian:
\begin{equation}
\label{lkmR}
\chi_{\Lambda'}^{\rm new}\!=\!C\chi_{\Lambda'},~~
\psi_{\Lambda'}^{\rm new}\!\!=\!C\psi_{\Lambda'},~~
K^{\rm new}\!\!=\frac{K}{C^2},~~
\mu^{\rm new}\!\!=\frac{\mu}{C^2},~~
\xi^{\rm new}\!\!=A(0)B(0)\xi,
\end{equation}
where $\mu\equiv m_H^2$, 
$$
C\equiv\left[1+\frac{Ka_1}{2}(\xi A(0)B(0))^2\left(1+\frac{{\Lambda'}^2}{{m_H}^2}
\right)^{\frac{K}{2\pi}}\right]^{1/2}\simeq$$
\vspace{-4mm}
\begin{equation}
\label{rgR}
\simeq\left[1+\frac{Ka_1}{2}(\xi A(0)B(0))^2\left(1+\frac{\Lambda^2}{m_H^2}\frac{K}{2\pi}\right)\right]^{1/2}.
\end{equation}
Besides that, we obtain the following shift of the free-energy density $F\equiv-\frac{\ln {\mathcal Z}'}{V}$:
\vspace{-4mm}
$$
F=F^{\rm new}-a_2(\xi A(0)B(0))^2\left(1+\frac{{\Lambda'}^2}{{m_H}^2}\right)^{\frac{K}{2\pi}}\simeq
$$
\vspace{-4mm}
\begin{equation}
\label{endensR}
\simeq F^{\rm new}-a_2(\xi A(0)B(0))^2\left(1+\frac{K}{2\pi}\frac{{\Lambda'}^2}{{m_H}^2}\right),
\end{equation}
where $V$ is the 2D-volume (i.e. area) of the system.

By making use of the relations~(\ref{abRA}), (\ref{abRB}), it is further straightforward to derive from 
Eqs.~(\ref{lkmR})-(\ref{endensR}) the RG equations in differential form. These are 
$$d\xi=-\frac{K\xi}{4\pi}\left(1-\frac{\Lambda^2}{\mu}\right)\frac{d\Lambda}{\Lambda}\ ,\quad
dK=-\frac{\alpha_1}{2}K^3\xi^2
\left[1+\left(\frac{K}{2\pi}+1\right)\frac{\Lambda^2}{\mu}\right]\frac{d\Lambda}{\Lambda^5}\ ,$$
\vspace{-4mm}
$$d\mu=-\frac{\alpha_1}{2}(K\xi)^2\mu
\left[1+\left(\frac{K}{2\pi}+1\right)\frac{\Lambda^2}{\mu}\right]\frac{d\Lambda}{\Lambda^5}\ ,$$
\vspace{-4mm}
$$
dF=\alpha_2K\xi^2\left[1+\left(\frac{K}{2\pi}+1\right)\frac{\Lambda^2}{\mu}\right]\frac{d\Lambda}{\Lambda^3}\ .$$
Now let us change from the momentum scale 
to the real-space one:
$\Lambda\to a\equiv1/\Lambda$, $d\Lambda\to-d\Lambda$. This modifies the above equations to
\begin{equation}
\label{xiR}
d\xi=-\frac{K\xi}{4\pi}\frac{da}{a}\left(1-\frac{1}{\mu a^2}\right),
\end{equation}
\vspace{-4mm}
\begin{equation}
\label{KRR}
dK=-\frac{\alpha_1}{2}K^3\xi^2a^3da\left[1+\left(\frac{K}{2\pi}+1\right)\frac{1}{\mu a^2}\right],
\end{equation}
\vspace{-4mm}
\begin{equation}
\label{muR}
d\mu=-\frac{\alpha_1}{2}(K\xi)^2\mu a^3da\left[1+\left(\frac{K}{2\pi}+1\right)\frac{1}{\mu a^2}\right],
\end{equation}
\vspace{-4mm}
\begin{equation}
\label{FR}
dF=\alpha_2K\xi^2ada\left[1+\left(\frac{K}{2\pi}+1\right)\frac{1}{\mu a^2}\right].
\end{equation}

Our main aim below is to derive from Eqs.~(\ref{xiR})-(\ref{muR}) the leading-order corrections in $(\mu a^2)^{-1}$
to the BKT RG flow in the vicinity of the critical point,
$K^{(0)}_c=8\pi$ (that clearly corresponds to $T_{\rm BKT}$), 
$y^{(0)}_c=0$, where $y\equiv \xi a^2$, and the subscript 
``${\,}{}^{(0)}{\,}$'' denotes
the zeroth order in the $(\mu a^2)^{-1}$-expansion. These values of $K^{(0)}_c$ and $y^{(0)}_c$
will be derived below. Additionally, it will be
demonstrated that, in the critical region, $\mu$ is evolving very slowly. Owing to this fact, the initial assumption
on the largeness of $\mu$ (namely, that it is of the order of $m_W^2$),
will be preserved by the RG flow, at least in that region.
This enables one to consider $\mu$ as almost a constant and seek corrections to the RG flow of $K^{(0)}$
in powers of $(\mu a^2)^{-1}$. The zeroth-order equation stemming from Eq.~(\ref{KRR}) then reads
\begin{equation}
\label{K0R}
dK^{(0)}=-\frac{\alpha_1}{2}K^{(0){\,}3}\xi^2a^3da.
\end{equation}
Respectively, the zeroth-order in $(\mu a^2)^{-1}$ equation for $y$ has the form
\begin{equation}
\label{y0R}
dy^{(0){\,}2}=2\frac{da}{a}y^{(0){\,}2}x,
\end{equation}
where $x\equiv 2-\frac{K^{(0)}}{4\pi}$.
Equations~(\ref{K0R}) and (\ref{y0R}) yield
the above-mentioned leading critical value of $K$, $K^{(0)}_c$. 
Next, with this value of $K^{(0)}_c$,
Eq.~(\ref{K0R}) can be rewritten in the vicinity of the critical point as
\begin{equation}
\label{xR}
dx=(8\pi)^2\alpha_1\xi^2a^3da.
\end{equation}
Introducing further instead of $y$ the new variable $z=(8\pi)^2\alpha_1y^2$ and performing the rescaling
$a^{\rm new}=a\sqrt{8\pi\alpha_1/\alpha_2}$
we get from Eqs.~(\ref{FR}), (\ref{y0R}), and~(\ref{xR}) the following system of equations:
\begin{equation}
\label{starR}
dz^{(0)}=2\frac{da}{a}xz^{(0)},~
dx=z^{(0)}\frac{da}{a},~
dF^{(0)}=z^{(0)}\frac{da}{a^3}.
\end{equation}
These equations yield the standard RG flow in the vicinity of the critical point, $x=z^{(0)}=0$,
which has the form~\cite{bkt, kogut} $z^{(0)}-x^2=\tau$, where
$\tau\propto(T_{\rm BKT}-T)/T_{\rm BKT}$ is some constant. In particular, $x\simeq\sqrt{z}$ at $T=T_{\rm BKT}-0$. 
Owing to the
first of Eqs.~(\ref{starR}), this relation yields
$\left(z_{\rm in}^{(0)}\right)^{-1/2}-
\left(z^{(0)}\right)^{-1/2}=\ln\left(a/a_{\rm in}\right)$, where the subscript ``${\,}{}_{\rm in}{\,}$''
means the initial value.
Taking into account that $z_{\rm in}^{(0)}$ is exponentially small, while
$z^{(0)}\sim 1$ (the value at which the growth of $z^{(0)}$ stops), we obtain in the case $x_{\rm in}\le\sqrt{\tau}$:
$\ln\left(a/a_{\rm in}\right)\sim\left(z_{\rm in}^{(0)}\right)^{-1/2}\sim\tau^{-1/2}$. According to this relation,
at $T=T_{\rm BKT}-0$, the correlation length diverges with an essential singularity as
$a(\tau)\sim\exp\left({\rm const}/\sqrt{\tau}\right)$. (At $T<T_{\rm BKT}$, $a\equiv d$, while at $T>T_{\rm BKT}$, the correlation
length becomes infinite due to the short-rangeness of molecular fields.)
As far as the leading part of the free-energy density is concerned, it scales as $F^{(0)}\sim a^{-2}$ and therefore
remains continuous in the critical region. Moreover, the correction to this behavior stemming from the finiteness
of the Higgs-boson mass [the last term on the r.h.s. of Eq.~(\ref{FR})] is clearly of the same functional form,
$\sim\exp\left(-{\rm const}'/\sqrt{\tau}\right)$, i.e. it is also continuous.

We are now in the position to address the leading-order [in $(\mu a^2)^{-1}$] corrections to the
above-discussed BKT RG flow of $K^{(0)}$ and $z^{(0)}$. To this end, let us represent $K$ and $z$ as
$K=K^{(0)}+K^{(1)}/(\mu a^2)$, $z=z^{(0)}+z^{(1)}/(\mu a^2)$
that, by virtue of Eqs.~(\ref{xiR}) and (\ref{KRR}), leads to the following novel equations:
\begin{equation}
\label{K1R}
dK^{(1)}-2K^{(1)}\frac{da}{a}=-4\pi\frac{da}{a}\left(z^{(0)}+\frac{z^{(1)}}{\mu a^2}\right)\left(1+\frac{K^{(0)}}{2\pi}+
\frac{3K^{(1)}}{K^{(0)}}\right),
\end{equation}
\vspace{-4mm}
\begin{equation}
\label{z1R}
dz^{(1)}-2z^{(1)}\frac{da}{a}=-2\frac{da}{a}\left[z^{(1)}\left(\frac{K^{(0)}}{4\pi}-2\right)+\frac{z^{(0)}}{4\pi}
\left(K^{(1)}-K^{(0)}\right)\right].
\end{equation}
In the vicinity of the critical point, we can
insert into Eq.~(\ref{z1R}) the above-obtained critical values of $K^{(0)}$ and $z^{(0)}$ to get 
\begin{equation}
\label{z11R}
dz^{(1)}=2z^{(1)}\frac{da}{a}\ .
\end{equation}
Therefore, $z^{(1)}=C_1a^2$, 
where $C_1$ is the integration constant of dimensionality $({\rm mass})^2$, $C_1\ll\mu$. Inserting further
this solution into Eq.~(\ref{K1R}), considered in the vicinity of the critical point, one obtains the following equation:
\begin{equation}
\label{K11R}
dK^{(1)}-2K^{(1)}\frac{da}{a}=-\frac{4\pi C_1}{\mu}\frac{da}{a}\left(\frac{3K^{(1)}}{8\pi}+5\right).
\end{equation}
Its integration is straightforward and yields
\begin{equation}
\label{C2R}
K^{(1)}=C_2\left(\mu a^2\right)^{1-\frac{3C_1}{4\mu}}+\frac{40\pi C_1}{4\mu-3C_1}\ ,
\end{equation}
where the dimensionless integration constant $C_2$ should be much smaller than $(\mu a^2)^{\frac{3C_1}{4\mu}}$. 
[Note that the last term 
in Eq.~(\ref{C2R}) is positive.] 
Therefore, the total correction, $K^{(1)}/(\mu a^2)$, approximately scales with $a$ in the critical region as
$\frac{40\pi C_1}{(4\mu-3C_1)\mu a^2}$. We see that, at the critical point, this expression
vanishes due to the divergence of the correlation length.
Note also that, for $z^{(1)}\ll\mu a^2$, Eq.~(\ref{C2R}) 
can obviously be rewritten as the following dependence of $K^{(1)}/(\mu a^2)$ on $z^{(1)}$: 
$$
\frac{K^{(1)}}{\mu a^2}=C_2\left(\mu a^2\right)^{-\frac{3z^{(1)}}{4\mu a^2}}+
\frac{40\pi z^{(1)}}{\left(4\mu a^2-3z^{(1)}\right)\mu a^2}\ .
$$
With the above-discussed critical behavior 
of the correlation length, $a(\tau)$, this relation determines the correction to the
BKT RG flow, $z^{(0)}-\left(2-\frac{K^{(0)}}{4\pi}\right)^2=\tau$.

Finally, in order to justify the approximation that $\mu$ was considered constant,
we should check that, under the RG flow, $\mu$ indeed evolves slowly. To this end, let us pass in
Eq.~(\ref{muR}), considered in the critical region, from the variable $\xi$ to the above-introduced variable $z$
and again perform the rescaling
$a\to a^{\rm new}$. This yields $\frac{d\mu}{\mu}=-\frac{z}{2}\frac{da}{a}$ or $d\mu=-\frac{C_1}{2}\frac{da}{a}$.
Since $C_1\ll\mu$, we conclude that $\frac{|d\mu|}{\mu}\ll\frac{da}{a}$. The obtained inequality means that, 
in the vicinity of the
BKT critical point, $\mu$ is really evolving slowly. 
This fact justifies our treatment of $\mu$ as a large (with respect to $\Lambda^2$)
constant quantity, approximately equal to its initial value, of the order of $m_W^2$.

Let us now proceed with a similar RG analysis of the SU($N$)-version of the theory~(\ref{ZMON})
given by Eq.~(\ref{zMON}). Applying to the respective 
dimensionally-reduced theory the above-described RG procedure, we arrive at the following analogue
of Eq.~(\ref{renR}):
$$
{\mathcal Z}_\Lambda=\int\limits_{0<p<\Lambda'}^{}{\mathcal D}\vec\chi({\bf p}){\mathcal D}\psi({\bf p})\exp\left\{
\int d^2x\left\{\frac12\,\psi_{\Lambda'}\left(\partial^2-m_H^2\right)
\psi_{\Lambda'}+\right.\right.$$
\vspace{-8mm}
$$+(\xi A(0)B(0))^2{\rm e}^{2\sqrt{K}\psi_{\Lambda'}}
\sum\limits_{ij}^{}a_2^{ij}\cos\left[\sqrt{K}\left(\vec q_i-\vec q_j\right)\vec\chi_{\Lambda'}\right]+
2\xi A(0)B(0)\times$$
\vspace{-8mm}
$$\times{\rm e}^{\sqrt{K}
\psi_{\Lambda'}}\sum\limits_{i}^{}\cos\left(\sqrt{K}\vec q_i\vec\chi_{\Lambda'}\right)-
\frac12\Biggl[\delta^{ab}+(\xi A(0)B(0))^2
\frac{K}{8}{\rm e}^{2\sqrt{K}\psi_{\Lambda'}}\times$$
\vspace{-8mm}
\begin{equation}
\label{ren1R}
\left.\left.\times\sum\limits_{ij}^{}a_1^{ij}\left(\vec q_i+\vec q_j\right)^{\alpha}
\left(\vec q_i+\vec q_j\right)^{\beta}
\cos\left[\sqrt{K}\left(\vec q_i-\vec q_j\right)\vec\chi_{\Lambda'}\right]\Biggr]\Bigl(
\partial_\mu\chi_{\Lambda'}^{\alpha}\Bigr)
\Bigl(\partial_\mu\chi_{\Lambda'}^{\beta}\Bigr)\right\}\right\}.
\end{equation}
Here, $\alpha,\beta=1,\ldots,(N-1)$, and we have introduced notations similar to (\ref{aaR}),
$$a_1^{ij}\equiv\int d^2rr^2\left[B^2({\bf r}){\rm e}^{K\vec q_i\vec q_jG_h({\bf r})}-1\right],~
a_2^{ij}\equiv\int d^2r\left[B^2({\bf r}){\rm e}^{K\vec q_i\vec q_jG_h({\bf r})}-1\right],$$
so that, at $\Lambda'=\Lambda-d\Lambda$,
$$a_1^{ij}=\alpha_1 K\frac{d\Lambda}{\Lambda^5}\left(\vec q_i\vec q_j+\frac{\Lambda^2}{m_H^2}\right),~
a_2^{ij}=\alpha_2 K\frac{d\Lambda}{\Lambda^3}\left(\vec q_i\vec q_j+\frac{\Lambda^2}{m_H^2}\right).$$
The main difference of Eq.~(\ref{ren1R}) from Eq.~(\ref{renR}) is due to the terms containing
$\cos\left[\sqrt{K}\left(\vec q_i-\vec q_j\right)\vec\chi_{\Lambda'}\right]$, which violate the
RG invariance. Nevertheless, approximately this invariance does hold, since the respective sums are dominated
by the terms with $i=j$. Working within this approximation and making use of the identity $\sum\limits_{i}^{}
q_i^\alpha q_i^\beta=\frac{N}{2}\delta^{\alpha\beta}$, we obtain
$$
{\mathcal Z}_\Lambda\simeq\exp\Bigg[a_2\frac{N(N-1)}{2}(\xi A(0)B(0))^2\left(
1+\frac{{\Lambda'}^{2}}{{m_H}^2}\right)^{\frac{K}{2\pi}}V\Bigg]\times$$
\vspace{-6mm}
$$
\times\int\limits_{0<p<\Lambda'}^{}{\mathcal D}\vec\chi({\bf p}){\mathcal D}\psi({\bf p})
\exp\Bigg\{
\int d^2x\Bigg\{\frac12\psi_{\Lambda'}\left(\partial^2-m_H^2\right)\psi_{\Lambda'}+$$
\vspace{-5mm}
$$
+2\xi A(0)B(0){\rm e}^{\sqrt{K}
\psi_{\Lambda'}}
\sum\limits_{i}^{}\cos\left(\sqrt{K}\vec q_i\vec\chi_{\Lambda'}\right)-
$$
\vspace{-6mm}
\begin{equation}
\label{doubleR}
-\frac12\Biggl[1+\left(
1+\frac{{\Lambda'}^{2}}{{m_H}^2}\right)^{\frac{K}{2\pi}}
\frac{NKa_1}{4}(\xi A(0)B(0))^2\Biggr]\left(\partial_\mu\vec\chi_{\Lambda'}
\right)^2\Bigg\}\Bigg\}\,.
\end{equation}
This expression has again been derived in the leading order of the cumulant
expansion applied in course of the average over $\psi$.

The shift of the free-energy density stemming from Eq.~(\ref{doubleR}) [cf. Eq.~(\ref{endensR})] reads
\begin{equation}
\label{en1R}
F^{\rm new}-F\simeq
a_2\frac{N(N-1)}{2}(\xi A(0)B(0))^2\left(
1+\frac{K}{2\pi}\frac{{\Lambda'}^{2}}{{m_H}^2}\right).
\end{equation}
As far as the renormalization of fields and coupling constants is concerned, it is given by Eq.~(\ref{lkmR}), where
the first equation should be modified as $\vec\chi_{\Lambda'}{}^{\rm new}=C\vec\chi_{\Lambda'}$, and
the parameter $C$ from Eq.~(\ref{rgR}) now reads
$$
C=\left[1+\frac{KNa_1}{4}(\xi A(0)B(0))^2\left(1+\frac{{\Lambda'}^2}{{m_H}^2}\right)^{\frac{K}{2\pi}}\right]^{1/2}
\simeq
$$
\vspace{-4mm}
\begin{equation}
\label{CNR}
\simeq\left[1+\frac{KNa_1}{4}(\xi A(0)B(0))^2\left(1+\frac{\Lambda^2}{m_H^2}\frac{K}{2\pi}\right)\right]^{1/2}.
\end{equation}
From Eqs.~(\ref{en1R}) and (\ref{CNR}) we deduce that, in the SU($N$)-case at $N>2$,
the RG flow of couplings and of the free-energy density is identical to that
of the SU(2)-case. Indeed, all the $N$-dependence can be absorbed into the constants $\alpha_{1,2}$
by rescaling them as $\bar\alpha_1\equiv N\alpha_1/2$, $\bar\alpha_2=N(N-1)\alpha_2/2$ and further redefining
[cf. the notations introduced after Eq.~(\ref{xR})]
$\bar z=(8\pi)^2\bar\alpha_1y^2$ and $\bar a^{\rm new}=a\sqrt{8\pi\bar\alpha_1/\bar\alpha_2}$.
In particular, the critical temperature $T_{\rm BKT}$
remains the same as in the SU(2)-case. (As has already been mentioned, 
this also follows from the estimate of the mean squared separation in the monopole-antimonopole
molecule, if one takes into account that the square of any root vector is equal to unity.)
Thus, the principal difference of the SU($N$)-case, $N>2$, from the
SU(2)-one is that, while in the SU(2)-case
the RG invariance is exact (modulo the negligibly small higher-order terms of the cumulant expansion applied
to the average over $\psi$), in the SU($N$)-case it is only approximate, even in the limit $m_H\to\infty$.

\subsection{Finite-temperature 3D compact QED with massless fundamental fermions}

Let us consider the extension of the model~(\ref{SGG}) by
fundamental dynamical quarks~\cite{quarks} (for the sake of simplicity,
we omit the summation over the flavor indices, although consider the
general case of an arbitrary number of flavors). The action is modified by 
$\Delta S=-i\int d^3x\bar\psi\left(\vec\gamma
\vec D+h\frac{\tau^a}{2}\Phi^a\right)\psi$, where 
the Yukawa coupling $h$ has the dimensionality $[{\rm mass}]^{1/2}$,
$D_\mu\psi=\left(\partial_\mu-ig\frac{\tau^a}{2}A_\mu^a
\right)\psi$, $\bar\psi=\psi^{\dag}\beta$, 
the Euclidean Dirac matrices are defined as $\vec\gamma=
-i\beta\vec\alpha$ with
$\beta=\left(
\begin{array}{cc}
1& 0\\
0& -1
\end{array}
\right)$,
$\vec\alpha=\left(
\begin{array}{cc}
0& \vec\tau\\
\vec\tau& 0
\end{array}
\right)$, and $\vec\tau$ denote the Pauli matrices.
As will be demonstrated, at $T>T_{\rm BKT}$,
quark zero modes in the monopole field lead to the additional attraction
of a monopole and an antimonopole ($M$ and $\bar M$ for shortness)
in the molecule.
In particular, when the number of these modes (equal to the
number of massless flavors) is sufficiently
large, the molecule shrinks so strongly that its size becomes of the order
of the inverse W-boson mass. Another factor which defines the size of the
molecule is the characteristic range of localization of zero modes. Namely, it can be shown that
the stronger zero modes are localized in the vicinity of the monopole center, the
smaller the molecular size is. We will consider the case when the Yukawa coupling $h$ of
quarks with the Higgs field vanishes,
and originally massless quarks do not acquire any mass. This means that
zero modes are maximally delocalized.
Such a weakness of the quark-mediated interaction of monopoles
opens the possibility for molecules to
undergo eventually the phase transition
into the plasma phase at the temperatures of the order of $T_{\rm BKT}$.
However, this will be shown to occur only provided that
the number of flavors is equal to one, whereas at any larger
number of flavors,
the critical temperature becomes exponentially small.
This means that the interaction mediated by such a number
of zero modes is already strong enough to maintain the molecular phase
at any larger temperature. 

For a short while, let us consider the general case $h\ne 0$.
One can then see that the Dirac equation in the field of the third
isotopic component of the 't Hooft--Polyakov monopole~\cite{mon} decomposes
into two equations for the components of the SU(2)-doublet $\psi$.
The masses of these components stemming from such equations are
equal to each other and read $m_q=h\eta/2$.
Next, the Dirac equation in the full monopole potential has been
shown~\cite{jr} to possess a zero mode, whose
asymptotic behavior at $r\equiv\left|\vec x{\,}\right|\gg m_q^{-1}$
has the form 
\begin{equation}
\label{N}
\chi_{\nu{\,}n}^{+}={\mathcal N}\,\frac{{\rm e}^{-m_qr}}{r}
\left(s_\nu^{+}s_n^{-}-s_\nu^{-}s_n^{+}\right),~~
\chi_{\nu{\,}n}^{-}=0.
\end{equation}
Here, $\chi^{\pm}_n$ are the upper and the lower components of the mode,
i.e. $\psi={\chi_n^{+}\choose \chi_n^{-}}$. Next,
$n=1,2$ is the isotopic index, $\nu=1,2$ is the Dirac index,
$s^{+}={1\choose 0}$, $s^{-}={0\choose 1}$, and ${\mathcal N}$
is the normalization constant.

It is a well known fact that, in 3D, the 't~Hooft--Polyakov monopole is actually
an instanton~\cite{polbook, pol}. Therefore, we can use the results of Ref.~\cite{lb} on the quark
contribution to the effective action of the instanton-antiinstanton molecule
in QCD. Let us thus recapitulate
the analysis of Ref.~\cite{lb} adopting it to our model.
To this end, we fix the gauge $\Phi^a=\eta\delta^{a3}$ and
define the analogue of the free propagator $S_0$ by the relation
$S_0^{-1}=-i\left(\vec\gamma\vec\partial+m_q\tau^3\right)$.
Next, we define the
propagator $S_M$ in the field of a monopole located at the origin,
$\vec A^{a{\,}M}$
[$A_i^{a{\,}M}\to\varepsilon^{aij}x^j/\left(gr^2\right)$ at $r\gg m_W^{-1}$],
by the formula $S_M^{-1}=S_0^{-1}-g\vec\gamma\frac{\tau^a}{2}\vec A^{a{\,}M}$.
Obviously, the propagator $S_{\bar M}$ in the field of an antimonopole
located at a certain point $\vec R$,
$\vec A^{a{\,}\bar M}\left(\vec x\right)=-\vec A^{a{\,}M}\left(\vec x-\vec R
\right)$, is defined by the equation for
$S_M^{-1}$ with the replacement $\vec A^{a{\,}M}\to \vec A^{a{\,}\bar M}$.
Finally, one can consider the molecule made out of these monopole
and antimonopole and define the total propagator $S$ in the field of such a
molecule, $\vec A^a=\vec A^{a{\,}M}+\vec A^{a{\,}\bar M}$,
by means of the equation for $S_M^{-1}$ with $\vec A^{a{\,}M}$
replaced by $\vec A^a$.

One can further
introduce the notation $\left|\psi_n\right>$, $n=0,1,2,\ldots$,
for the eigenfunctions of the operator $-i\vec\gamma\vec D$ defined
at the field of the molecule, namely $-i\vec\gamma\vec D
\left|\psi_n\right>=\lambda_n\left|\psi_n\right>$, where
$\lambda_0= 0$. This yields the following formal spectral
representation for the total propagator $S$:

$$S\left(\vec x, \vec y\right)=\sum\limits_{n=0}^{\infty}
\frac{\left|\psi_n(\vec x)\right>\left<\psi_n(\vec y)\right|}
{\lambda_n-im_q\tau^3}\ .$$
Next, it is worth recalling the mean-field approximation, discussed in section~3,
within which zero modes dominate in the quark propagator, i.e.
\begin{equation}
\label{domin}
S\left(\vec x, \vec y\right)\simeq
\frac{\left|\psi_0(\vec x)\right>\left<\psi_0(\vec y)\right|}
{-im_q\tau^3}+S_0\left(\vec x, \vec y\right).
\end{equation}
The approximation~(\ref{domin}) remains valid
for the molecular phase near the phase transition (i.e. when the
temperature approaches $T_{\rm BKT}$ from above), we will be interested in.
That is merely due to the fact that, in this regime, molecules become very much
inflated being about to dissociate.

Within the notations adopted, one now has
$S=\left(S_M^{-1}+S_{\bar M}^{-1}-S_0^{-1}\right)^{-1}=
S_{\bar M}{\mathcal S}^{-1}S_M$, where
$${\mathcal S}= S_0-\left(S_M-S_0\right)S_0^{-1}
\left(S_{\bar M}-S_0\right)=S_0-
\frac{\big|\psi_0^M\big>\big<\psi_0^M\big|}{-im_q\tau^3}
\,S_0^{-1}\frac{\big|\psi_0^{\bar M}\big>
\big<\psi_0^{\bar M}\big|}{-im_q\tau^3}\ ,$$
and $\big|\psi_0^M\big>$, $\big|\psi_0^{\bar M}\big>$ are the
zero modes of the operator $-i\vec\gamma\vec D$ defined in the field of
a monopole and an antimonopole, respectively. Denoting further
$a=\big<\psi_0^{\bar M}\big|g\vec\gamma\frac{\tau^a}{2}
\vec A^{a{\,}M}\big|\psi_0^M\big>$,
it is straightforward to see, by the definition of the zero mode,
that $a=\big<\psi_0^{\bar M}\big|-i\vec\gamma\,\vec\partial
\big|\psi_0^M\big>=\big<\psi_0^{\bar M}\big|S_0^{-1}
\big|\psi_0^M\big>$. This yields ${\mathcal S}=S_0+({a^{*}}/{m_q^2})
\big|\psi_0^M\big>\big<\psi_0^{\bar M}\big|$, where the star
stands for complex conjugation, and therefore $\det {\mathcal S}=
\big[1+\left({|a|}/{m_q}\right)^2\big]\cdot\det S_0$. Finally,
defining the desired effective action as $\Gamma=\ln
[{\det S^{-1}}/{\det S_0^{-1}}]$, we obtain 
$\Gamma={\,}{\rm const}{\,}+N_f\ln\left(m_q^2+|a|^2\right)$ in the
general case with $N_f$ flavors.
The constant in this formula, standing for the
sum of effective actions defined at the monopole and at the antimonopole,
cancels out in the normalized expression for the mean squared
separation in the $M \bar M$-molecule.

Let us now set $h$ equal to zero, and so $m_q$ is equal to zero as well.
Notice first of all that, although in this case the direct Yukawa
interaction of the Higgs boson with quarks is absent, they keep
interacting with each other
via the gauge field. Owing to this fact, the problem of
finding a quark zero mode in the monopole field is still legitimate. 
[Note that, according to Eq.~(\ref{N}), this mode will be non-normalizable
in the sense of a discrete spectrum.
However, it is clear that, in the gapless case $m_q=0$ under discussion, the zero mode,
which lies exactly on the border of the two contiguous Dirac seas, should
be treated not as an isolated state of a discrete spectrum, but rather
as a state of a continuum spectrum. 
 This means that it should be
understood as follows: $\left|\psi_0(\vec x)\right>
\sim\lim\limits_{p\to 0}^{}\left({\rm e}^{ipr}/r\right)$, where $p=
\left|\vec p{\,}\right|$. Once considered in this way, zero modes
are normalizable by the standard condition of normalization of the
radial parts of spherical waves, $R_{pl}$, which reads~\cite{zeromode1}
$\int\limits_{0}^{\infty}drr^2R_{p'l}R_{pl}=2\pi\delta(p'-p)$.]
The dependence of the
absolute value of the matrix element $a$ on the distance $R$ between
a monopole and an antimonopole
can now be readily found. Indeed, we have
$|a|\propto\int d^3r/\big(r^2\big|\vec r-\vec R\big|\big)\simeq
-4\pi\ln(\mu R)$, where $\mu$ stands for the infrared cutoff.

We can now switch on the temperature are explore a possible modification of the
BKT critical temperature, $T_{\rm BKT}$,
due to the zero-mode mediated interaction. As has been discussed, this can be done upon the evaluation of
the mean squared separation in the $M \bar M$-molecule 
and further finding the temperature below which it starts diverging.
In this way we should take into account that, in the dimensionally-reduced
theory, the usual Coulomb interaction of monopoles (Without 
loss of generality, we consider the molecule with the temporal component
of $\vec R$ equal to zero.)
$R^{-1}=
\sum\limits_{n=-\infty}^{+\infty}\left({\mathcal R}^2+(\beta n)^2\right)^{-1/2}$
goes over into $-2T\ln(\mu{\mathcal R})$, where ${\mathcal R}$ denotes the
absolute value of the 2-vector ${\bf R}$. This statement can be checked
e.g. by virtue of the Euler -- MacLaurin formula.
As far as the novel logarithmic interaction,
$\ln(\mu R)= \sum\limits_{n=-\infty}^{+\infty}\ln\left[\mu
\left({\mathcal R}^2+(\beta n)^2\right)^{1/2}\right]$, is concerned, it
transforms into
\begin{equation}
\label{sumlog}
\pi T{\mathcal R}+\ln\left[1-\exp(-2\pi T{\mathcal R})\right]-\ln 2\ .
\end{equation}
Let us prove this statement.
To this end, we make use of the following formula~\cite{gr}:
$$
\sum\limits_{n=1}^{\infty}\frac{1}{n^2+x^2}=\frac{1}{2x}
\left[\pi\coth(\pi x)-\frac{1}{x}\right].
$$
This yields
$$
x\sum\limits_{n=-\infty}^{+\infty}\frac{1}{x^2+\left({2\pi n}/{a}\right)^2}=
\frac{1}{x}+\frac{xa^2}{2\pi^2}\sum\limits_{n=1}^{\infty}\frac{1}{n^2+
\left({xa}/{2\pi}\right)^2}=\frac{a}{2}\coth\left(\frac{ax}{2}\right).
$$
On the other hand, the l.h.s. of this expression can be written as
$$\frac{1}{2}\,
\frac{d}{dx}\sum\limits_{n=-\infty}^{+\infty}\ln\left[x^2+\left(\frac{2\pi n}{a}\right)^2
\right].
$$
Integrating over $x$, we get
$$
\sum\limits_{n=-\infty}^{+\infty}\ln\left[x^2+\left(\frac{2\pi n}{a}\right)^2
\right]=a\int dx\coth\left(\frac{ax}{2}\right)=$$
\vspace{-5mm}
$$=2\ln\sinh\left(\frac{ax}{2}\right)=ax+
2\ln\left(1-{\rm e}^{-ax}\right)-2\ln 2\ .
$$
Setting $\frac{2\pi}{a}=\mu \beta$ and $x=\mu {\mathcal R} $ we arrive at Eq.~(\ref{sumlog}).

Thus, the statistical weight of the quark-mediated $M\bar M$-interaction in the
molecule at high temperatures has the form
$\exp({-2N_f\ln|a|})\propto\left[
\pi T{\mathcal R}+\ln\left[1-\exp(-2\pi T{\mathcal R})\right]-\ln 2
\right]^{-2N_f}$. Accounting for both, (former) logarithmic and Coulomb,
interactions we eventually arrive at the following expression for the
mean squared separation $\left<L^2\right>$ in the molecule as a function of
$T$, $g$, and $N_f$:
\begin{equation}
\label{lsqq}
\left<L^2\right>=\frac{
\int\limits_{m_W^{-1}}^{\infty}d{\mathcal R}{\mathcal R}^{3-\frac{8\pi T}{g^2}}
\left[\pi T{\mathcal R}+\ln\left[1-\exp(-2\pi T{\mathcal R})\right]-\ln 2\right]^{-2N_f}}
{\int\limits_{m_W^{-1}}^{\infty}d{\mathcal R}{\mathcal R}^{1-\frac{8\pi T}{g^2}}
\left[\pi T{\mathcal R}+\ln\left[1-\exp(-2\pi T{\mathcal R})\right]-\ln 2\right]^{-2N_f}}\ .
\end{equation}
At large ${\mathcal R}$, $\ln 2 \ll \pi T{\mathcal R}$ and
$\bigl|\ln\left[1-\exp(-2\pi T{\mathcal R})\right]\bigr|\simeq
\exp(-2\pi T{\mathcal R})\ll
\pi T{\mathcal R}$. Consequently, we see that $\left<L^2\right>$ is finite at
$T>T_{\rm BKT}^{N_f}=(2-N_f)g^2/(4\pi)$, that reproduces $T_{\rm BKT}$
at $N_f=0$. For $N_f=1$, the plasma phase
is still present at $T<g^2/(4\pi)$. Instead, 
when $N_f\ge 2$, at whatever temperatures, which are parametrically larger than 
the temperature of dimensional reduction, ${\mathcal O}(\zeta^{1/3})$, the monopole ensemble exists only
in the molecular phase. At such temperatures, massless dynamical quarks produce their own deconfinement.
They also destroy the monopole-based confinement of any other fundamental matter.
Clearly, the obtained temperature $T_{\rm BKT}^{(N_f=1)}$ is exponentially larger than  
$\zeta^{1/3}$ just as $T_{\rm BKT}$ is, that fully validates
the idea of dimensional reduction. Note also that, at $N_f\gg \max\left\{1,{4\pi T}/{g^2}\right\}$,
$\sqrt{\left<L^2\right>}\to m_W^{-1}$,
which means that such a large number of zero modes shrinks the molecule
to the minimal admissible size. In conclusion, notice that, in the presence of W-bosons, the influence 
of {\it heavy} fundamental scalar matter to the dynamics of the deconfining phase transition has been 
studied in Ref.~\cite{dkn}.

\subsection{3D GG model and its supersymmetric generalization at finite temperature}

The crucial difference in the finite-temperature behavior of the 3D GG model~\cite{dkkt} from that of 
3D compact QED~\cite{az} stems from the presence in the former of the heavy charged $W^{\pm}$-bosons. 
Although being practically irrelevant at zero temperature due to their heaviness, at finite temperature these bosons,
due to their thermal excitation, form a plasma whose density is compatible to that of monopoles.
The density of the $W^{\pm}$-plasma can be evaluated as follows (see e.g. Ref.~\cite{feyn}):
$$\left.\rho_W=-6T\left\{\frac{\partial}{\partial\tilde\mu}\int\frac{d^2p}{(2\pi)^2}\ln\left[1-
{\rm e}^{\beta\left(\tilde\mu-\varepsilon({\bf p})\right)}\right]\right\}\right|_{\tilde\mu=0}=
6\int\frac{d^2p}{(2\pi)^2}\frac{1}{{\rm e}^{\beta\varepsilon({\bf p})}-1}=$$
\vspace{-4mm}
$$=\frac{3m_W^2}{\pi}\int\limits_{1}^{\infty}\frac{dzz}{{\rm e}^{m_W\beta z}-1}\simeq
\frac{3m_W^2}{\pi}\int\limits_{1}^{\infty}dzz{\rm e}^{-m_W\beta z}=
\frac{3m_WT}{\pi}\left(1+\frac{T}{m_W}\right){\rm e}^{-m_W\beta}.$$
Here, $\tilde\mu$ stands for the chemical potential, $\varepsilon({\bf p})=\sqrt{{\bf p}^2+m_W^2}$, and the 
factor ``6'' represents the total number of spin states of $W^{+}$- and $W^{-}$-bosons. We have also
denoted $z\equiv\varepsilon({\bf p})/m_W$ and taken into account that, since we want to stay in the 
plasma phase of monopoles and the temperatures under consideration
should not exceed $T_{\rm BKT}$, $T\ll m_W$. Therefore, up to small corrections, 
$\rho_W=\frac{3m_WT}{\pi}{\rm e}^{-m_W\beta}$.

It has been found in~\cite{dkkt}, further elaborated in~\cite{dkn, higgs}, etc., and 
reviewed in~\cite{ai} that, when the finite-temperature 3D compact QED is promoted, upon the incorporation of 
W-bosons, to the finite-temperature 3D GG model, the critical temperature $T_{\rm BKT}$ 
and the U(1) universality class of the 
deconfining phase transition are changed to $T_c=\frac{g^2}{4\pi\epsilon}$ 
and the 2D-Ising universality class, respectively. In what follows, we will survey
the main ideas of Ref.~\cite{dkkt}, proceeding further to the analysis 
of the supersymmetric 3D GG model at finite temperature~\cite{susy}.

The Lagrangian of 3D compact QED, rewritten 
in terms of the vortex operator $V={\rm e}^{-ig_m\chi/2}$, is 
$${\mathcal L}_{\rm 3D}=
\frac12(\partial_\mu\chi)^2-2\zeta\cos(g_m\chi)=\frac{2}{g_m^2}\left|\partial_\mu V\right|^2-
\zeta\left[V^2+(V^{*})^2\right].$$
In this form it has 
explicit magnetic $Z_2$-symmetry~\cite{Alik}. 
At zero temperature, this symmetry is 
spontaneously broken, since $\left<V(\vec x)V^{*}(0)\right>\stackrel{|\vec x|\to\infty}{\longrightarrow}1$.
It can be shown~\cite{Alik} that the breakdown of the magnetic symmetry implies confinement.
Since this symmetry is inherited in the 3D compact QED from the initial 3D GG model, 
the deconfinement phase transition, occurring in the latter at a certain temperature (equal to $T_c$), 
should be associated with the 
restoration of this very symmetry. Already this fact alone indicates that the 
phase transition in the finite-temperature 3D GG model
should be of the same kind as that of the ($Z_2$-invariant) 2D Ising model, rather than the BKT
phase transition of compact QED. It has been shown quantitatively in Ref.~\cite{dkkt} that this is indeed
the case if W-bosons are additionally included in the compact-QED Lagrangian. At finite temperature, 
this can be done by noting that W-bosons are nothing but 
vortices of the $\chi$-field and they can be incorporated into the dimensionally-reduced Lagrangian,
${\mathcal L}_{\rm 2D}=\frac12(\partial_\mu\chi)^2-2\xi\cos(\sqrt{K}\chi)$,
by adding to it the term $-2\mu\cos\tilde\chi$ (from now on, $\chi$ will denote the dual-photon 
field in the {\it dimensionally-reduced} 
theory). The field $\tilde\chi$, dual to the field $\chi$, is defined 
through the relation $i\partial_\mu\tilde\chi=g\sqrt{\beta}\varepsilon_{\mu\nu}\partial_\nu\chi$. The fugacity 
of W-bosons, $\mu$, is proportional to their density $\rho_W$, therefore $\mu\propto m_WT{\rm e}^{-m_W\beta}$.
[Note also that an alternative way to introduce $W$-bosons into ${\mathcal L}_{\rm 2D}$ has been proposed in 
Ref.~\cite{kson}. Within that approach, the above definition of $\tilde\chi$ through $\chi$ is abolished.
Instead,
an extra interaction of these, now independent, fields of the type $i(\partial_{x_1}\chi)
(\partial_{x_2}\tilde\chi)$ appears.]

Owing to the novel cosine term, even in the absence of the monopole plasma, the dual photon never
becomes massless~\cite{dkkt}. Rather, its mass $m_D$ increases, and the vacuum correlation length
$d$ decreases, with the increase of the temperature. Consequently, contrary to what we had in the case of the inverse BKT phase 
transition, where the correlation length was becoming infinite at $T>T_{\rm BKT}$, now it never becomes infinite.
This result parallels the general expectation that, with the increase of the temperature in any 
local field theory, the degree of disorder becomes higher, and the correlation length decreases.
Indeed, in the absence of monopoles (i.e. at $\xi=0$), 
\begin{equation}
\label{l2d}
{\mathcal L}_{\rm 2D}=\frac12(\partial_\mu\phi)^2-2\mu
\cos\left(g\sqrt{\beta}\phi\right),
\end{equation} 
where $\phi\equiv\sqrt{T}\tilde\chi/g$. Therefore, $m_D^2(\beta)=2\mu\beta g^2\propto
m_Wg^2{\rm e}^{-m_W\beta}$. We see that $m_D$ grows with the 
decrease of $\beta$, i.e. with the increase of $T$, proving the above statement.

Evaluating further the mean squared separation 
in the $W^{+}W^{-}$-molecule in the same way as we did earlier for 
the $M\bar M$-molecule [cf. Eqs.~(\ref{lsquared}) and~(\ref{lsqq})], we have 
$\left<L^2\right>\propto\int d^2{\bf R}{\mathcal R}^{2-\frac{g^2}{2\pi T}}$.
The convergence of this integral at $T<T_{\rm XY}\equiv\frac{g^2}{8\pi}$ means that, in the absence of monopoles, 
with the increase of temperature W-bosons pass from the molecular phase into the plasma one 
at $T=T_{\rm XY}$. In the same case of absence of dynamical
monopoles, a {\it static} $M\bar M$-pair becomes linearly confined at $T>T_{\rm XY}$. Indeed, the $M\bar M$-potential, 
${\mathcal V}$, is 
related to the correlation function of two vortex operators as 
$\left<V({\bf x})V^{*}({\bf y})\right>={\rm e}^{-\beta {\mathcal V}({\bf x}-
{\bf y})}$. In the phase where static monopoles are confined, one has 
\begin{equation}
\label{vorvor}
\left<V({\bf x})V^{*}({\bf y})\right>\stackrel{|{\bf x}-{\bf y}|\to\infty}{\longrightarrow}{\rm e}^{-\beta\sigma
|{\bf x}-{\bf y}|}\ ,
\end{equation} 
and the magnetic $Z_2$-symmetry is restored. Let us clarify the origin of Eq.~(\ref{vorvor}) and  
get an idea of the mass parameter which appears in 
$\sigma$. To this end, notice that we are dealing with the situation 
dual to that of finite-temperature 
3D compact QED below $T_{\rm BKT}$~\cite{az}. In the latter case, external static quarks, represented by 
Polyakov loops, are linearly confined. Therefore, 
just as the correlation function of two Polyakov loops in that theory 
decreases with the distance at the inverse soliton mass, so does the 
correlation function of two vortex operators in our case.  
In fact, it can be shown that vortex operators create solitons in the theory~(\ref{l2d}).
Solitons in this theory carry a unit topological charge, corresponding to the U(1) symmetry generated by the 
topological current $j_\mu=\frac{g\sqrt{\beta}}{2\pi}\varepsilon_{\mu\nu}\partial_\nu\phi$. 
In the dimensionally-reduced theory, 
the 3D vortex operator, $V(\vec x)={\rm e}^{-ig_m\chi(\vec x)/2}$, is replaced by 
$V({\bf x})={\rm e}^{-i\sqrt{K}\chi({\bf x})/2}$, or equivalently, 
\begin{equation}
\label{vx}
V({\bf x})=\exp\left(\frac{\sqrt{K}}{2}\varepsilon_{\mu\nu}\int\limits_{\bf x}^{}dx_\mu\partial_\nu\phi\right).
\end{equation}
The integration contour here goes either to infinity or to a point where a vortex operator of the opposite magnetic 
charge, 
\begin{equation}
\label{vy}
V^{*}({\bf y})=\exp\left(\frac{\sqrt{K}}{2}\varepsilon_{\mu\nu}\int\limits_{}^{\bf y}dx_\mu\partial_\nu\phi\right),
\end{equation}
is defined. One can further demonstrate that the operators~(\ref{vx}) and~(\ref{vy}) have topological charges $1$ and 
$-1$ and are nothing but the creation operators of a soliton and an antisoliton respectively. This fact stems from the relations
$\left<QV(0)\right>=\left<V(0)\right>$ and $\left<QV^{*}(0)\right>=-\left<V^{*}(0)\right>$, where 
$Q=\varepsilon_{\mu\nu}\oint_{C}^{}dx_\mu
j_\nu({\bf x})$ is the topological-charge operator with the contour $C$ encircling the origin anticlockwise.
These relations stem from others, 
$$\left<j_\mu({\bf x})V(0)\right>\simeq-\frac{x_\mu}{2\pi{\bf x}^2}\left<V(0)\right>,\qquad
\left<j_\mu({\bf x})V^{*}(0)\right>\simeq\frac{x_\mu}{2\pi{\bf x}^2}\left<V^{*}(0)\right>$$
(where ``$\simeq$'' means ``neglecting correlation functions of topological currents higher than the two-point one''),   
that themselves can be obtained from the two-point correlation function of topological currents,
$$\left<j_\mu({\bf x}j_\nu(0)\right>=\frac{g^2\beta}{(2\pi)^3}\left(2\frac{x_\mu x_\nu}{{\bf x}^2}-\delta_{\mu\nu}
\right)\frac{1}{{\bf x}^2}\ .$$
Owing to the fact that the operator~(\ref{vx}) [(\ref{vy})] creates a soliton [antisoliton], the large-distance 
behavior of the correlation function~(\ref{vorvor}) is determined by the lightest state with unit 
topological charge, i.e. by the soliton. One can therefore identify $\beta\sigma$ on the r.h.s. of Eq.~(\ref{vorvor})
with the soliton mass $m_{\rm sol}$, therefore $\sigma=Tm_{\rm sol}$. 

We have considered above the two idealistic situations, namely when either dynamical monopoles or W-bosons are absent.
In reality, when both are present and interact with each other, the deconfining phase transition temperature 
is neither $T_{\rm BKT}$ nor $T_{\rm XY}$, but is rather $T_c$, which in fact lies between the two. 
The deconfining phase transition takes place when the densities of monopoles and W-bosons become equal, i.e. $2\xi=\rho_W$.
Up to inessential preexponential factors, this happens when the monopole action $S_0$ is equal to $m_W\beta$, 
i.e. at $T=T_c$. At this temperature, the 
thickness of the string connecting a $W^{+}$- and a $W^{-}$-boson is 
of the order of its length. As has been discussed [cf. e.g. after Eq.~(\ref{Sig})], 
the thickness of the string is  
the inverse Debye mass of the dual photon, $d\propto\zeta^{-1/2}$. The length of the 
string is clearly of the order of the average distance between W's, that is $\mu^{-1/2}$. Therefore, again up to 
preexponential factors, the thickness and the length of the string are equal at $T=T_c$.
Note that these qualitative arguments can be formalized by the RG procedure~\cite{dkkt}. In the presence 
of W-bosons, the RG equations possess three fixed points. The first two are the zero- and infinite-temperature ones.
The third fixed point of the RG flow is a nontrivial infrared unstable one, $T=T_c$, $\xi=\mu$, 
which corresponds to the phase transition. [In this point, however, both fugacities become infinite.
It has been demonstrated in Ref.~\cite{kson} that 
a correcting factor $(\epsilon+2)/(2\epsilon+1)$, by which 
$T_c$ should be multiplied, appears if one demands that the RG flow of the 
fugacities should stop when at least one of them becomes of the order of unity.]

Note that an independent condition of the deconfining phase transition stems from the 
coincidence of the scaling dimensions of the operators $:\!\cos\big(\sqrt{K}\chi\big)\!:$ and  
$:\!\cos\left(g\sqrt{\beta}\phi\right)\!:$ at the critical temperature. 
The scaling dimensions are equal $\frac{K}{4\pi}$ and $\frac{g^2\beta}{4\pi}$
respectively, so that the monopole cosine term is relevant at $T<T_{\rm BKT}$, 
whereas the cosine term of the W-bosons 
is relevant at $T>T_{\rm XY}$. The two scaling dimensions become equal when $g_m^2T=g^2\beta$, 
that yields the critical temperature 
$\frac{g^2}{4\pi}$. As we see, this value is indeed equal to $T_c$, up to the factor $\frac{1}{\epsilon}$, which is of 
the order of unity and is 
generated by the loop corrections. At the temperature $\frac{g^2}{4\pi}$, both scaling dimensions are equal to unity,
therefore both cosine terms in ${\mathcal L}_{\rm 2D}$ are relevant at this temperature. In fact, in the whole 
region of temperatures $T_{\rm XY}<T<T_{\rm BKT}$, both terms are relevant.

The supersymmetric version of the 3D GG model~\cite{susy} contains, in addition to the dual photon,
its superpartner -- an adjoint fermion, which we will call the photino. This model possesses a discrete 
parity symmetry, which should lead to
the masslessness of photinos. At zero temperature, this is however not the case, since the parity is
spontaneously broken via a nonvanishing photino condensate. Thus,
at finite temperature, one can anticipate two phase transitions --
one related to the vanishing of the photino condensate and the
other one due to deconfinement. These two transitions could either
be distinct and happen at different temperatures, or could
coincide. In this respect, the model is similar to QCD with adjoint
quarks, where a similar question can be asked about the
(non-)coincidence of deconfinement and restoration of discrete
chiral symmetry.

The Lagrangian of the model
contains the bosonic fields of the nonsupersymmetric
GG model, that are the massless photon, the heavy
$W^\pm$ vector bosons and the massive Higgs field, as well as
their superpartners -- photino, winos and Higgsino. It has been shown
in~\cite{ahw} that, just like in the GG model, the monopole effects
render the dual photon massive, although the mass in this case is
parametrically smaller, since it is due to the contribution from a
two-monopole sector, rather than a single-monopole sector as in
the GG model. The low-energy
sector of the theory is described by the supersymmetric sine-Gordon model.
Its Euclidean action in the superfield notation is [We
adopt here the notations of Ref.~\cite{MZ}, in particular $\int
d^2\theta\bar\theta\theta=1\,$.]
\begin{equation}
\label{0su}
S=-\int d^3xd^2\theta\left[\frac12\,\Phi\bar D_\alpha D_\alpha\Phi+\bar\zeta
\cos\left(g_m\Phi\right)\right].
\end{equation}
In this equation, the scalar supermultiplet and supercovariant
derivatives have the form

$$
\Phi(\vec x, \theta)=\chi+\bar\theta\lambda+\frac12\bar\theta\theta F,~~
D_\alpha=\frac{\partial}{\partial\bar\theta_\alpha}-(\hat\partial\theta)_\alpha,~~
\bar D_\alpha=\frac{\partial}{\partial\theta_\alpha}-(\bar\theta\hat\partial)_\alpha\ .
$$
Here, $\chi$ again denotes the dual-photon field (real scalar), $\lambda$ is
the photino field, which is the two-component Majorana spinor
($\bar\lambda=\lambda^T\tau^2$), $F$ is an auxiliary scalar
field, $\hat\partial\equiv\gamma_i\partial_i$, and 
$\vec\gamma=\vec\tau$. The
monopole fugacity $\bar\zeta$ has dimensionality $[{\rm mass}]^{2}$ and is
exponentially small. In terms of the disorder operator, the action~(\ref{0su}) in component 
notations can be rewritten, up to an inessential constant, as
$$
S=\!\!\int\!\!
d^3x\left[\frac{2}{g_m^2}\left|\partial_\mu V\right|^2
\!-\frac{1}{2}\,\bar\lambda\hat\partial\lambda
-{g_m^2\zeta\over 2} \left(V^2\!+V^{*2}\right)\bar\lambda\lambda-
{(g_m\zeta)^2\over 2}\left(V^4\!+V^{*4}\right)\right],
$$
where $\zeta=\bar\zeta/4$. Besides the magnetic $Z_2$-symmetry, this action 
has an additional discrete parity symmetry
inherited from the full  supersymmetric GG action,
$$V(x_1,x_2,x_3)\rightarrow iV(-x_1,x_2,x_3),~~   
\lambda(x_1,x_2,x_3)\rightarrow\tau^3\lambda(-x_1,x_2,x_3)\ .
$$
The photino mass term is odd under the
parity transformation. Thus, the photino can acquire a
mass only if parity is spontaneously broken, that is indeed the case. 
The breaking of parity results in a
nonvanishing photino condensate
$\left<\bar\lambda\lambda\right>\sim g_m^2\zeta m_W$
and leads to a nonvanishing photino mass
$m=2g_m^2\zeta$ (equal to the Debye mass of the dual photon). One can prove~\cite{susy} that 
the equality of the dual-photon and the photino masses is preserved on the quantum level as well.

At finite temperatures, one can integrate photinos out in the dimensionally-reduced theory.
Again including W-bosons, we arrive at a theory similar to the one we had in the nonsupersymmetric case. Its action reads 
\begin{equation}
\label{susyDR}
S_{\rm d.-r.}=\int d^2x\left[\frac12(\partial_\mu\chi)^2-2\bar\xi\cos\left(2\sqrt{K}\chi\right)-
2\mu\cos\left(g\sqrt{\beta}\phi\right)\right],
\end{equation}
where $\bar\xi\propto\beta\zeta^2$ is a certain positive and exponentially small fugacity.
The crucial difference from the nonsupersymmetric GG model is due to the factor ``2'' in 
the term $\cos(2\sqrt{K}\chi)$, which makes the model~(\ref{susyDR}) $Z_4$-invariant.
Similarly to the nonsupersymmetric case, up to higher-loop corrections, 
the condition for the determination of the deconfinement phase transition temperature reads 
$4g_m^2T=g^2\beta$, that yields $T_c=\frac{g^2}{8\pi}$. The universality class of the phase transition 
is therefore $Z_4$ rather than $Z_2$, as it was in the nonsupersymmetric case. 
Further, it has been argued in Ref.~\cite{susy} that the parity order parameter, 
$\left<V^2\right>+\left<V^{*2}\right>$, vanishes at $T>T_c$.
Therefore, the parity restoring phase transition takes place at the same temperature $T_c$
as the deconfining transition. While this is an interesting
phenomenon, it seems to be somewhat non-generic. In particular, in
4D gauge theory with adjoint fermions there is no reason to
expect the deconfining and chiral-symmetry restoring phase
transitions to coincide. The physical order parameter for
deconfinement is the 't~Hooft loop $V$~\cite{thooftloop}, while
for chiral symmetry it is the fermionic bilinear form
$\bar\lambda\lambda$. In four dimensions, the two have a very 
different nature. While $\bar\lambda\lambda$ is a local field, $V$
is a string-like object. It is thus difficult to imagine these two order parameters
combining into a single one as it is the case in the
3D theory discussed above. The lattice results
indeed indicate that, at least in the SU(3)-theory, in four
dimensions the two transitions are distinct~\cite{adjoint}.

\setcounter{equation}{0}
\section{Summary}

Below we would like to mention once again some specific issues discussed in this review.

\begin{itemize}
\item
3D GG model at $T=0$, $N\ge 2$
\begin{itemize}
\item
$m_H<\infty$ $\Rightarrow$ $N$ must be smaller than $\exp[({\rm certain}~ {\rm constant})\times m_W/g^2]$, 
otherwise the Higgs vacuum is not stochastic;
\item
$m_H=\infty$ $\Rightarrow$ the string tension of the flat Wilson loop in the fundamental 
representation is obtained; it possesses an ambiguity in the numerical factor due to the exponentially large 
thickness of the string;
\item
the Kalb--Ramond field, which incorporates both monopoles and free photons $\Rightarrow$
the numerical factor at all the string coupling constants (in particular, at the string tension)
is fixed for an arbitrarily shaped surface,
in the weak-field (low-density) approximation; 
a generalization of the so-obtained theory of confining strings to the adjoint case 
and to $k$-strings: $\frac{\sigma_{\rm adj}}{\sigma_{\rm fund}}\simeq 2$ at $N\gg 1$;
$\frac{\sigma_k}{\sigma_{\rm fund}}=\frac{k(N-k)}{N-1}$ (Casimir scaling) in the low-density
approximation; the leading non-diluteness correction is derived, such that, at $k\sim N\gg 1$, 
it can significantly distort the Casimir scaling;
\item
4D-case with the field-theoretical $\theta$-term in the strong-coupling regime $\Rightarrow$
the string $\theta$-term; fundamental, adjoint, and $k$-case critical values of $\theta$, at which 
crumpling might disappear; a modification of the vacuum structure due to the $\theta$-term.
\end{itemize}
\item
3D GG model at $T>T_{\rm d.r.}\sim\zeta^{1/3}$, effects of W-bosons are neglected
\begin{itemize}
\item
$N=2$, $m_H<\infty$ $\Rightarrow$ $\frac{g^2}{m_W}$ may not be larger than a certain function of
$\frac{m_H}{m_W}$, otherwise the Higgs vacuum is not stochastic;
\item
Higgs-induced corrections to the BKT RG flow in the leading order in $m_H^{-1}$; $m_H$ itself evolves 
very slowly in the vicinity of the BKT critical point; $N>2$ $\Rightarrow$ unlike the SU(2)-case, the RG invariance 
holds only modulo the approximation $\sum\limits_{ij}^{}a_{ij}\cos[(\vec q_i-\vec q_j)\vec b]\simeq
\sum\limits_{i}^{}a_{ii}$, even at $m_H\to\infty$;
\item
in the presence of $N_f$ dynamical fundamental quark flavors, at $N_f=1$ and $m_q=0$, $T_{\rm BKT}=\frac{g^2}{4\pi}$;
at $N_f\ge 2$ and/or $m_q\ne 0$, any fundamental matter (including quarks themselves) is deconfined at $T>T_{\rm d.r.}$;
\end{itemize}
\item
W-bosons are taken into account, supersymmetric generalization $\Rightarrow$ the deconfining and the discrete-parity
restoring phase transitions occur at the same temperature, $\frac{g^2}{8\pi}$; the universality class
of the deconfining phase transition is the same as in $Z_4$-invariant spin models.
\end{itemize}
\begin{itemize}
\item
confining string  
\begin{itemize}
\item
the presence of the negative-stiffness term forces the introduction of a term into
the string action that suppresses the formation of spikes and, thus, prevents the
crumpling of the world sheet. This action, in the large-D limit, has an infrared-stable 
fixed point at zero stiffness, that corresponds to a tensionless smooth
string with Hausdorff dimension 2;
\item
the effective theory describing the infrared behavior of the confining string is
a conformal field theory with central charge $c=1$;
\item 
at high temperature, the free energy of the confining string goes like
$F^2(\beta) \propto -1/\beta^2$ with $\beta = 1/T$ and it agrees in sign,
temperature behavior, and the reality property with the large-$N$ QCD result
obtained by Polchinski.

\end{itemize}
\end{itemize}

\section*{Acknowledgments} 
D.A. is grateful to Profs.~A.~Di~Giacomo and D.~Ebert, as well as to the staffs of the Physics
Departments of the University of Pisa and of the Humboldt University of Berlin, where his part of this work has been 
respectively initiated and finalized,
for cordial hospitality. 
He also acknowledges the financial support of INFN and of the Alexander~von~Humboldt foundation
in the initial and final stages of this work, respectively. M.C.D. thanks CERN, where part of this review was written,
for kind hospitality.

\end{document}